\documentclass{elsart}

\usepackage{graphicx}

\usepackage{bm}% bold math
\usepackage{braket}% bra-ket (Dirac) notation

\usepackage{amssymb,amsbsy}
\usepackage{subfigure}% Sub-figures

\begin{document}

%\pdfpageheight\paperheight
%\pdfpagewidth\paperwidth
% You may need to change the horizontal offset to do what you
% want.  Setting \hoffset to a negative value moves all printed
% material to the left on all pages; setting it to a positive value
% moves all printed material to the right on all pages; not setting
% it keeps all printed material in it's default position.  \voffset
% is the vertical offset: use negative value for up; don't set if
% you want to use default position; use positive for down.

%\hoffset = +0.10truein
%\voffset = +0.10truein

\begin{frontmatter}



\title{On the Three-dimensional Central Moment Lattice Boltzmann Method}

\author[label1]{Kannan N. Premnath\corauthref{cor1}}
\ead{knandhap@uwyo.edu}
\author[label2]{Sanjoy Banerjee}
\address[label1]{Department of Mechanical Engineering, University of Wyoming, Laramie, WY 82071}
\address[label2]{Department of Chemical Engineering, City College of New York, City University of New York, New York, NY 10031}
\corauth[cor1]{Corresponding author.}

\author{}

\begin{abstract}
A three-dimensional (3D) lattice Boltzmann method based on central moments is derived. Two main elements are the local attractors
in the collision term and the source terms representing the effect of external and/or self-consistent internal forces. For suitable choices of the orthogonal moment basis for the three-dimensional, twenty seven velocity (D3Q27), and, its subset, fifteen velocity (D3Q15) lattice models, attractors are expressed in terms of factorization of lower order moments as suggested in an earlier work;
the corresponding source terms are specified to correctly influence lower order hydrodynamic fields, while avoiding aliasing effects for higher order moments. These are achieved by successively matching the corresponding continuous and discrete central moments at various orders, with the final expressions written in terms of raw moments via a transformation based on the binomial theorem.
Furthermore, to alleviate the discrete effects with the source terms, they are treated to be temporally semi-implicit
and second-order, with the implicitness subsequently removed by means of a transformation. As a result, the approach is frame-invariant by construction and its emergent dynamics describing fully 3D fluid motion in the presence of force fields
is Galilean invariant. Numerical experiments for a set of benchmark problems demonstrate its accuracy.
\end{abstract}


\end{frontmatter}

% main text
\section{Introduction} \label{section_introduction}
The use of discrete velocity models based on kinetic theory is a powerful theoretical approach and forms the basis of a modern
computational method for fluid mechanics. While the work of Broadwell~\cite{broadwell64} represents an early effort in this direction, careful exploitation of symmetries and local conservation laws to construct such models for discrete configuration spaces
underpinned the recent approaches, starting from the work of Frisch \emph{et al}~\cite{frisch86}. The latter led to the development
of the lattice Boltzmann method (LBM)~\cite{mcnamara88}, albeit without any direct connection to kinetic theory in its initial stages. Indeed, formal demonstration of this approach as a simplified model for the continuous Boltzmann equation~\cite{abe97,he97,shan98},
provided much impetus for recent developments, particularly for complex fluids~\cite{luo00,he02,asinari08} and for representation beyond continuum description~\cite{shan06}, among others (see~\cite{benzi92,chen98,succi01,yu03} for general reviews on the LBM).

The basic procedure involved in the LBM is represented by the synchronous free-streaming of particle distribution functions along discrete directions followed by collision, represented as a relaxation process. The latter has major influence on the physical fidelity as well as numerical stability. A popular approach is based on the single-relaxation time (SRT) model~\cite{qian92,chen92}. While it is successful in many applications, it is prone to numerical instability for situations with relatively low viscosities and is inadequate for representing certain physical phenomena (e.g. viscoelasticity and thermal transport) and in correctly accounting for kinetic layers near boundaries. In contrast, the use of multiple relaxation time (MRT) models~\cite{dhumieres92}, which are simplified versions of the relaxation LBM~\cite{higuera89a,higuera89b}, have addressed these aspects significantly. Its characteristic feature is that the relaxation process is carried out in moment space~\cite{grad49}. In particular, the relaxation times for the kinetic modes can be independently adjusted by means of a linear stability analysis to improve numerical stability~\cite{lallemand00,dhumieres02}.
Furthermore, based on the notion of duality between hydrodynamic and kinetic modes, a procedure for construction of matrix based LBM has been proposed recently~\cite{adhikari08}. From a different standpoint, non-linear stability can be enforced with a discrete H-theorem locally in the collision step using the SRT model~\cite{karlin99,ansumali02,succi02}. In this Entropic LBM, minimization of a convex H-function with hydrodynamic conservation constraints yields transcendental local attractors. It was also shown that the choice of the H-function in this context can be determined by enforcing Galilean invariance~\cite{boghosian03}. The construction based on the minimization of a convex function has been generalized to include a larger set of constraints that includes second-order moments yielding quasi-equilibrium attractors and thus allowing for a two-relaxation time Entropic LBM via a continuous H-theorem~\cite{asinari09,asinari10}. A theoretical basis for such an approach based on factorization symmetry considerations has been presented in~\cite{karlin10}. This work, along with others~\cite{chen08,chikatamarla09,shan10}, also provides rational procedures for constructing higher order models.

For general applicability of models and numerical schemes, it is necessary that their description of fluid behavior be the same in all inertial frames of reference. This important physical requirement of Galilean invariance, when not met can also lead to numerical instability in the context of the LBM. The latter arises from the fact that the degeneracies due to the finiteness of the standard lattice velocity sets can lead to linear dependence of higher order moments in terms of those at lower orders, which, in turn, can result in negative dependence of viscosity on fluid velocity~\cite{qian98}. Thus, it becomes necessary to consider large lattice velocity sets, which, however, by themselves do not guarantee in strictly observing Galilean invariance, as they can only lead to kinematically complete models~\cite{rubinstein08}. Proper selection of the collision model provides the sufficient or the dynamically complete condition in this regard to recover the correct physics, such as the Navier-Stokes equations. This can be seen by the use of unwieldy fitting of parameters~\cite{qian98} or elaborate construction procedures~\cite{chikatamarla09} for the attractors in the collision model with such extended lattice sets. Thus, the collision process still needs to be carefully designed and has an important role to play in the proper observation of Galilean invariance.

In this context, relaxation in a moving frame of reference, i.e. in terms of moments obtained by shifting the particle velocity with the local hydrodynamic velocity, or central moments~\cite{geier06}, provides a natural setting and a simple construction procedure to maintain Galilean invariance for a given velocity set. That is, the relaxation process is constructed to observe inertial frame invariance to a degree as permitted by the chosen lattice velocity set. We consider this specific meaning when we use the term Galilean invariance in this paper. Also, when different central moments are relaxed at different rates, i.e. formulated as a MRT model, it can enhance numerical stability by providing additional numerical dissipation similar to standard MRT models based on raw moments. It is noted that the ideas and procedures based on central moment relaxation are not restricted to standard lattice velocity sets, but can be used for any extended or kinematically complete velocity sets as well. The central moment approach exhibits a cascaded structure, which was shown to be equivalent to considering a generalized equilibrium in the lattice or rest frame of reference~\cite{asinari08a}. Furthermore, to further improve the physical fidelity, the local attractors for the central moments given in terms of their factorization into lower order moments has been proposed~\cite{geier09}. To incorporate the effect of force fields, which are important for numerous physical applications, a new approach for the source terms based on central moments was recently developed for a two-dimensional (2D) lattice~\cite{premnath09d}. In addition, a detailed theoretical basis for the central moment method, including a consistency analysis of the emergent fluid motion, was also provided~\cite{premnath09d}.

The objective of this work is the derivation and validation of a 3D central moment lattice Boltzmann method, with a particular focus on deriving Galilean invariant source terms, which are important, for example, in situations involving multiphase/multicomponent flows or turbulence modeling. In this regard, three-dimensional, twenty seven velocity (D3Q27) and its minimal subset, i.e. fifteen velocity (D3Q15) velocity lattices that can recover Navier-Stokes behavior are considered, and the overall procedure and notations used in~\cite{premnath09d} are adopted. The D3Q27 lattice is chosen so that our results provide the forcing scheme based on central moments to the overall formulation considered in~\cite{geier06}. It is noted that the notations and the details provided in that work~\cite{geier06} are cumbersome even for the collision model without forcing for implementation. On the other hand, in practice, the computational complexity is considerably reduced with the use of the D3Q15 lattice. Hence, the details with both the lattices are provided, with the smaller lattice set used in most of the computations in our validation studies. The overall procedure is as follows. Starting from suitable choices of the orthogonal moment basis for these lattice velocity sets, the continuous and discrete central moments of the local attractors and source terms at different orders are successively matched. The results are then transformed in terms of raw moments by means of the binomial theorem. To maintain physical coherence for the discrete velocity set, factorized local attractors for higher order central moments and temporally second-order accurate treatment of source terms are considered. This construction yields Galilean invariant representation of 3D fluid dynamics in the presence of general external or self-consistent internal forces. The computational approach thus derived is then assessed by comparison of its results for a set of canonical problems involving forcing for which analytical solutions are available.

The paper is organized as follows. Its main body containing the derivation focuses only on the essential steps involved in the derivation, choosing the D3Q27 lattice as an example, with the attendant details presented in various appendices (see Appendices \ref{app:d3q27matrix}-\ref{app:d3q15formulation}; the computational scheme for the D3Q15 lattice is presented in Appendix~\ref{app:d3q15formulation}). Section~\ref{sec:discreteparticlevelocity} discusses the choice made for the orthogonal moment basis corresponding to the D3Q27 lattice. The ansatz for the continuous central moments for the distribution functions, local attractors and sources due to the force fields are presented in Secs.~\ref{sec:ccentralmomentsfandfeq}. Section~\ref{sec:cascadedLBEforcing} provides the corresponding 3D lattice Boltzmann equation (LBE) with source terms based on central moments. Various discrete central moments needed for the construction of the central moment method are defined and the matching principle to preserve Galilean invariance is stated in Sec.~\ref{sec:discretecentralmoments}. Section~\ref{sec:rawmoments} obtains expressions for various discrete raw moments using the matching principle via the binomial theorem, including the derivation of the source terms in particle velocity space. The construction of the collision kernel is presented in Sec.~\ref{sec:cascadedcollisionforcing} and the overall procedure of the central moment LBM is provided in Sec.~\ref{sec:computationalprocedure}. Validation studies involving various canonical problems are discussed in Sec.~\ref{section_numerical_tests}. The conclusions are finally summarized in Sec.~\ref{sec:summary}.

\section{\label{sec:discreteparticlevelocity}Selection of Moment Basis}
\begin{figure}[h]
\begin{center}
\includegraphics[width = 140mm]{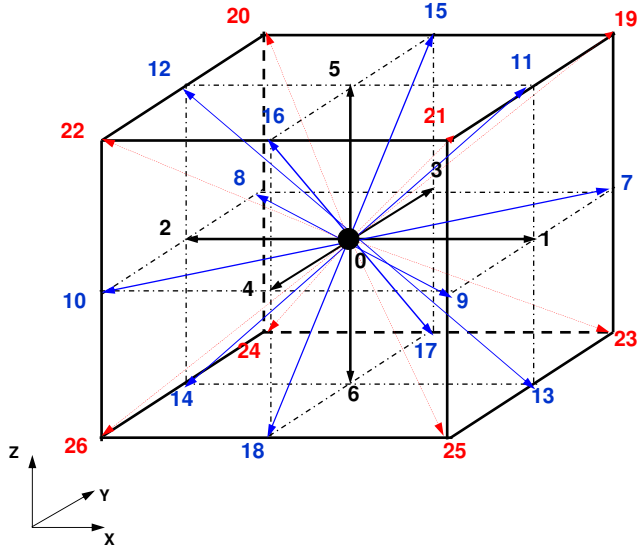}% Here is how to import EPS art
\caption{\label{fig:d3q27} Three-dimensional, twenty seven particle velocity (D3Q27) lattice.}
\end{center}
\end{figure}

We now discuss the moment basis, which is an important element on which the central moment LBM is constructed, corresponding to the three-dimensional, twenty seven velocity (D3Q27) lattice model (see Fig.~\ref{fig:d3q27}). The particle velocity for this lattice model $\overrightarrow{e}_{\alpha}$ is given by
\begin{equation}
\overrightarrow{e_{\alpha}} = \left\{\begin{array}{ll}
   {(0,0,0),}&{\alpha=0}\\
   {(\pm 1,0,0), (0,\pm 1,0), (0,0,\pm 1),}&{\alpha=1,\cdots,6}\\
   {(\pm 1,\pm 1,0), (\pm 1,0,\pm 1), (0,\pm 1,\pm 1),}&{\alpha=7,\cdots,18}\\
   {(\pm 1,\pm 1,\pm 1),}&{\alpha=19,\cdots,26}
\end{array} \right.
\label{eq:velocityd3qq27}
\end{equation}
%%%%% FIGURE %%%%%
%%%%% FIGURE %%%%%
For convenience, as in~\cite{premnath09d}, we use Dirac's bra-ket notion to represent the basis vectors, and Greek and Latin subscripts for particle velocity directions and Cartesian coordinate directions, respectively. By definition, the moments in the LBM are discrete integral properties of the distribution function $f_{\alpha}$, i.e. $\sum_{\alpha=0}^{26}e_{\alpha x}^m e_{\alpha y}^n e_{\alpha z}^pf_{\alpha}$, where $m$, $n$ and $p$ are integers, in 3D. As a result, we begin with the following twenty-seven non-orthogonal independent basis vectors obtained by combining monomials $e_{\alpha x}^m e_{\alpha y}^n e_{\alpha z}^p$ and arranged in an ascending order. First, the nominal basis for the conserved (zeroth and first order) moments follows immediately:
\begin{eqnarray}
\ket{T_{0}}&=&\ket{\rho}\equiv \ket{|\overrightarrow{e}_{\alpha}|^{0}},\nonumber\\
\ket{T_{1}}&=&\ket{e_{\alpha x}},\nonumber\\
\ket{T_{2}}&=&\ket{e_{\alpha y}},\nonumber\\
\ket{T_{3}}&=&\ket{e_{\alpha z}}.\nonumber
\end{eqnarray}
The basis for second-order moments are chosen such that it allows correct representation of the momentum flux (based on the Maxwell distribution, see below in Sec.~\ref{sec:ccentralmomentsfandfeq}) in the hydrodynamic equations. Three off-diagonal components ($\ket{T_{4}}$--$\ket{T_{6}}$) and three diagonal components ($\ket{T_{7}}$--$\ket{T_{9}}$) with $max(m,n,p)=1$ and $max(m,n,p)=2$,
respectively, while satisfying $m+n+p=2$ are considered:
\begin{eqnarray}
\ket{T_{4}}&=&\ket{e_{\alpha x}e_{\alpha y}},\nonumber\\
\ket{T_{5}}&=&\ket{e_{\alpha x}e_{\alpha z}},\nonumber\\
\ket{T_{6}}&=&\ket{e_{\alpha y}e_{\alpha z}},\nonumber\\
\ket{T_{7}}&=&\ket{e_{\alpha x}^2-e_{\alpha y}^2},\nonumber\\
\ket{T_{8}}&=&\ket{e_{\alpha x}^2-e_{\alpha z}^2},\nonumber\\
\ket{T_{9}}&=&\ket{e_{\alpha x}^2+e_{\alpha y}^2+e_{\alpha z}^2}.\nonumber
\end{eqnarray}
The following six third-order basis vectors for moments are chosen ($\ket{T_{10}}$--$\ket{T_{15}}$ with $max(m,n,p)=2$ and $\ket{T_{16}}$ with $max(m,n,p)=1$, while satisfying $m+n+p=3$):
\begin{eqnarray}
\ket{T_{10}}&=&\ket{e_{\alpha x}e_{\alpha y}^2+e_{\alpha x}e_{\alpha z}^2},\nonumber\\
\ket{T_{11}}&=&\ket{e_{\alpha x}^2e_{\alpha y}+e_{\alpha y}e_{\alpha z}^2},\nonumber\\
\ket{T_{12}}&=&\ket{e_{\alpha x}^2e_{\alpha z}+e_{\alpha y}^2e_{\alpha z}},\nonumber\\
\ket{T_{13}}&=&\ket{e_{\alpha x}e_{\alpha y}^2-e_{\alpha x}e_{\alpha z}^2},\nonumber\\
\ket{T_{14}}&=&\ket{e_{\alpha x}^2e_{\alpha y}-e_{\alpha y}e_{\alpha z}^2},\nonumber\\
\ket{T_{15}}&=&\ket{e_{\alpha x}^2e_{\alpha z}-e_{\alpha y}^2e_{\alpha z}},\nonumber\\
\ket{T_{16}}&=&\ket{e_{\alpha x}e_{\alpha y}e_{\alpha z}}.\nonumber
\end{eqnarray}
For the fourth-order basis vectors, we consider the following six of them ($\ket{T_{17}}$--$\ket{T_{22}}$ with $max(m,n,p)=2$ for $m+n+p=4$):
\begin{eqnarray}
\ket{T_{17}}&=&\ket{e_{\alpha x}^2e_{\alpha y}^2+e_{\alpha x}^2e_{\alpha z}^2+e_{\alpha y}^2e_{\alpha z}^2},\nonumber\\
\ket{T_{18}}&=&\ket{e_{\alpha x}^2e_{\alpha y}^2+e_{\alpha x}^2e_{\alpha z}^2-e_{\alpha y}^2e_{\alpha z}^2},\nonumber\\
\ket{T_{19}}&=&\ket{e_{\alpha x}^2e_{\alpha y}^2-e_{\alpha x}^2e_{\alpha z}^2},\nonumber\\
\ket{T_{20}}&=&\ket{e_{\alpha x}^2e_{\alpha y}e_{\alpha z}},\nonumber\\
\ket{T_{21}}&=&\ket{e_{\alpha x}e_{\alpha y}^2e_{\alpha z}},\nonumber\\
\ket{T_{22}}&=&\ket{e_{\alpha x}e_{\alpha y}e_{\alpha z}^2}.\nonumber
\end{eqnarray}
Finally, three fifth-order basis vectors ($\ket{T_{23}}$--$\ket{T_{25}}$) and one sixth-order basis vector ($\ket{T_{26}}$) are considered to complete moment basis corresponding to the D3Q27 model. In the above, in each case $max(m,n,p)=2$, with $m+n+p=5$ and $m+n+p=6$, respectively. Thus,
\begin{eqnarray}
\ket{T_{23}}&=&\ket{e_{\alpha x}e_{\alpha y}^2e_{\alpha z}^2},\nonumber\\
\ket{T_{24}}&=&\ket{e_{\alpha x}^2e_{\alpha y}e_{\alpha z}^2},\nonumber\\
\ket{T_{25}}&=&\ket{e_{\alpha x}^2e_{\alpha y}^2e_{\alpha z}},\nonumber\\
\ket{T_{26}}&=&\ket{e_{\alpha x}^2e_{\alpha y}^2e_{\alpha z}^2}.
\end{eqnarray}
Note that due to the finiteness of the particle velocity set, higher order longitudinal moments, i.e. $\ket{e_{\alpha i}^m}$ with $m \geq 3$ are eliminated from consideration in the above. The components of the basis vectors for the conserved moments corresponding to Eq.~(\ref{eq:velocityd3qq27}) may be written as
\begin{eqnarray}
\ket{\rho}\equiv \ket{|\overrightarrow{e}_{\alpha}|^{0}} &=&\left(1,1,1,1,1,1,1,1,1,1,1,1,1,1,1,1,1,1,1,1,1,\right.\nonumber\\
                &&\left.1,1,1,1,1,1\right)^\dagger, \nonumber \\
\ket{e_{\alpha x}} &=&\left(0,1,-1,0,0,0,0,1,-1,1,-1,1,-1,1,-1,0,0,0,0,1,-1,\right.\nonumber\\
                &&\left.1,-1,1,-1,1,-1\right)^\dagger, \nonumber \\
\ket{e_{\alpha y}} &=&\left(0,0,0,1,-1,0,0,1,1,-1,-1,0,0,0,0,1,-1,1,-1,1,1,\right.\nonumber\\
                &&\left.-1,-1,1,1,-1,-1\right)^\dagger,\nonumber \\
\ket{e_{\alpha z}} &=&\left(0,0,0,0,0,1,-1,0,0,0,0,1,1,-1,-1,1,1,-1,-1,1,1,\right.\nonumber\\
                &&\left.1,1,-1,-1,-1,-1\right)^\dagger.\nonumber
\end{eqnarray}
Here and henceforth, the superscript `$\dagger$' represents the transpose operator. The next key step is to transform the above non-orthogonal nominal basis set into an equivalent orthogonal set that would allow an efficient implementation~\cite{dhumieres92}.
This is accomplished by means of the standard Gram-Schmidt procedure for the above arrangement, i.e. in the increasing order of the monomials of the products of the Cartesian components of the particle velocities. As a result the components of the orthogonal basis
vectors are given by
\begin{eqnarray}
\ket{K_{0}}&=&\ket{\rho}\equiv \ket{|\overrightarrow{e}_{\alpha}|^{0}},\nonumber\\
\ket{K_{1}}&=&\ket{e_{\alpha x}},\nonumber\\
\ket{K_{2}}&=&\ket{e_{\alpha y}},\nonumber\\
\ket{K_{3}}&=&\ket{e_{\alpha z}},\nonumber\\
\ket{K_{4}}&=&\ket{e_{\alpha x}e_{\alpha y}},\nonumber\\
\ket{K_{5}}&=&\ket{e_{\alpha x}e_{\alpha z}},\nonumber\\
\ket{K_{6}}&=&\ket{e_{\alpha y}e_{\alpha z}},\nonumber\\
\ket{K_{7}}&=&\ket{e_{\alpha x}^2-e_{\alpha y}^2},\nonumber\\
\ket{K_{8}}&=&\ket{e_{\alpha x}^2+e_{\alpha y}^2+e_{\alpha z}^2}-3\ket{e_{\alpha z}^2},\nonumber\\
\ket{K_{9}}&=&\ket{e_{\alpha x}^2+e_{\alpha y}^2+e_{\alpha z}^2}-2\ket{\rho},\nonumber\\
\ket{K_{10}}&=&3\ket{e_{\alpha x}e_{\alpha y}^2+e_{\alpha x}e_{\alpha z}^2}-4\ket{e_{\alpha x}},\nonumber\\
\ket{K_{11}}&=&3\ket{e_{\alpha x}^2e_{\alpha y}+e_{\alpha y}e_{\alpha z}^2}-4\ket{e_{\alpha y}},\nonumber\\
\ket{K_{12}}&=&3\ket{e_{\alpha x}^2e_{\alpha z}+e_{\alpha y}^2e_{\alpha z}}-4\ket{e_{\alpha z}},\nonumber\\
\ket{K_{13}}&=&\ket{e_{\alpha x}e_{\alpha y}^2-e_{\alpha x}e_{\alpha z}^2},\nonumber\\
\ket{K_{14}}&=&\ket{e_{\alpha x}^2e_{\alpha y}-e_{\alpha y}e_{\alpha z}^2},\nonumber\\
\ket{K_{15}}&=&\ket{e_{\alpha x}^2e_{\alpha z}-e_{\alpha y}^2e_{\alpha z}},\nonumber\\
\ket{K_{16}}&=&\ket{e_{\alpha x}e_{\alpha y}e_{\alpha z}},\nonumber\\
\ket{K_{17}}&=&3\ket{e_{\alpha x}^2e_{\alpha y}^2+e_{\alpha x}^2e_{\alpha z}^2+e_{\alpha y}^2e_{\alpha z}^2}-4\ket{e_{\alpha x}^2+e_{\alpha y}^2+e_{\alpha z}^2}+4\ket{\rho},\nonumber\\
\ket{K_{18}}&=&3\ket{e_{\alpha x}^2e_{\alpha y}^2+e_{\alpha x}^2e_{\alpha z}^2-2e_{\alpha y}^2e_{\alpha z}^2}-2\ket{2e_{\alpha x}^2-e_{\alpha y}^2-e_{\alpha z}^2},\nonumber\\
\ket{K_{19}}&=&3\ket{e_{\alpha x}^2e_{\alpha y}^2-e_{\alpha x}^2e_{\alpha z}^2}-2\ket{e_{\alpha y}^2-e_{\alpha z}^2},\nonumber\\
\ket{K_{20}}&=&3\ket{e_{\alpha x}^2e_{\alpha y}e_{\alpha z}}-2\ket{e_{\alpha y}e_{\alpha z}},\nonumber\\
\ket{K_{21}}&=&3\ket{e_{\alpha x}e_{\alpha y}^2e_{\alpha z}}-2\ket{e_{\alpha x}e_{\alpha z}},\nonumber\\
\ket{K_{22}}&=&3\ket{e_{\alpha x}e_{\alpha y}e_{\alpha z}^2}-2\ket{e_{\alpha x}e_{\alpha y}},\nonumber\\
\ket{K_{23}}&=&9\ket{e_{\alpha x}e_{\alpha y}^2e_{\alpha z}^2}-6\ket{e_{\alpha x}e_{\alpha y}^2+e_{\alpha x}e_{\alpha z}^2}+4\ket{e_{\alpha x}},\nonumber\\
\ket{K_{24}}&=&9\ket{e_{\alpha x}^2e_{\alpha y}e_{\alpha z}^2}-6\ket{e_{\alpha x}^2e_{\alpha y}+e_{\alpha y}e_{\alpha z}^2}+4\ket{e_{\alpha y}},\nonumber\\
\ket{K_{25}}&=&9\ket{e_{\alpha x}^2e_{\alpha y}^2e_{\alpha z}}-6\ket{e_{\alpha x}^2e_{\alpha z}+e_{\alpha y}^2e_{\alpha z}}+4\ket{e_{\alpha z}},\nonumber\\
\ket{K_{26}}&=&27\ket{e_{\alpha x}^2e_{\alpha y}^2e_{\alpha z}^2}-18\ket{e_{\alpha x}^2e_{\alpha y}^2+e_{\alpha x}^2e_{\alpha z}^2+e_{\alpha y}^2e_{\alpha z}^2}\nonumber\\
&&+12\ket{e_{\alpha x}^2+e_{\alpha y}^2+e_{\alpha z}^2}-8\ket{\rho}.
\end{eqnarray}
This can be explicitly written in terms of a orthogonal matrix of moment basis $\mathcal{K}$ given by
\begin{eqnarray}
\mathcal{K}&=&\left[\ket{K_{0}},\ket{K_{1}},\ket{K_{2}},\ket{K_{3}},\ket{K_{4}},\ket{K_{5}},\ket{K_{6}},\ket{K_{7}},\ket{K_{8}}\right.\nonumber\\
           &&\left.\ket{K_{9}},\ket{K_{10}},\ket{K_{11}},\ket{K_{12}},\ket{K_{13}},\ket{K_{14}},\ket{K_{15}},\ket{K_{16}},\ket{K_{17}}\right].\nonumber\\
           &&\left.\ket{K_{18}},\ket{K_{19}},\ket{K_{20}},\ket{K_{21}},\ket{K_{22}},\ket{K_{23}},\ket{K_{24}},\ket{K_{25}},\ket{K_{26}}\right]
\label{eq:collisionmatrix1}
\end{eqnarray}
whose components are presented in Appendix~\ref{app:d3q27matrix}. Note that unlike the standard MRT formulation based on raw moments~\cite{dhumieres02}, which orders the basis vectors by considering the character of moments, i.e. increasing powers of their tensorial orders (i.e. scalars, vectors, tensors of different ranks,\ldots), the central moment basis vectors considered here are ordered according to their ascending powers (i.e. zeroth order moment, first order moments, second order moments,\ldots). Furthermore, the details of the basis vectors considered in this paper are different from those provided in~\cite{geier06}.

\section{\label{sec:ccentralmomentsfandfeq}Continuous Central Moments: Distribution Function, its Local Attractor and Forcing}
The central moment LBM, which is defined at the discrete level, should preserve certain continuous integral properties of the
distribution function $f$ given in terms of its central moments, i.e. those shifted by the macroscopic fluid velocity. In this regard, we first define \emph{continuous} central moment of $f$ of order $(m+n+p)$ as
\begin{equation}
\widehat{\Pi}_{x^my^nz^p}=\int_{-\infty}^{\infty}\int_{-\infty}^{\infty}\int_{-\infty}^{\infty}f(\xi_x-u_x)^m(\xi_y-u_y)^n(\xi_z-u_z)^pd\xi_xd\xi_yd\xi_z.
\label{eq:centralmomentfdefinition}
\end{equation}
Here, and in the rest of this paper, the use of ``hat" over a symbol represents quantities in the space of moments. The effect of collision
is to relax the distribution function, or equivalently, its central moments, towards its local attractor. The corresponding central moment local attractor may be written as
\begin{equation}
\widehat{\Pi}_{x^my^nz^p}^{at}=\int_{-\infty}^{\infty}\int_{-\infty}^{\infty}\int_{-\infty}^{\infty}f^{at}(\xi_x-u_x)^m(\xi_y-u_y)^n(\xi_z-u_z)^pd\xi_xd\xi_yd\xi_z.
\label{eq:centralmomentfattractor}
\end{equation}
Here $f^{at}$ is as yet unknown, and its effect on the dynamics will be determined in what follows. Similarly, the continuous
central moments due to sources may be written as
\begin{equation}
\widehat{\Gamma}_{x^my^nz^p}^{F}=\int_{-\infty}^{\infty}\int_{-\infty}^{\infty}\int_{-\infty}^{\infty}\Delta f^{F}(\xi_x-u_x)^m(\xi_y-u_y)^n(\xi_z-u_z)^pd\xi_xd\xi_yd\xi_z,
\label{eq:centralmomentforce}
\end{equation}
where $\Delta f^{F}$ is the change in the distribution function due to forcing, which will be specified later. One possibility is to consider the local Maxwellian as the attractor~\cite{geier06}. That is, consider
\begin{equation}
f^\mathcal{M}\equiv
f^\mathcal{M}(\rho,\overrightarrow{u},\xi_x,\xi_y,\xi_z)=\frac{\rho}{2\pi c_s^2}\exp\left[-\frac{\left(\overrightarrow{\xi}-\overrightarrow{u}\right)^2}{2c_s^2}\right],\label{eq:Maxwellian}
\end{equation}
where $c_s^2=1/3$, which yields corresponding continuous Maxwellian central moments as
\begin{equation}
\widehat{\Pi}^{\mathcal{M}}_{x^my^nz^p}=\int_{-\infty}^{\infty}\int_{-\infty}^{\infty}\int_{-\infty}^{\infty}f^{\mathcal{M}}(\xi_x-u_x)^m(\xi_y-u_y)^n(\xi_z-u_z)^p d\xi_xd\xi_yd\xi_z.\label{eq:centralmomentfeqdefinition}
\end{equation}
By virtue of the fact that $f^\mathcal{M}$ being an even function, $\widehat{\Pi}^{\mathcal{M}}_{x^my^nz^p}\neq0$ when $m$, $n$ and $p$ are even and $\widehat{\Pi}^{\mathcal{M}}_{x^my^nz^p}=0$ when $m$ or $n$ or $p$ is odd. Here and henceforth, the subscripts $x^my^nz^p$ mean
$xxx\cdots m\mbox{-times}$, $yyy\cdots n\mbox{-times}$ and $zzz\cdots p\mbox{-times}$. Thus,
\begin{eqnarray}
\widehat{\Pi}^{\mathcal{M}}_{0}&=&\rho, \nonumber\\
\widehat{\Pi}^{\mathcal{M}}_{i}&=&0, \nonumber\\
\widehat{\Pi}^{\mathcal{M}}_{ii}&=&c_s^2\rho,\nonumber\\
\widehat{\Pi}^{\mathcal{M}}_{ij}&=&0,\quad i \neq j,\nonumber\\
\widehat{\Pi}^{\mathcal{M}}_{ijj}&=&0,\quad i \neq j,\nonumber\\
\widehat{\Pi}^{\mathcal{M}}_{ijk}&=&0,\quad i \neq j \neq k,\nonumber\\
\widehat{\Pi}^{\mathcal{M}}_{iijj}&=&c_s^4\rho,\quad i \neq j,\nonumber\\
\widehat{\Pi}^{\mathcal{M}}_{iijk}&=&0,\quad i \neq j \neq k,\nonumber\\
\widehat{\Pi}^{\mathcal{M}}_{ijjkk}&=&0,\quad i \neq j \neq k,\nonumber\\
\widehat{\Pi}^{\mathcal{M}}_{iijjkk}&=&c_s^6\rho,\quad i \neq j \neq k.
\end{eqnarray}
Now, as discussed in~\cite{geier09} using $\widehat{\Pi}^{at}_{x^my^nz^p}=\widehat{\Pi}^{\mathcal{M}}_{x^my^nz^p}$ for all orders
leads to some inconsistency in recovering the macroscopic fluid equations. To circumvent this issue, we use a factorized
form for (central moment) attractors proposed in~\cite{geier09}. Essentially, in addition to satisfying Galilean invariance, the Maxwellian (equilibrium) satisfies the factorization property, i.e. independence along Cartesian coordinate directions, which
immediately applies to its central moments. In the factorized central moment formulation, this property is extended to model
non-equilibrium process, i.e. relaxation towards equilibrium. In other words, the higher order central moment attractors
are given as its factorization in terms of lower order central moments that are not yet in equilibrium~\cite{geier09}. To proceed
further, let us define the following post-collision continuous central moment of order $(m+n+p)$:
\begin{equation}
\widetilde{\widehat{\Pi}}_{x^my^nz^p}=\int_{-\infty}^{\infty}\int_{-\infty}^{\infty}\int_{-\infty}^{\infty}\widetilde{f}(\xi_x-u_x)^m(\xi_y-u_y)^n(\xi_z-u_z)^pd\xi_xd\xi_yd\xi_z.
\label{eq:centralmomentfdefinitionpostcollision}
\end{equation}
Then, we consider the factorized form for attractors as
\begin{eqnarray}
\widehat{\Pi}^{at}_{i}&=&\widetilde{\widehat{\Pi}}_{i}=0, \nonumber\\
\widehat{\Pi}^{at}_{ij}&=&\widetilde{\widehat{\Pi}}_{i}\widetilde{\widehat{\Pi}}_{j}=0, \nonumber\\
\widehat{\Pi}^{at}_{iij}&=&\widetilde{\widehat{\Pi}}_{ii}\widetilde{\widehat{\Pi}}_{j}=0, \nonumber\\
\widehat{\Pi}^{at}_{ijk}&=&\widetilde{\widehat{\Pi}}_{i}\widetilde{\widehat{\Pi}}_{j}\widetilde{\widehat{\Pi}}_{k}=0,\nonumber\\
\widehat{\Pi}^{at}_{iijj}&=&\widetilde{\widehat{\Pi}}_{ii}\widetilde{\widehat{\Pi}}_{jj},\nonumber\\
\widehat{\Pi}^{at}_{iijk}&=&\widetilde{\widehat{\Pi}}_{ii}\widetilde{\widehat{\Pi}}_{jk},\nonumber\\
\widehat{\Pi}^{at}_{iijjk}&=&\widetilde{\widehat{\Pi}}_{ii}\widetilde{\widehat{\Pi}}_{jj}\widetilde{\widehat{\Pi}}_{k}=0,\nonumber\\
\widehat{\Pi}^{at}_{iijjkk}&=&\widetilde{\widehat{\Pi}}_{ii}\widetilde{\widehat{\Pi}}_{jj}\widetilde{\widehat{\Pi}}_{kk}.
\end{eqnarray}
Now, however, to correctly recover the momentum flux and pressure tensor in the macroscopic fluid dynamical equations, the diagonal components of the second-order central moments should preserve those obtained from the Maxwellian. That is, we set $\widehat{\Pi}^{at}_{ii}=c_s^2\rho$. Thus, the $27$ independent components of the local factorized central moment attractors
can be written as
\begin{eqnarray}
\widehat{\Pi}^{at}_{0}&=&0,\widehat{\Pi}^{at}_{x}=\widehat{\Pi}^{at}_{y}=\widehat{\Pi}^{at}_{z}=0, \nonumber\\
\widehat{\Pi}^{at}_{xx}&=&\widehat{\Pi}^{at}_{yy}=\widehat{\Pi}^{at}_{zz}=c_s^2\rho, \nonumber\\
\widehat{\Pi}^{at}_{xy}&=&\widehat{\Pi}^{at}_{xz}=\widehat{\Pi}^{at}_{yz}=0,\nonumber\\
\widehat{\Pi}^{at}_{xyy}&=&\widehat{\Pi}^{at}_{xzz}=\widehat{\Pi}^{at}_{xxy}=\widehat{\Pi}^{at}_{yzz}=\widehat{\Pi}^{at}_{xxz}=\widehat{\Pi}^{at}_{yyz}=\widehat{\Pi}^{at}_{xyz}=0,\nonumber\\
\widehat{\Pi}^{at}_{xxyy}&=&\widetilde{\widehat{\Pi}}_{xx}\widetilde{\widehat{\Pi}}_{yy},\nonumber\\
\widehat{\Pi}^{at}_{xxzz}&=&\widetilde{\widehat{\Pi}}_{xx}\widetilde{\widehat{\Pi}}_{zz},\nonumber\\
\widehat{\Pi}^{at}_{yyzz}&=&\widetilde{\widehat{\Pi}}_{yy}\widetilde{\widehat{\Pi}}_{zz},\nonumber\\
\widehat{\Pi}^{at}_{xxyz}&=&\widehat{\Pi}^{at}_{xyyz}=\widehat{\Pi}^{at}_{xyzz}=0,\nonumber\\
\widehat{\Pi}^{at}_{xyyzz}&=&\widehat{\Pi}^{at}_{xxyzz}=\widehat{\Pi}^{at}_{xxyyz}=0,\nonumber\\
\widehat{\Pi}^{at}_{xxyyzz}&=&\widetilde{\widehat{\Pi}}_{xx}\widetilde{\widehat{\Pi}}_{yy}\widetilde{\widehat{\Pi}}_{zz}.
\end{eqnarray}
In essence, for the D3Q27 lattice, the fourth-order and sixth-order moments are factorized in terms of longitudinal second-order moments. It may be noted that symmetries in the factorization of the Maxwellian have been exploited to construct other types of quasi-equilibrium
attractors recently~\cite{karlin10}.

Similarly for the continuous source central moments due to force fields, one possible choice is obtained by choosing that based on
the local Maxwellian, i.e. $\Delta f^{F}=\frac{\overrightarrow{F}}{\rho}\cdot\frac{(\overrightarrow{\xi}-\overrightarrow{u})}{c_s^2}f^{\mathcal{M}}$,
which, however, leads to aliasing effects for higher order moments~\cite{premnath09d}. To circumvent this issue, a simple choice
involves de-aliasing higher order moments while preserving its necessary effect on the first-order central moments~\cite{premnath09d}
which is extended to 3D in this work. Thus, we specify the continuous source central moments as
\begin{equation}
\widehat{\Gamma}_{x^my^nz^p}^{F} = \left\{\begin{array}{ll}
   {F_x,}&{\quad m=1, n=0, p=0}\\
   {F_y,}&{\quad m=0, n=1, p=0}\\
   {F_z,}&{\quad m=0, n=0, p=1}\\
   {0,}&{\quad \mbox{Otherwise.}}
\end{array} \right.
\label{eq:forcingaliased}
\end{equation}

\section{\label{sec:cascadedLBEforcing}Central Moment Lattice-Boltzmann Equation with Forcing Terms}
Let us now write the central moment lattice Boltzmann equation (LBE) with forcing terms by first defining a \emph{discrete} distribution function supported by the discrete particle velocity set $\overrightarrow{e}_{\alpha}$ as
$\mathbf{f}=\ket{f_{\alpha}}=(f_0,f_1,f_2,\ldots,f_{26})^\dagger$, a collision operator as
$\mathbf{\Omega}^{c}=\ket{\Omega_{\alpha}^{c}}=(\Omega_{0}^{c},\Omega_{1}^{c},\Omega_{2}^{c},\ldots,\Omega_{26}^{c})$ as well as a source term as $\mathbf{S}=\ket{S_{\alpha}}=(S_0,S_1,S_2,\ldots,S_{26})^\dagger$ based on central moments. The LBE can then be obtained as a
discrete version of the continuous Boltzmann equation by temporally integrating along particle characteristics as~\cite{premnath09d}
\begin{equation}
f_{\alpha}(\overrightarrow{x}+\overrightarrow{e}_{\alpha},t+1)=f_{\alpha}(\overrightarrow{x},t)+\Omega_{{\alpha}(\overrightarrow{x},t)}^{c}+
\int_{t}^{t+1}S_{{\alpha}(\overrightarrow{x}+\overrightarrow{e}_{\alpha}\theta,t+\theta)}d\theta.
\label{eq:cascadedLBE1}
\end{equation}
In Eq.~(\ref{eq:cascadedLBE1}), the collision operator can be written in terms of the unknown collision kernel $\mathbf{\widehat{g}}$ projected to the orthogonal matrix of the moment basis as~\cite{geier06}
\begin{equation}
\Omega_{\alpha}^{c}\equiv \Omega_{\alpha}^{c}(\mathbf{f},\mathbf{\widehat{g}})=(\mathcal{K}\cdot \mathbf{\widehat{g}})_{\alpha},
\label{eq:cascadecollision1}
\end{equation}
where $\mathbf{\widehat{g}}=\ket{\widehat{g}_{\alpha}}=(\widehat{g}_0,\widehat{g}_1,\widehat{g}_2,\ldots,\widehat{g}_{26})^{\dagger}$, which will be derived later.
The macroscopic conserved moments, i.e. the local density and momentum, are obtained from the distribution function as
\begin{eqnarray}
\rho&=&\sum_{\alpha=0}^{26}f_{\alpha}=\braket{f_{\alpha}|\rho},\\
\rho u_i&=&\sum_{\alpha=0}^{26}f_{\alpha} e_{\alpha i}=\braket{f_{\alpha}|e_{\alpha i}}, i \in {x,y,z}.
\end{eqnarray}
We consider a semi-implicit representation for the source term, i.e. the last term in the above equation, Eq.~(\ref{eq:cascadedLBE1}), to provide second-order accuracy~\cite{premnath09d}, i.e.
$\int_{t}^{t+1}S_{{\alpha}(\overrightarrow{x}+\overrightarrow{e}_{\alpha}\theta,t+\theta)}d\theta=\frac{1}{2}\left[S_{{\alpha}(\overrightarrow{x},t)}+S_{{\alpha}(\overrightarrow{x}+\overrightarrow{e}_{\alpha},t+1)}\right]$. This equation is then made effectively explicit by using the transformation
$\overline{f}_{\alpha}=f_{\alpha}-\frac{1}{2}S_{\alpha}$ to reduce it to~\cite{premnath09d}
\begin{equation}
\overline{f}_{\alpha}(\overrightarrow{x}+\overrightarrow{e}_{\alpha},t+1)=\overline{f}_{\alpha}(\overrightarrow{x},t)+\Omega_{{\alpha}(\overrightarrow{x},t)}^{c}+
S_{{\alpha}(\overrightarrow{x},t)}.
\label{eq:cascadedLBE3}
\end{equation}
The explicit expressions for the source term $S_{\alpha}$ as well as the collision kernel $\mathbf{\widehat{g}}$ will be derived so as to rigorously enforce Galilean invariance through a matching principle and a binomial transformation. These are discussed in Secs.~\ref{sec:rawmoments} and ~\ref{sec:cascadedcollisionforcing}, respectively.

\section{\label{sec:discretecentralmoments}Various Discrete Central Moments and Galilean Invariance Matching Principle}
To facilitate the determination of the structure of the collision operator kernel $\mathbf{\widehat{g}}$ and the source terms $S_{\alpha}$,
we now define the following \emph{discrete} central moments of the distribution function, Maxwellian, and source term, respectively:
\begin{eqnarray}
\widehat{\kappa}_{x^m y^n z^p}&=&\braket{(e_{\alpha x}-u_x)^m(e_{\alpha y}-u_y)^n(e_{\alpha z}-u_z)^p|f_{\alpha}},\label{eq:centralmomentdistributionfunction1}\nonumber\\
\widehat{\kappa}_{x^m y^n z^p}^{at}&=&\braket{(e_{\alpha x}-u_x)^m(e_{\alpha y}-u_y)^n(e_{\alpha z}-u_z)^p|f_{\alpha}^{at}},\label{eq:centralmomentMaxwelldistribution1}\nonumber\\
\widehat{\sigma}_{x^m y^n z^p}&=&\braket{(e_{\alpha x}-u_x)^m(e_{\alpha y}-u_y)^n(e_{\alpha z}-u_z)^p|S_{\alpha}}\label{eq:centralmomentforcingterm1}.
\end{eqnarray}
Furthermore, the following definitions involving discrete central moments based on post-collision ($\widetilde{f}_{\alpha}$) and transformed ($\overline{f}_{\alpha}$) distribution functions, and its combination $\widetilde{\overline{f}}_{\alpha}$, are useful for further simplifications:
\begin{eqnarray}
\widetilde{\widehat{\kappa}}_{x^m y^n z^p}&=&\braket{(e_{\alpha x}-u_x)^m(e_{\alpha y}-u_y)^n(e_{\alpha z}-u_z)^p|\widetilde{f}_{\alpha}},\label{eq:postcentralmomentdistributionfunction1}\nonumber\\
\widehat{\overline{\kappa}}_{x^m y^n z^p}&=&\braket{(e_{\alpha x}-u_x)^m(e_{\alpha y}-u_y)^n(e_{\alpha z}-u_z)^p|\overline{f}_{\alpha}},\label{eq:transformedcentralmomentMaxwelldistribution1}\nonumber\\
\widetilde{\widehat{\overline{\kappa}}}_{x^m y^n z^p}&=&\braket{(e_{\alpha x}-u_x)^m(e_{\alpha y}-u_y)^n(e_{\alpha z}-u_z)^p|\widetilde{\overline{f}}_{\alpha}}\label{eq:posttransformedcentralmomentdistributionfunction1}.
\end{eqnarray}
Based on the definition of the transformed distribution function as given in the last section, it immediately
follows that
\begin{equation}
\widehat{\overline{\kappa}}_{x^m y^n z^p}=\widehat{\kappa}_{x^m y^n z^p}-\frac{1}{2}\widehat{\sigma}_{x^m y^n z^p}.
\label{eq:transformedcentralmomentsrelation}
\end{equation}

In order to preserve the main physical characteristic, i.e. Galilean invariance at the discrete level, we now invoke
the key matching principle, which is to set the \emph{discrete} central moments of the attractors
of the distribution function and the source terms, defined above, equal to their corresponding \emph{continuous} central moments,
whose forms are known exactly from the ansatz derived in Sec.~\ref{sec:ccentralmomentsfandfeq}. In other words,
\begin{eqnarray}
\widehat{\kappa}_{x^m y^n z^p}^{at}&=&\widehat{\Pi}^{at}_{x^m y^n z^p},\\
\widehat{\sigma}_{x^m y^n z^p}&=&\widehat{\Gamma}^{F}_{x^m y^n z^p}.
\end{eqnarray}
This yields the following expressions for the discrete local central moment attractors
\begin{eqnarray}                                                                                                                                                                                                                                                      \widehat{\kappa}^{at}_{0}&=&0,\widehat{\kappa}^{at}_{x}=\widehat{\kappa}^{at}_{y}=\widehat{\kappa}^{at}_{z}=0, \nonumber\\
\widehat{\kappa}^{at}_{xx}&=&\widehat{\kappa}^{at}_{yy}=\widehat{\kappa}^{at}_{zz}=c_s^2\rho, \nonumber\\
\widehat{\kappa}^{at}_{xy}&=&\widehat{\kappa}^{at}_{xz}=\widehat{\kappa}^{at}_{yz}=0,\nonumber\\
\widehat{\kappa}^{at}_{xyy}&=&\widehat{\kappa}^{at}_{xzz}=\widehat{\kappa}^{at}_{xxy}=\widehat{\kappa}^{at}_{yzz}=\widehat{\kappa}^{at}_{xxz}=\widehat{\kappa}^{at}_{yyz}=\widehat{\kappa}^{at}_{xyz}=0,\nonumber\\
\widehat{\kappa}^{at}_{xxyy}&=&\widetilde{\widehat{\kappa}}_{xx}\widetilde{\widehat{\kappa}}_{yy},\nonumber\\
\widehat{\kappa}^{at}_{xxzz}&=&\widetilde{\widehat{\kappa}}_{xx}\widetilde{\widehat{\kappa}}_{zz},\nonumber\\
\widehat{\kappa}^{at}_{yyzz}&=&\widetilde{\widehat{\kappa}}_{yy}\widetilde{\widehat{\kappa}}_{zz},\nonumber\\
\widehat{\kappa}^{at}_{xxyz}&=&\widehat{\kappa}^{at}_{xyyz}=\widehat{\kappa}^{at}_{xyzz}=0,\nonumber\\
\widehat{\kappa}^{at}_{xyyzz}&=&\widehat{\kappa}^{at}_{xxyzz}=\widehat{\kappa}^{at}_{xxyyz}=0,\nonumber\\
\widehat{\kappa}^{at}_{xxyyzz}&=&\widetilde{\widehat{\kappa}}_{xx}\widetilde{\widehat{\kappa}}_{yy}\widetilde{\widehat{\kappa}}_{zz}.
\end{eqnarray}                                                                                                                                                                                                                                                        In addition, the discrete source central moments satisfy the following
\begin{eqnarray}
\widehat{\sigma}_{0}=0,\widehat{\sigma}_{x}=F_x,\widehat{\sigma}_{y}=F_y,\widehat{\sigma}_{z}&=&F_z, \nonumber\\
\widehat{\sigma}_{x^m y^n z^p}&=&0,\quad m,n,p>1.\label{eq:discretecentralsources}
\end{eqnarray}
Thus, finally, in view of Eq.~(\ref{eq:transformedcentralmomentsrelation}), the attractors in terms of the transformed central moments can be written as
\begin{eqnarray}                                                                                                                                                                                                                                                      \widehat{\overline{\kappa}}^{at}_{0}&=&0,\widehat{\overline{\kappa}}^{at}_{x}=-\frac{1}{2}F_x,
\widehat{\overline{\kappa}}^{at}_{y}=-\frac{1}{2}F_y,\widehat{\overline{\kappa}}^{at}_{z}=-\frac{1}{2}F_z, \nonumber\\
\widehat{\overline{\kappa}}^{at}_{xx}&=&\widehat{\overline{\kappa}}^{at}_{yy}=\widehat{\overline{\kappa}}^{at}_{zz}=c_s^2\rho, \nonumber\\
\widehat{\overline{\kappa}}^{at}_{xy}&=&\widehat{\overline{\kappa}}^{at}_{xz}=\widehat{\overline{\kappa}}^{at}_{yz}=0,\nonumber\\
\widehat{\overline{\kappa}}^{at}_{xyy}&=&\widehat{\overline{\kappa}}^{at}_{xzz}=\widehat{\overline{\kappa}}^{at}_{xxy}=\widehat{\overline{\kappa}}^{at}_{yzz}=\widehat{\overline{\kappa}}^{at}_{xxz}=\widehat{\overline{\kappa}}^{at}_{yyz}=\widehat{\overline{\kappa}}^{at}_{xyz}=0,\nonumber\\
\widehat{\overline{\kappa}}^{at}_{xxyy}&=&\widetilde{\widehat{\kappa}}_{xx}\widetilde{\widehat{\kappa}}_{yy},\nonumber\\
\widehat{\overline{\kappa}}^{at}_{xxzz}&=&\widetilde{\widehat{\kappa}}_{xx}\widetilde{\widehat{\kappa}}_{zz},\nonumber\\
\widehat{\overline{\kappa}}^{at}_{yyzz}&=&\widetilde{\widehat{\kappa}}_{yy}\widetilde{\widehat{\kappa}}_{zz},\nonumber\\
\widehat{\overline{\kappa}}^{at}_{xxyz}&=&\widehat{\overline{\kappa}}^{at}_{xyyz}=\widehat{\overline{\kappa}}^{at}_{xyzz}=0,\nonumber\\
\widehat{\overline{\kappa}}^{at}_{xyyzz}&=&\widehat{\overline{\kappa}}^{at}_{xxyzz}=\widehat{\overline{\kappa}}^{at}_{xxyyz}=0,\nonumber\\
\widehat{\overline{\kappa}}^{at}_{xxyyzz}&=&\widetilde{\widehat{\kappa}}_{xx}\widetilde{\widehat{\kappa}}_{yy}\widetilde{\widehat{\kappa}}_{zz}.\label{eq:transformedcentralmomentattractors}
\end{eqnarray}

\section{\label{sec:rawmoments}Various Discrete Raw Moments and Source Terms in Particle Velocity Space}
In order to construct an executable central moment LBM, the above information based on the central moments need
to be related to the raw moments, i.e. those in the usual lattice or rest frame of reference. This can be
readily accomplished by means of the binomial theorem applied to the orthogonal products of the discrete
quantities supported by the particle velocity set~\cite{premnath09d}. In this regard, the following notations
that specify various \emph{discrete raw} moments will be useful:
\begin{eqnarray}
\widehat{\kappa}_{x^m y^n z^p}^{'}&=&\sum_{\alpha}f_{\alpha}e_{\alpha x}^m e_{\alpha y}^n e_{\alpha z}^p=\braket{e_{\alpha x}^m e_{\alpha y}^n e_{\alpha z}^p|f_{\alpha}},\label{eq:rawmomentdistributionfunction1}\\
\widehat{\overline{\kappa}}_{x^m y^n z^p}^{'}&=&\sum_{\alpha}\overline{f}_{\alpha}e_{\alpha x}^m e_{\alpha y}^n e_{\alpha z}^p=\braket{e_{\alpha x}^m e_{\alpha y}^n e_{\alpha z}^p|\overline{f}_{\alpha}},\label{eq:rawmomenttransformeddistribution1}\\
\widehat{\sigma}_{x^m y^n z^p}^{'}&=&\sum_{\alpha}S_{\alpha}e_{\alpha x}^m e_{\alpha y}^n e_{\alpha z}^p=\braket{e_{\alpha x}^m e_{\alpha y}^n e_{\alpha z}^p|S_{\alpha}}\label{eq:rawmomentforcingterm1}.
\end{eqnarray}
Here and in what follows, the superscript ``prime'' ($'$) is used to distinguish the raw moments from the central moments that are designated without the primes. Furthermore, similar to Eq.~(\ref{eq:transformedcentralmomentsrelation}), the relation $\widehat{\overline{\kappa}}_{x^m y^n}^{'}=\widehat{\kappa}_{x^m y^n z^p}^{'}-\frac{1}{2}\widehat{\sigma}_{x^m y^n z^p}^{'}$ is satisfied.
Based on the above, first, we write the raw moments of the distribution function of different orders supported by the particle velocity
set $\widehat{\overline{\kappa}}_{x^m y^n z^p}^{'}=\braket{\overline{f}_{\alpha}|e_{\alpha x}^m e_{\alpha y}^n e_{\alpha z}^p}$ in terms of the known quantities. To obtain a compact description of results, the following operator notation is helpful~\cite{premnath09d}:
\begin{eqnarray}
a(\overline{f}_{\alpha_1}+\overline{f}_{\alpha_3}+\overline{f}_{\alpha_3}+\cdots)&+&
b(\overline{f}_{\beta_1}+\overline{f}_{\beta_2}+\overline{f}_{\beta_3}+\cdots)+\cdots\nonumber\\
&&=\left(a\sum_{\alpha}^A+b\sum_{\alpha}^B\cdots\right)\otimes \overline{f}_{\alpha},\label{eq:summationoperator}
\end{eqnarray}
where $A=\left\{\alpha_1,\alpha_2,\alpha_3,\cdots\right\}$,~$B=\left\{\beta_1,\beta_2,\beta_3,\cdots\right\}$,$\cdots$.
First, the conserved transformed raw moments follows directly from the definition as
\begin{eqnarray}
\widehat{\overline{\kappa}}_{0}^{'}&=&\braket{\overline{f}_{\alpha}|\rho}=\rho, \qquad \qquad \qquad \quad
\widehat{\overline{\kappa}}_{x}^{'}=\braket{\overline{f}_{\alpha}|e_{\alpha x}}=\rho u_x-\frac{1}{2}F_x,\nonumber\\
\widehat{\overline{\kappa}}_{y}^{'}&=&\braket{\overline{f}_{\alpha}|e_{\alpha y}}=\rho u_y-\frac{1}{2}F_y,\qquad
\widehat{\overline{\kappa}}_{z}^{'}=\braket{\overline{f}_{\alpha}|e_{\alpha z}}=\rho u_z-\frac{1}{2}F_z.\label{eq:rawmomenttransformedconserved}
\end{eqnarray}
The non-conserved transformed raw moments can be written, using the above operator notation (Eq.~(\ref{eq:summationoperator})), in terms of the subsets of the particle velocity directions, which are presented in Appendix~\ref{app:nonconservedrawmoments}.

The next step is to transform the central moments of the source terms (Eq.~(\ref{eq:discretecentralsources})) in terms of raw moments by using the definitions, i.e. Eq.~(\ref{eq:centralmomentforcingterm1}) and (\ref{eq:rawmomentforcingterm1}), which by the binomial theorem, readily yields the expressions that are enumerated in Appendix~\ref{app:rawsourcemoments}. These moments should be related to the discrete source terms in particle velocity space so that an operational Galilean invariant approach can be derived. To accomplish this, we first obtain a set of intermediate quantities $\widehat{m}^{s}_{\beta}$, which are the projections of the source terms to the orthogonal matrix of the moment basis $\mathcal{K}$, i.e. $\widehat{m}^{s}_{\beta}=\braket{K_{\beta}|S_{\alpha}}$, $\beta=0,1,2,\ldots, 26$, which can be obtained from the above using Eqs.~(\ref{eq:collisionmatrix1}) and (\ref{eq:rawmomentsourceterm}). The details of $\widehat{m}^{s}_{\beta}$ are provided in Appendix~\ref{app:rawsourcemomentsorthogonalrojections}.

It is noted that $\widehat{m}^{s}_{\beta}$ is equivalent to the following matrix formulation
\begin{eqnarray}
\mathcal{K}^\dagger\mathbf{S}=(\mathcal{K}\cdot\mathbf{S})_{\alpha}&=&(\braket{K_0|S_{\alpha}},\braket{K_1|S_{\alpha}},\braket{K_2|S_{\alpha}},\ldots,\braket{K_{26}|S_{\alpha}}) \nonumber \\
&=& (\widehat{m}^{s}_{0},\widehat{m}^{s}_{1},\widehat{m}^{s}_{2},\ldots,\widehat{m}^{s}_{26})^{\dagger}\equiv \ket{\widehat{m}^{s}_{\alpha}}, \label{eq:sourceformulation1}
\end{eqnarray}
which can be exactly inverted by using the following orthogonal property of $\mathcal{K}$, i.e.
$\mathcal{K}^{-1}=\mathcal{K}^\dagger \cdot \mathcal{D}^{-1}$, where $\mathcal{D}$ is the diagonal matrix given by
$\mathcal{D}=\mbox{diag}(\braket{K_0|K_0},\braket{K_1|K_1},\braket{K_2|K_2},\ldots,\braket{K_{26}|K_{26}})$~\cite{premnath09d}. Exploiting this fact, the linear system (Eq.~(\ref{eq:sourceformulation1})) can be solved exactly to yield the expressions for the
Galilean invariant source terms in velocity space $S_{\alpha}$ in terms of $\widehat{m}^{s}_{\beta}$, or equivalently the force $\overrightarrow{F}$ and velocity fields $\overrightarrow{u}$. The final results of  $S_{\alpha}$, where $\alpha=0,1,2,\ldots, 26$ are summarized in Appendix~\ref{app:sourcetermsvelocityspace}.

Finally, to obtain the collision kernel $\widehat{g}_{\beta}$ in the next section, we need to evaluate the expressions for its raw moments of various orders projected to the moment basis matrix $\mathcal{K}$, i.e.
\begin{equation}
\sum_{\alpha}(\mathcal{K}\cdot \mathbf{\widehat{g}})_{\alpha}e_{\alpha x}^m e_{\alpha y}^n e_{\alpha z}^p= \sum_{\beta} \braket{K_{\beta}|e_{\alpha x}^m e_{\alpha y}^n e_{\alpha z}^p}\widehat{g}_{\beta}.
\end{equation}
For conserved moments, it follows by definition that $\widehat{g}_{0}=\widehat{g}_{1}=\widehat{g}_{2}=\widehat{g}_{3}=0$.
Again, exploiting the orthogonal property of $\mathcal{K}$, the moments of the collision kernel can be obtained which are presented in Appendix~\ref{app:momentscollisionkernel}.

The central moment LBE given in Eq.~(\ref{eq:cascadedLBE3}) can be rewritten in terms of the collision and streaming steps, respectively, as
\begin{eqnarray}
\widetilde{\overline{f}}_{\alpha}(\overrightarrow{x},t)&=&\overline{f}_{\alpha}(\overrightarrow{x},t)+\Omega_{{\alpha}(\overrightarrow{x},t)}^{c}+
S_{{\alpha}(\overrightarrow{x},t)},\label{eq:cascadedcollision1}\\
\overline{f}_{\alpha}(\overrightarrow{x}+\overrightarrow{e}_{\alpha},t+1)&=&\widetilde{\overline{f}}_{\alpha}(\overrightarrow{x},t), \label{eq:cascadedstreaming1}
\end{eqnarray}
where the symbol ``tilde" ($\sim$) in the first equation refers to the post-collision state. Furthermore, the conserved local fluid density and momentum are finally written in terms of the moments of the transformed distribution functions as
\begin{eqnarray}
\rho&=&\sum_{\alpha=0}^{26}\overline{f}_{\alpha}=\braket{\overline{f}_{\alpha}|\rho},\label{eq:densitycalculation}\\
\rho u_i&=&\sum_{\alpha=0}^{26}\overline{f}_{\alpha} e_{\alpha i}+\frac{1}{2}F_i=\braket{\overline{f}_{\alpha}|e_{\alpha i}}+\frac{1}{2}F_i, \qquad i \in {x,y,z}. \label{eq:velocitycalculation}
\end{eqnarray}

\section{\label{sec:cascadedcollisionforcing}Structure of the Central Moment Collision Operator}
We are now in a position to obtain the expressions for the collision kernel of the 3D central moment LBM in the presence of source terms. In essence, the procedure begins with the lowest order (i.e. second-order, off-diagonal) post-collision central moments (i.e. $\widetilde{\widehat{\overline{\kappa}}}_{xy}, \widetilde{\widehat{\overline{\kappa}}}_{xz}$ and $\widetilde{\widehat{\overline{\kappa}}}_{yz}$), which are successively set equal to the corresponding attractors given in Eq.~(\ref{eq:transformedcentralmomentattractors}) (i.e. $\widehat{\overline{\kappa}}_{xy}^{at}, \widehat{\overline{\kappa}}_{xz}^{at}$ and $\widehat{\overline{\kappa}}_{yz}^{at}$, respectively). This intermediate step is based on an equilibrium assumption. Dropping this modeling assumption to represent collision as a relaxation process by multiplying with a corresponding relaxation parameter results in the collision kernels $\widehat{g}_{\alpha}$ for a given order~\cite{geier06}. Here, the relaxation parameter needs to be carefully applied to only those terms that are not yet in post-collision states, i.e. those that do not contain $\widehat{g}_{\beta}$, where $\beta=0,1,2,\ldots,\alpha-1$ for a given $\widehat{g}_{\alpha}$. Then the results are transformed in terms of raw moments via the binomial theorem to yield expressions
useful for computations. The details of various elements in obtaining the collision kernel are presented in~\cite{premnath09d}. To simplify exposition, let us introduce the following notation:
\begin{equation}
\widehat{\overline{\eta}}_{x^my^nz^p}^{'}=\widehat{\overline{\kappa}}_{x^my^nz^p}^{'}+\widehat{\sigma}_{x^my^nz^p}^{'},
\end{equation}
where the expressions for $\widehat{\overline{\kappa}}_{x^my^nz^p}^{'}$ and $\widehat{\sigma}_{x^my^nz^p}^{'}$ are known from
Sec.~\ref{sec:rawmoments}. In the following, for brevity, we present only the final results. For the above three off-diagonal central
moments, we get
\begin{equation}
\widehat{g}_4=\frac{\omega_4}{12}\left[-\widehat{\overline{\eta}}_{xy}^{'}+\rho u_xu_y+\frac{1}{2}(\widehat{\sigma}_{x}^{'}u_y+\widehat{\sigma}_{y}^{'}u_x)\right],\label{eq:collisionkernelg4}
\end{equation}
\begin{equation}
\widehat{g}_5=\frac{\omega_5}{12}\left[-\widehat{\overline{\eta}}_{xz}^{'}+\rho u_xu_z+\frac{1}{2}(\widehat{\sigma}_{x}^{'}u_z+\widehat{\sigma}_{z}^{'}u_x)\right],\label{eq:collisionkernelg5}
\end{equation}
\begin{equation}
\widehat{g}_6=\frac{\omega_6}{12}\left[-\widehat{\overline{\eta}}_{yz}^{'}+\rho u_yu_z+\frac{1}{2}(\widehat{\sigma}_{y}^{'}u_z+\widehat{\sigma}_{z}^{'}u_y)\right].\label{eq:collisionkernelg6}
\end{equation}
where $\omega_4$, $\omega_5$ and $\omega_6$ are relaxation parameters. Similarly, applying the procedure for the remaining three
second-order diagonal components with $\widehat{\overline{\kappa}}_{xx}^{at}=\widehat{\overline{\kappa}}_{yy}^{at}=\widehat{\overline{\kappa}}_{zz}^{at}=c_s^2\rho$,
which preserves the Maxwellian values to provide the correct momentum flux and pressure tensor, yields\begin{equation}
\widehat{g}_7=\frac{\omega_7}{12}\left[-(\widehat{\overline{\eta}}_{xx}^{'}-\widehat{\overline{\eta}}_{yy}^{'})+\rho (u_x^2-u_y^2)+(\widehat{\sigma}_{x}^{'}u_x-\widehat{\sigma}_{y}^{'}u_y)\right],\label{eq:collisionkernelg7}
\end{equation}
\begin{eqnarray}
\widehat{g}_8&=&\frac{\omega_8}{36}\left[-(\widehat{\overline{\eta}}_{xx}^{'}+\widehat{\overline{\eta}}_{yy}^{'}-2\widehat{\overline{\eta}}_{zz}^{'})+
\rho (u_x^2+u_y^2-2u_z^2)\right.\nonumber\\
&&\left.+(\widehat{\sigma}_{x}^{'}u_x+\widehat{\sigma}_{y}^{'}u_y-2\widehat{\sigma}_{z}^{'}u_z)\right],\label{eq:collisionkernelg8}
\end{eqnarray}
\begin{eqnarray}
\widehat{g}_9&=&\frac{\omega_9}{18}\left[-(\widehat{\overline{\eta}}_{xx}^{'}+\widehat{\overline{\eta}}_{yy}^{'}+\widehat{\overline{\eta}}_{zz}^{'})+
\rho (u_x^2+u_y^2+u_z^2)\right.\nonumber\\
&&\left.+(\widehat{\sigma}_{x}^{'}u_x+\widehat{\sigma}_{y}^{'}u_y+\widehat{\sigma}_{z}^{'}u_z)+\rho\right].\label{eq:collisionkernelg9}
\end{eqnarray}
Next, carrying out the above matching, transformation, and relaxation steps (with the last of this applicable only for the pre-collision terms) successively to all the seven components of the third-order moments we get
\begin{eqnarray}
\widehat{g}_{10}&=&\frac{\omega_{10}}{24}\left[-(\widehat{\overline{\eta}}_{xyy}^{'}+\widehat{\overline{\eta}}_{xzz}^{'})+
2(u_y\widehat{\overline{\eta}}_{xy}^{'}+u_z\widehat{\overline{\eta}}_{xz}^{'})+
u_x(\widehat{\overline{\eta}}_{yy}^{'}+\widehat{\overline{\eta}}_{zz}^{'})\right.\nonumber\\
&&\left.-2\rho u_x(u_y^2+u_z^2)-\frac{1}{2}\widehat{\sigma}_{x}^{'}(u_y^2+u_z^2)
-u_x(\widehat{\sigma}_{y}^{'}u_y+\widehat{\sigma}_{z}^{'}u_z)\right]\nonumber\\
&&+(u_y\widehat{g}_4+u_z\widehat{g}_5)+\frac{1}{4}u_x(-\widehat{g}_7-\widehat{g}_8+2\widehat{g}_9),\label{eq:collisionkernelg10}
\end{eqnarray}
\begin{eqnarray}
\widehat{g}_{11}&=&\frac{\omega_{11}}{24}\left[-(\widehat{\overline{\eta}}_{xxy}^{'}+\widehat{\overline{\eta}}_{yzz}^{'})+
2(u_x\widehat{\overline{\eta}}_{xy}^{'}+u_z\widehat{\overline{\eta}}_{yz}^{'})+
u_y(\widehat{\overline{\eta}}_{xx}^{'}+\widehat{\overline{\eta}}_{zz}^{'})\right.\nonumber\\
&&\left.-2\rho u_y(u_x^2+u_z^2)-\frac{1}{2}\widehat{\sigma}_{y}^{'}(u_x^2+u_z^2)
-u_y(\widehat{\sigma}_{x}^{'}u_x+\widehat{\sigma}_{z}^{'}u_z)\right]\nonumber\\
&&+(u_x\widehat{g}_4+u_z\widehat{g}_6)+\frac{1}{4}u_y(\widehat{g}_7-\widehat{g}_8+2\widehat{g}_9),\label{eq:collisionkernelg11}
\end{eqnarray}
\begin{eqnarray}
\widehat{g}_{12}&=&\frac{\omega_{12}}{24}\left[-(\widehat{\overline{\eta}}_{xxz}^{'}+\widehat{\overline{\eta}}_{yyz}^{'})+
2(u_x\widehat{\overline{\eta}}_{xz}^{'}+u_y\widehat{\overline{\eta}}_{yz}^{'})+
u_z(\widehat{\overline{\eta}}_{xx}^{'}+\widehat{\overline{\eta}}_{yy}^{'})\right.\nonumber\\
&&\left.-2\rho u_z(u_x^2+u_y^2)-\frac{1}{2}\widehat{\sigma}_{z}^{'}(u_x^2+u_y^2)
-u_z(\widehat{\sigma}_{x}^{'}u_x+\widehat{\sigma}_{y}^{'}u_y)\right]\nonumber\\
&&+(u_x\widehat{g}_5+u_y\widehat{g}_6)+\frac{1}{2}u_z(\widehat{g}_8+\widehat{g}_9),\label{eq:collisionkernelg12}
\end{eqnarray}
\begin{eqnarray}
\widehat{g}_{13}&=&\frac{\omega_{13}}{8}\left[-(\widehat{\overline{\eta}}_{xyy}^{'}-\widehat{\overline{\eta}}_{xzz}^{'})+
2(u_y\widehat{\overline{\eta}}_{xy}^{'}-u_z\widehat{\overline{\eta}}_{xz}^{'})+
u_x(\widehat{\overline{\eta}}_{yy}^{'}-\widehat{\overline{\eta}}_{zz}^{'})\right.\nonumber\\
&&\left.-2\rho u_x(u_y^2-u_z^2)-\frac{1}{2}\widehat{\sigma}_{x}^{'}(u_y^2-u_z^2)
-u_x(\widehat{\sigma}_{y}^{'}u_y-\widehat{\sigma}_{z}^{'}u_z)\right]\nonumber\\
&&+3(u_y\widehat{g}_4-u_z\widehat{g}_5)+\frac{3}{4}u_x(-\widehat{g}_7+3\widehat{g}_8),\label{eq:collisionkernelg13}
\end{eqnarray}
\begin{eqnarray}
\widehat{g}_{14}&=&\frac{\omega_{14}}{8}\left[-(\widehat{\overline{\eta}}_{xxy}^{'}-\widehat{\overline{\eta}}_{yzz}^{'})+
2(u_x\widehat{\overline{\eta}}_{xy}^{'}-u_z\widehat{\overline{\eta}}_{yz}^{'})+
u_y(\widehat{\overline{\eta}}_{xx}^{'}-\widehat{\overline{\eta}}_{zz}^{'})\right.\nonumber\\
&&\left.-2\rho u_y(u_x^2-u_z^2)-\frac{1}{2}\widehat{\sigma}_{y}^{'}(u_x^2-u_z^2)
-u_y(\widehat{\sigma}_{x}^{'}u_x-\widehat{\sigma}_{z}^{'}u_z)\right]\nonumber\\
&&+3(u_x\widehat{g}_4-u_z\widehat{g}_6)+\frac{3}{4}u_y(\widehat{g}_7+3\widehat{g}_8),\label{eq:collisionkernelg14}
\end{eqnarray}
\begin{eqnarray}
\widehat{g}_{15}&=&\frac{\omega_{15}}{8}\left[-(\widehat{\overline{\eta}}_{xxz}^{'}-\widehat{\overline{\eta}}_{yyz}^{'})+
2(u_x\widehat{\overline{\eta}}_{xz}^{'}-u_y\widehat{\overline{\eta}}_{yz}^{'})+
u_z(\widehat{\overline{\eta}}_{xx}^{'}-\widehat{\overline{\eta}}_{yy}^{'})\right.\nonumber\\
&&\left.-2\rho u_z(u_x^2-u_y^2)-\frac{1}{2}\widehat{\sigma}_{z}^{'}(u_x^2-u_y^2)
-u_z(\widehat{\sigma}_{x}^{'}u_x-\widehat{\sigma}_{y}^{'}u_y)\right]\nonumber\\
&&+3(u_x\widehat{g}_5-u_y\widehat{g}_6)+\frac{3}{2}u_z\widehat{g}_7,\label{eq:collisionkernelg15}
\end{eqnarray}
\begin{eqnarray}
\widehat{g}_{16}&=&\frac{\omega_{16}}{8}\left[-\widehat{\overline{\eta}}_{xyz}^{'}+u_x\widehat{\overline{\eta}}_{yz}^{'}+
u_y\widehat{\overline{\eta}}_{xz}^{'}+u_z\widehat{\overline{\eta}}_{xy}^{'}-2\rho u_xu_yu_z\right.\nonumber\\
&&\left.-\frac{1}{2}(\widehat{\sigma}_{x}^{'}u_yu_z+\widehat{\sigma}_{y}^{'}u_xu_z+\widehat{\sigma}_{z}^{'}u_xu_y)\right]\nonumber\\
&&+\frac{3}{2}(u_z\widehat{g}_4+u_y\widehat{g}_5+u_z\widehat{g}_6),\label{eq:collisionkernelg16}
\end{eqnarray}

Notice that the cascaded structure is evident for the collision kernel starting from the third-order moments. Now, the next three diagonal components of the fourth-order central moments needs to carefully consider the non-zero factorized attractors given in terms of second-order components, i.e.
$\widehat{\overline{\kappa}}^{at}_{xxyy}=\widetilde{\widehat{\kappa}}_{xx}\widetilde{\widehat{\kappa}}_{yy}$,
$\widehat{\overline{\kappa}}^{at}_{xxzz}=\widetilde{\widehat{\kappa}}_{xx}\widetilde{\widehat{\kappa}}_{zz}$, and
$\widehat{\overline{\kappa}}^{at}_{yyzz}=\widetilde{\widehat{\kappa}}_{yy}\widetilde{\widehat{\kappa}}_{zz}$ (see Eq.~(\ref{eq:transformedcentralmomentattractors})). This yields the corresponding collision kernels as
\begin{eqnarray}
\widehat{g}_{17}&=&\frac{\omega_{17}}{12}\left[-(\widehat{\overline{\eta}}_{xxyy}^{'}+\widehat{\overline{\eta}}_{xxzz}^{'}+\widehat{\overline{\eta}}_{yyzz}^{'})
+2\left(u_x(\widehat{\overline{\eta}}_{xyy}^{'}+\widehat{\overline{\eta}}_{xzz}^{'})+
        u_y(\widehat{\overline{\eta}}_{yzz}^{'}+\widehat{\overline{\eta}}_{xxy}^{'})+\right.\right.\nonumber\\
        &&\left.\left.u_z(\widehat{\overline{\eta}}_{xxz}^{'}+\widehat{\overline{\eta}}_{yyz}^{'})\right)
        -u_x^2(\widehat{\overline{\eta}}_{yy}^{'}+\widehat{\overline{\eta}}_{zz}^{'})
        -u_y^2(\widehat{\overline{\eta}}_{xx}^{'}+\widehat{\overline{\eta}}_{zz}^{'})
        -u_z^2(\widehat{\overline{\eta}}_{xx}^{'}+\widehat{\overline{\eta}}_{yy}^{'})\right.\nonumber\\
&&\left.-4(u_xu_y\widehat{\overline{\eta}}_{xy}^{'}+u_xu_z\widehat{\overline{\eta}}_{xz}^{'}+u_yu_z\widehat{\overline{\eta}}_{yz}^{'})+
(\widetilde{\widehat{\kappa}}_{xx}\widetilde{\widehat{\kappa}}_{yy}+\widetilde{\widehat{\kappa}}_{xx}\widetilde{\widehat{\kappa}}_{zz}+
\widetilde{\widehat{\kappa}}_{yy}\widetilde{\widehat{\kappa}}_{zz})\right.\nonumber\\
&&\left.+3\rho(u_x^2u_y^2+u_x^2u_z^2+u_y^2u_z^2)+u_x^2(u_y\widehat{\sigma}_{y}^{'}+u_z\widehat{\sigma}_{z}^{'})+
                                                 u_y^2(u_x\widehat{\sigma}_{x}^{'}+u_z\widehat{\sigma}_{z}^{'})\right.\nonumber\\
&&\left.+u_z^2(u_x\widehat{\sigma}_{x}^{'}+u_y\widehat{\sigma}_{y}^{'})\right]
-4u_xu_y\widehat{g}_4-4u_xu_z\widehat{g}_5-4u_yu_z\widehat{g}_6\nonumber\\
&&+\frac{1}{2}(u_x^2-u_y^2)\widehat{g}_7+\frac{1}{2}(u_x^2+u_y^2-2u_z^2)\widehat{g}_8
+\frac{1}{2}(-2u_x^2-2u_y^2-u_z^2-4)\widehat{g}_9\nonumber\\
&&+4u_x\widehat{g}_{10}+4u_y\widehat{g}_{11}+4u_z\widehat{g}_{12},\label{eq:collisionkernelg17}
\end{eqnarray}
\begin{eqnarray}
\widehat{g}_{18}&=&\frac{\omega_{18}}{24}\left[-(\widehat{\overline{\eta}}_{xxyy}^{'}+\widehat{\overline{\eta}}_{xxzz}^{'}-2\widehat{\overline{\eta}}_{yyzz}^{'})
+2\left(u_x\widehat{\overline{\eta}}_{xyy}^{'}+u_x\widehat{\overline{\eta}}_{xzz}^{'}+
        u_y\widehat{\overline{\eta}}_{xxy}^{'}+u_z\widehat{\overline{\eta}}_{xxz}^{'}\right.\right.\nonumber\\
        &&\left.\left.-2(u_y\widehat{\overline{\eta}}_{yzz}^{'}+u_z\widehat{\overline{\eta}}_{yyz}^{'})\right)
        -u_x^2\widehat{\overline{\eta}}_{yy}^{'}-u_x^2\widehat{\overline{\eta}}_{zz}^{'}
        -u_y^2\widehat{\overline{\eta}}_{xx}^{'}-u_z^2\widehat{\overline{\eta}}_{xx}^{'}
        +2u_y^2\widehat{\overline{\eta}}_{zz}^{'}+\right.\nonumber\\
&&\left.+2u_z^2\widehat{\overline{\eta}}_{yy}^{'}-4(u_xu_y\widehat{\overline{\eta}}_{xy}^{'}+u_xu_z\widehat{\overline{\eta}}_{xz}^{'}-2u_yu_z\widehat{\overline{\eta}}_{yz}^{'})+
(\widetilde{\widehat{\kappa}}_{xx}\widetilde{\widehat{\kappa}}_{yy}+\widetilde{\widehat{\kappa}}_{xx}\widetilde{\widehat{\kappa}}_{zz}
-2\widetilde{\widehat{\kappa}}_{yy}\widetilde{\widehat{\kappa}}_{zz})\right.\nonumber\\
&&\left.+3\rho(u_x^2u_y^2+u_x^2u_z^2-2u_y^2u_z^2)+u_x^2u_y\widehat{\sigma}_{y}^{'}+u_x^2u_z\widehat{\sigma}_{z}^{'}+
                                                 u_y^2u_x\widehat{\sigma}_{x}^{'} +u_z^2u_x\widehat{\sigma}_{x}^{'}\right.\nonumber\\
&&\left.-2u_y^2u_z\widehat{\sigma}_{z}^{'}-2u_z^2u_y\widehat{\sigma}_{y}^{'})\right]
-2u_xu_y\widehat{g}_4-2u_xu_z\widehat{g}_5+4u_yu_z\widehat{g}_6\nonumber\\
&&+\frac{1}{4}(u_x^2-u_y^2-3u_z^2-2)\widehat{g}_7+\frac{1}{4}(u_x^2-5u_y^2+u_z^2-2)\widehat{g}_8
+\frac{1}{4}(-2u_x^2+u_y^2\nonumber\\
&&+2u_z^2)\widehat{g}_9
+2u_x\widehat{g}_{10}-u_y\widehat{g}_{11}-u_z\widehat{g}_{12}+u_y\widehat{g}_{14}+u_z\widehat{g}_{15},\label{eq:collisionkernelg18}
\end{eqnarray}
\begin{eqnarray}
\widehat{g}_{19}&=&\frac{\omega_{19}}{8}\left[-(\widehat{\overline{\eta}}_{xxyy}^{'}-\widehat{\overline{\eta}}_{xxzz}^{'})
+2\left(u_x\widehat{\overline{\eta}}_{xyy}^{'}-u_x\widehat{\overline{\eta}}_{xzz}^{'}+
        u_y\widehat{\overline{\eta}}_{xxy}^{'}-u_z\widehat{\overline{\eta}}_{xxz}^{'}\right)\right.\nonumber\\
        &&\left.-(u_x^2\widehat{\overline{\eta}}_{yy}^{'}-u_x^2\widehat{\overline{\eta}}_{zz}^{'}+
                  u_y^2\widehat{\overline{\eta}}_{xx}^{'}-u_z^2\widehat{\overline{\eta}}_{xx}^{'})
        -4(u_xu_y\widehat{\overline{\eta}}_{xy}^{'}-u_xu_z\widehat{\overline{\eta}}_{xz}^{'})\right.\nonumber\\
&&\left.+(\widetilde{\widehat{\kappa}}_{xx}\widetilde{\widehat{\kappa}}_{yy}-\widetilde{\widehat{\kappa}}_{xx}\widetilde{\widehat{\kappa}}_{zz})
+3\rho(u_x^2u_y^2-u_x^2u_z^2)\right.\nonumber\\
&&+\left.(u_x^2u_y\widehat{\sigma}_{y}^{'}-u_x^2u_z\widehat{\sigma}_{z}^{'}+u_y^2u_x\widehat{\sigma}_{x}^{'}-u_z^2u_x\widehat{\sigma}_{x}^{'})\right]
-6u_xu_y\widehat{g}_4+6u_xu_z\widehat{g}_5\nonumber\\
&&+\frac{1}{4}(3u_x^2-3u_y^2+3u_z^2+2)\widehat{g}_7+\frac{1}{4}(-9u_x^2-3u_y^2+3u_z^2-6)\widehat{g}_8\nonumber\\
&&+\frac{1}{4}(-3u_y^2-8)\widehat{g}_9+3u_y\widehat{g}_{11}-3u_z\widehat{g}_{12}+2u_x\widehat{g}_{13}+u_y\widehat{g}_{14}-u_z\widehat{g}_{15}.\label{eq:collisionkernelg19}
\end{eqnarray}
For calculating $\widehat{g}_{17}$ through $\widehat{g}_{19}$ in the above equations, we need the post collision states $\widetilde{\widehat{\kappa}}_{xx}$,
$\widetilde{\widehat{\kappa}}_{yy}$ and $\widetilde{\widehat{\kappa}}_{zz}$. These can be obtained from Eq.~(\ref{eq:transformedcentralmomentsrelation}) as follows.
\begin{eqnarray}
\widetilde{\widehat{\kappa}}_{xx}&=&\widetilde{\widehat{\overline{\kappa}}}_{xx}+\frac{1}{2}\widehat{\sigma}_{xx},\nonumber\\
\widetilde{\widehat{\kappa}}_{yy}&=&\widetilde{\widehat{\overline{\kappa}}}_{yy}+\frac{1}{2}\widehat{\sigma}_{yy},\nonumber\\
\widetilde{\widehat{\kappa}}_{zz}&=&\widetilde{\widehat{\overline{\kappa}}}_{zz}+\frac{1}{2}\widehat{\sigma}_{zz},\nonumber
\end{eqnarray}
where the second-order transformed central moments, in turn, can be related to corresponding raw moments, which are known, as
\begin{eqnarray}
\widetilde{\widehat{\overline{\kappa}}}_{xx}&=&\widetilde{\widehat{\overline{\kappa}}}_{xx}^{'}-\rho u_x^2-F_xu_x,\nonumber\\
\widetilde{\widehat{\overline{\kappa}}}_{yy}&=&\widetilde{\widehat{\overline{\kappa}}}_{yy}^{'}-\rho u_y^2-F_yu_y,\nonumber\\
\widetilde{\widehat{\overline{\kappa}}}_{zz}&=&\widetilde{\widehat{\overline{\kappa}}}_{zz}^{'}-\rho u_z^2-F_zu_z.\nonumber
\end{eqnarray}
Note that in terms of $\widehat{\overline{\eta}}_{x^my^nz^p}^{'}$ these can also be written as
\begin{eqnarray}
\widetilde{\widehat{\overline{\kappa}}}_{xx}&=&\left[\widehat{\overline{\eta}}_{xx}^{'}+6\widehat{g}_7+6\widehat{g}_8+6\widehat{g}_9\right]-\rho u_x^2-F_xu_x,\nonumber\\
\widetilde{\widehat{\overline{\kappa}}}_{yy}&=&\left[\widehat{\overline{\eta}}_{yy}^{'}-6\widehat{g}_7+6\widehat{g}_8+6\widehat{g}_9\right]-\rho u_y^2-F_yu_y,\nonumber\\
\widetilde{\widehat{\overline{\kappa}}}_{zz}&=&\left[\widehat{\overline{\eta}}_{xx}^{'}-12\widehat{g}_8+6\widehat{g}_9\right]-\rho u_z^2-F_zu_z.\nonumber
\end{eqnarray}

Proceeding further for the remaining three fourth-order central moments using $\widehat{\kappa}^{at}_{xxyz}=\widehat{\kappa}^{at}_{xyyz}=\widehat{\kappa}^{at}_{xyzz}=0$, we get
\begin{eqnarray}
\widehat{g}_{20}&=&\frac{\omega_{20}}{8}\left[-\widehat{\overline{\eta}}_{xxyz}^{'}+u_z\widehat{\overline{\eta}}_{xxy}^{'}
+u_y\widehat{\overline{\eta}}_{xxz}^{'}+2u_x\widehat{\overline{\eta}}_{xyz}^{'}-u_yu_z\widehat{\overline{\eta}}_{xx}^{'}
-2u_xu_z\widehat{\overline{\eta}}_{xy}^{'}\right.\nonumber\\
&&\left.-2u_xu_y\widehat{\overline{\eta}}_{xz}^{'}-u_x^2\widehat{\overline{\eta}}_{yz}^{'}+3\rho u_x^2u_yu_z
+\widehat{\sigma}_{x}^{'}u_xu_yu_z+\frac{1}{2}u_x^2(\widehat{\sigma}_{y}^{'}u_z+\widehat{\sigma}_{z}^{'}u_y)\right]\nonumber\\
&&-3u_xu_z\widehat{g}_{4}-3u_xu_y\widehat{g}_{5}-\left(\frac{3}{2}u_x^2+1\right)\widehat{g}_{6}-\frac{3}{4}u_yu_z\widehat{g}_{7}
-\frac{3}{4}u_yu_z\widehat{g}_{8}\nonumber\\
&&-\frac{3}{4}u_yu_z\widehat{g}_{9}+\frac{3}{2}u_z\widehat{g}_{11}+\frac{3}{2}u_y\widehat{g}_{12}
+\frac{1}{2}u_z\widehat{g}_{14}+\frac{1}{2}u_y\widehat{g}_{15}+2u_x\widehat{g}_{16},\label{eq:collisionkernelg20}
\end{eqnarray}
\begin{eqnarray}
\widehat{g}_{21}&=&\frac{\omega_{21}}{8}\left[-\widehat{\overline{\eta}}_{xyyz}^{'}+2u_y\widehat{\overline{\eta}}_{xyz}^{'}
+u_z\widehat{\overline{\eta}}_{xyy}^{'}+u_x\widehat{\overline{\eta}}_{yyz}^{'}-u_y^2\widehat{\overline{\eta}}_{xz}^{'}
-2u_yu_z\widehat{\overline{\eta}}_{xy}^{'}\right.\nonumber\\
&&\left.-2u_xu_y\widehat{\overline{\eta}}_{yz}^{'}-u_xu_z\widehat{\overline{\eta}}_{yy}^{'}+3\rho u_xu_y^2u_z
+\widehat{\sigma}_{y}^{'}u_xu_yu_z+\frac{1}{2}u_y^2(\widehat{\sigma}_{x}^{'}u_z+\widehat{\sigma}_{z}^{'}u_x)\right]\nonumber\\
&&-3u_yu_z\widehat{g}_{4}-\left(\frac{3}{2}u_y^2+1\right)\widehat{g}_{5}-3u_xu_y\widehat{g}_{6}+\frac{3}{4}u_xu_z\widehat{g}_{7}
-\frac{3}{4}u_xu_z\widehat{g}_{8}\nonumber\\
&&-\frac{3}{4}u_xu_z\widehat{g}_{9}+\frac{3}{2}u_z\widehat{g}_{10}+\frac{3}{2}u_x\widehat{g}_{12}
+\frac{1}{2}u_z\widehat{g}_{13}-\frac{1}{2}u_x\widehat{g}_{15}+2u_y\widehat{g}_{16},\label{eq:collisionkernelg21}
\end{eqnarray}
\begin{eqnarray}
\widehat{g}_{22}&=&\frac{\omega_{22}}{8}\left[-\widehat{\overline{\eta}}_{xyzz}^{'}+2u_z\widehat{\overline{\eta}}_{xyz}^{'}
+u_y\widehat{\overline{\eta}}_{xzz}^{'}+u_x\widehat{\overline{\eta}}_{yzz}^{'}-u_z^2\widehat{\overline{\eta}}_{xy}^{'}
-2u_yu_z\widehat{\overline{\eta}}_{xz}^{'}\right.\nonumber\\
&&\left.-2u_xu_z\widehat{\overline{\eta}}_{yz}^{'}-u_xu_y\widehat{\overline{\eta}}_{zz}^{'}+3\rho u_xu_yu_z^2
+\widehat{\sigma}_{z}^{'}u_xu_yu_z+\frac{1}{2}u_z^2(\widehat{\sigma}_{x}^{'}u_y+\widehat{\sigma}_{y}^{'}u_x)\right]\nonumber\\
&&-\left(\frac{3}{2}u_z^2+1\right)\widehat{g}_{4}-3u_yu_z\widehat{g}_{5}-3u_xu_z\widehat{g}_{6}+\frac{3}{2}u_xu_y\widehat{g}_{8}\nonumber\\
&&-\frac{3}{4}u_xu_y\widehat{g}_{9}+\frac{3}{2}u_y\widehat{g}_{10}+\frac{3}{2}u_x\widehat{g}_{11}
-\frac{1}{2}u_y\widehat{g}_{13}-\frac{1}{2}u_x\widehat{g}_{14}+2u_z\widehat{g}_{16}.\label{eq:collisionkernelg22}
\end{eqnarray}

The collision kernels for the three fifth-order central moments follow similarly from
$\widehat{\overline{\kappa}}^{at}_{xyyzz}=\widehat{\overline{\kappa}}^{at}_{xxyzz}=\widehat{\overline{\kappa}}^{at}_{xxyyz}=0$ as
\begin{eqnarray}
\widehat{g}_{23}&=&\frac{\omega_{23}}{8}\left[-\widehat{\overline{\eta}}_{xyyzz}^{'}+u_x\widehat{\overline{\eta}}_{yyzz}^{'}+
2u_y\widehat{\overline{\eta}}_{xyzz}^{'}+2u_z\widehat{\overline{\eta}}_{xyyz}^{'}-u_y^2\widehat{\overline{\eta}}_{xzz}^{'}
-u_z^2\widehat{\overline{\eta}}_{xyy}^{'}\right.\nonumber\\
&&\left.-2u_xu_y\widehat{\overline{\eta}}_{yzz}^{'}-2u_xu_z\widehat{\overline{\eta}}_{yyz}^{'}
-4u_yu_z\widehat{\overline{\eta}}_{xyz}^{'}+u_xu_z^2\widehat{\overline{\eta}}_{yy}^{'}+
u_xu_y^2\widehat{\overline{\eta}}_{zz}^{'}\right.\nonumber\\
&&+\left.2u_yu_z^2\widehat{\overline{\eta}}_{xy}^{'}+2u_y^2u_z\widehat{\overline{\eta}}_{xz}^{'}
+4u_xu_yu_z\widehat{\overline{\eta}}_{yz}^{'}-4\rho u_xu_y^2u_z^2-\frac{1}{2}u_y^2u_z^2\widehat{\sigma}_{x}^{'}\right.\nonumber\\
&&\left.-u_x(u_z^2u_y\widehat{\sigma}_{y}^{'}+u_y^2u_z\widehat{\sigma}_{z}^{'})\right]+(3u_yu_z^2+2u_y)\widehat{g}_{4}
+(3u_y^2u_z+2u_z)\widehat{g}_{5}\nonumber\\
&&+6u_xu_yu_z\widehat{g}_{6}+\left(\frac{3}{4}u_xu_z^2-\frac{1}{4}u_x\right)\widehat{g}_{7}+
\left(\frac{3}{4}u_xu_z^2-\frac{3}{2}u_xu_y^2-\frac{1}{2}u_x\right)\widehat{g}_{8}\nonumber\\
&&+\left(\frac{3}{4}u_xu_z^2+\frac{3}{4}u_xu_y^2+u_x\right)\widehat{g}_{9}+\left(-\frac{3}{2}u_y^2-\frac{3}{2}u_z^2-2\right)\widehat{g}_{10}
-3u_xu_y\widehat{g}_{11}\nonumber\\
&&-3u_xu_z\widehat{g}_{12}+\left(\frac{1}{2}u_y^2-\frac{1}{2}u_z^2\right)\widehat{g}_{13}+u_xu_y\widehat{g}_{14}
+u_xu_z\widehat{g}_{15}-4u_yu_z\widehat{g}_{16}\nonumber\\
&&+\frac{1}{2}u_x\widehat{g}_{17}-u_x\widehat{g}_{18}+2u_z\widehat{g}_{21}+2u_y\widehat{g}_{22},\label{eq:collisionkernelg23}
\end{eqnarray}
\begin{eqnarray}
\widehat{g}_{24}&=&\frac{\omega_{24}}{8}\left[-\widehat{\overline{\eta}}_{xxyzz}^{'}+2u_x\widehat{\overline{\eta}}_{xyzz}^{'}+
2u_z\widehat{\overline{\eta}}_{xxyz}^{'}+u_y\widehat{\overline{\eta}}_{xxzz}^{'}-u_x^2\widehat{\overline{\eta}}_{yzz}^{'}
-u_z^2\widehat{\overline{\eta}}_{xxy}^{'}\right.\nonumber\\
&&\left.-2u_xu_y\widehat{\overline{\eta}}_{xzz}^{'}-2u_yu_z\widehat{\overline{\eta}}_{xxz}^{'}
-4u_xu_z\widehat{\overline{\eta}}_{xyz}^{'}+u_yu_z^2\widehat{\overline{\eta}}_{xx}^{'}+
u_x^2u_y\widehat{\overline{\eta}}_{zz}^{'}\right.\nonumber\\
&&+\left.2u_xu_z^2\widehat{\overline{\eta}}_{xy}^{'}+2u_x^2u_z\widehat{\overline{\eta}}_{yz}^{'}
+4u_xu_yu_z\widehat{\overline{\eta}}_{xz}^{'}-4\rho u_x^2u_yu_z^2-\frac{1}{2}u_x^2u_z^2\widehat{\sigma}_{y}^{'}\right.\nonumber\\
&&\left.-u_y(u_xu_z^2\widehat{\sigma}_{x}^{'}+u_x^2u_z\widehat{\sigma}_{z}^{'})\right]+(3u_xu_z^2+2u_x)\widehat{g}_{4}
+6u_xu_yu_z\widehat{g}_{5}\nonumber\\
&&+(3u_x^2u_z+2u_z)\widehat{g}_{6}+\left(\frac{3}{4}u_yu_z^2+\frac{1}{2}u_y\right)\widehat{g}_{7}+
\left(\frac{3}{4}u_yu_z^2-\frac{3}{2}u_x^2u_y-\frac{1}{2}u_y\right)\widehat{g}_{8}\nonumber\\
&&+\left(\frac{3}{4}u_yu_z^2+\frac{3}{4}u_x^2u_y+u_y\right)\widehat{g}_{9}-3u_xu_y\widehat{g}_{10}
+\left(-\frac{3}{2}u_x^2-\frac{3}{2}u_z^2-2\right)\widehat{g}_{11}\nonumber\\
&&-3u_yu_z\widehat{g}_{12}+u_xu_y\widehat{g}_{13}+\left(\frac{1}{2}u_x^2-\frac{1}{2}u_z^2\right)\widehat{g}_{14}
-u_yu_z\widehat{g}_{15}-4u_xu_z\widehat{g}_{16}\nonumber\\
&&+\frac{1}{2}u_y\widehat{g}_{17}+\frac{1}{2}u_y\widehat{g}_{18}-\frac{1}{2}u_y\widehat{g}_{19}+2u_z\widehat{g}_{20}+
2u_x\widehat{g}_{22},\label{eq:collisionkernelg24}
\end{eqnarray}
\begin{eqnarray}
\widehat{g}_{25}&=&\frac{\omega_{25}}{8}\left[-\widehat{\overline{\eta}}_{xxyyz}^{'}+2u_x\widehat{\overline{\eta}}_{xyyz}^{'}+
2u_y\widehat{\overline{\eta}}_{xxyz}^{'}+u_z\widehat{\overline{\eta}}_{xxyy}^{'}-u_x^2\widehat{\overline{\eta}}_{yyz}^{'}
-u_y^2\widehat{\overline{\eta}}_{xxz}^{'}\right.\nonumber\\
&&\left.-2u_xu_z\widehat{\overline{\eta}}_{xyy}^{'}-2u_yu_z\widehat{\overline{\eta}}_{xxy}^{'}
-4u_xu_y\widehat{\overline{\eta}}_{xyz}^{'}+u_y^2u_z\widehat{\overline{\eta}}_{xx}^{'}+
u_x^2u_z\widehat{\overline{\eta}}_{yy}^{'}\right.\nonumber\\
&&+\left.2u_xu_y^2\widehat{\overline{\eta}}_{xz}^{'}+2u_x^2u_y\widehat{\overline{\eta}}_{yz}^{'}
+4u_xu_yu_z\widehat{\overline{\eta}}_{xy}^{'}-4\rho u_x^2u_y^2u_z-\frac{1}{2}u_x^2u_y^2\widehat{\sigma}_{z}^{'}\right.\nonumber\\
&&\left.-u_z(u_xu_y^2\widehat{\sigma}_{x}^{'}+u_x^2u_y\widehat{\sigma}_{y}^{'})\right]+6u_xu_yu_z\widehat{g}_{4}
+(3u_xu_y^2+2u_x)\widehat{g}_{5}\nonumber\\
&&+(3u_x^2u_y+2u_y)\widehat{g}_{6}+\left(\frac{3}{4}u_y^2u_z-\frac{3}{4}u_x^2u_z\right)\widehat{g}_{7}+
\left(\frac{3}{4}u_y^2u_z+\frac{3}{4}u_x^2u_z+u_z\right)\widehat{g}_{8}\nonumber\\
&&+\left(\frac{3}{4}u_y^2u_z+\frac{3}{4}u_x^2u_z+u_z\right)\widehat{g}_{9}-3u_xu_z\widehat{g}_{10}-3u_yu_z\widehat{g}_{11}\nonumber\\
&&+\left(-\frac{3}{2}u_x^2-\frac{3}{2}u_y^2-2\right)\widehat{g}_{12}   -u_xu_z\widehat{g}_{13}-u_yu_z\widehat{g}_{14}
+\left(\frac{1}{2}u_x^2-\frac{1}{2}u_y^2\right)\widehat{g}_{15}\nonumber\\
&&-6u_xu_y\widehat{g}_{16}+\frac{1}{2}u_z\widehat{g}_{17}+\frac{1}{2}u_z\widehat{g}_{18}+\frac{1}{2}u_z\widehat{g}_{19}+2u_y\widehat{g}_{20}+
2u_x\widehat{g}_{21}.\label{eq:collisionkernelg25}
\end{eqnarray}

Finally, for the one sixth-order component, we obtain the collision kernel based on the non-zero factorized attractor
(see Eq.~(\ref{eq:transformedcentralmomentattractors})) as
\begin{eqnarray}
\widehat{g}_{26}&=&\frac{\omega_{26}}{8}\left[-\widehat{\overline{\eta}}_{xxyyzz}^{'}+
2\left(u_x\widehat{\overline{\eta}}_{xyyzz}^{'}+u_y\widehat{\overline{\eta}}_{xxyzz}^{'}
+u_z\widehat{\overline{\eta}}_{xxyyz}^{'}\right)
-\left(u_x^2\widehat{\overline{\eta}}_{yyzz}^{'}\right.\right.\nonumber\\
&&+\left.\left.u_y^2\widehat{\overline{\eta}}_{xxzz}^{'}
+u_z^2\widehat{\overline{\eta}}_{xxyy}^{'}\right)
-4\left(u_xu_y\widehat{\overline{\eta}}_{xyzz}^{'}+u_xu_z\widehat{\overline{\eta}}_{xyyz}^{'}
+u_yu_z\widehat{\overline{\eta}}_{xxyz}^{'}\right)\right.\nonumber\\
&&\left.+2\left(u_x^2u_y\widehat{\overline{\eta}}_{yzz}^{'}+u_xu_y^2\widehat{\overline{\eta}}_{xzz}^{'}
+u_x^2u_z\widehat{\overline{\eta}}_{yyz}^{'}+u_xu_z^2\widehat{\overline{\eta}}_{xyy}^{'}
+u_y^2u_z\widehat{\overline{\eta}}_{xxz}^{'}\right.\right.\nonumber\\
&&\left.\left.+u_yu_z^2\widehat{\overline{\eta}}_{xxy}^{'}\right)+8u_xu_yu_z\widehat{\overline{\eta}}_{xyz}^{'}
-\left(u_y^2u_z^2\widehat{\overline{\eta}}_{xx}^{'}+u_x^2u_z^2\widehat{\overline{\eta}}_{yy}^{'}
+u_x^2u_y^2\widehat{\overline{\eta}}_{zz}^{'}\right)\right.\nonumber\\
&&\left.-4u_xu_yu_z\left(u_z\widehat{\overline{\eta}}_{xy}^{'}+u_y\widehat{\overline{\eta}}_{xz}^{'}
+u_x\widehat{\overline{\eta}}_{yz}^{'}\right)+5\rho u_x^2u_y^2u_z^2+\right.\nonumber\\
&&\left.\widetilde{\widehat{\kappa}}_{xx}\widetilde{\widehat{\kappa}}_{yy}\widetilde{\widehat{\kappa}}_{zz}
+u_xu_yu_z\left(u_xu_y\widehat{\sigma}_{z}^{'}+u_xu_z\widehat{\sigma}_{y}^{'}+u_yu_z\widehat{\sigma}_{x}^{'}\right)\right]\nonumber\\
&&+\left(-4u_xu_y-6u_xu_yu_z^2\right)\widehat{g}_{4}+\left(-4u_xu_z-6u_xu_y^2u_z\right)\widehat{g}_{5}\nonumber\\
&&+\left(-4u_yu_z-6u_x^2u_yu_z\right)\widehat{g}_{6}+\left(\frac{1}{2}u_x^2-\frac{1}{2}u_y^2+\frac{3}{4}u_x^2u_z^2
-\frac{3}{4}u_y^2u_z^2\right)\widehat{g}_{7}\nonumber\\
&&+\left(\frac{1}{2}u_x^2+\frac{1}{2}u_y^2-u_z^2-\frac{3}{4}u_y^2u_z^2-\frac{3}{4}u_x^2u_z^2+\frac{3}{2}u_x^2u_y^2\right)\widehat{g}_{8}\nonumber\\
&&+\left(-u_x^2-u_y^2-u_z^2-\frac{3}{4}u_x^2u_y^2-\frac{3}{4}u_x^2u_z^2-\frac{3}{4}u_y^2u_z^2-1\right)\widehat{g}_{9}\nonumber\\
&&+\left(3u_xu_y^2+3u_xu_z^2+4u_x\right)\widehat{g}_{10}+\left(3u_x^2u_y+3u_yu_z^2+4u_y\right)\widehat{g}_{11}\nonumber\\
&&+\left(3u_x^2u_z+3u_y^2u_z+4u_z\right)\widehat{g}_{12}+\left(u_xu_z^2-u_xu_y^2\right)\widehat{g}_{13}\nonumber\\
&&+\left(u_yu_z^2-u_x^2u_y\right)\widehat{g}_{14}+\left(u_y^2u_z-u_x^2u_z\right)\widehat{g}_{15}+8u_xu_yu_z\widehat{g}_{16}\nonumber\\
&&+\left(-\frac{1}{2}u_x^2-\frac{1}{2}u_y^2-\frac{1}{2}u_z^2-1\right)\widehat{g}_{17}
+\left(u_x^2-\frac{1}{2}u_y^2-\frac{1}{2}u_z^2\right)\widehat{g}_{18}\nonumber\\
&&+\left(\frac{1}{2}u_y^2-\frac{1}{2}u_z^2\right)\widehat{g}_{19}-4u_yu_z\widehat{g}_{20}-4u_xu_z\widehat{g}_{21}
-4u_xu_y\widehat{g}_{22}+2u_x\widehat{g}_{23}\nonumber\\
&&+2u_y\widehat{g}_{24}+2u_z\widehat{g}_{25}.\label{eq:collisionkernelg26}
\end{eqnarray}
Note that the transformed raw moments of various orders, i.e. $\widehat{\overline{\kappa}}_{x^m y^n z^p}^{'}$
and raw source moments, i.e. $\widehat{\sigma}_{x^m y^n z^p}^{'}$ needed for $\widehat{\overline{\eta}}_{x^m y^n z^p}^{'}$
for various $m$, $n$ and $p$ combinations can be obtained from Eqs.~(\ref{eq:rawmomenttransformedconserved}) and (\ref{eq:rawmomenttransformednonconserved}) and Eq.~(\ref{eq:rawmomentsourceterm}), respectively, which are given in Sec.~\ref{sec:rawmoments}. Similar to the 2D central moment LBM with source terms~\cite{premnath09d}, we can apply the Chapman-Enskog expansion to the above 3D formulation to show that its emergent dynamics corresponds to the Navier-Stokes equations representing fluid motion in the presence of general force fields. Some of the relaxation parameters in the collision model can be related to the transport coefficients. For example, those corresponding to the second-order moments control shear viscosity $\nu$ of the fluid. That is, $\nu=c_s^2\left(\frac{1}{\omega^\nu}-\frac{1}{2}\right)$ where $\omega^{\nu}=\omega_{j}$ where $j=4,5,6,7,8$. The rest of the parameters can be set either to $1$ (i.e. equilibration) or adjusted independently to carefully control and improve numerical stability by means of a linear stability analysis, while all satisfying the usual bounds $0<\omega_{\beta}<2$.

\section{\label{sec:computationalprocedure}Operational Steps of the Central Moment LBM}
To provide explicit expressions for the collision step in the central moment LBM as a stream-and-collide procedure (i.e. Eq.~(\ref{eq:cascadedcollision1}) and (\ref{eq:cascadedstreaming1})), we first expand the elements of the matrix
multiplication of $\mathcal{K}$ with $\widehat{\mathbf{g}}$ in Eq.~(\ref{eq:cascadecollision1}). This yields the post-collision
values of all the 27 components of the transformed distribution function in terms of the Galilean invariant collision kernel $\widehat{g}_{\beta}$ (see Sec.~\ref{sec:cascadedcollisionforcing}) and source terms $S_{\beta}$ (see Eq.~(\ref{eq:vsourceterms}) in Appendix.~\ref{app:sourcetermsvelocityspace}) which can be summarized as follows:
\begin{eqnarray}
\widetilde{\overline{f}}_{0}&=&\overline{f}_{0}+\left[\widehat{g}_0-2\widehat{g}_9+4\widehat{g}_{17}-8\widehat{g}_{26}\right]+S_0, \nonumber\\
\widetilde{\overline{f}}_{1}&=&\overline{f}_{1}+\left[\widehat{g}_0+\widehat{g}_1+\widehat{g}_7+\widehat{g}_8-\widehat{g}_9
-4\widehat{g}_{10}-4\widehat{g}_{18}+4\widehat{g}_{23}+4\widehat{g}_{26}\right]+S_1, \nonumber\\
\widetilde{\overline{f}}_{2}&=&\overline{f}_{2}+\left[\widehat{g}_0-\widehat{g}_1+\widehat{g}_7+\widehat{g}_8-\widehat{g}_9
+4\widehat{g}_{10}-4\widehat{g}_{18}-4\widehat{g}_{23}+4\widehat{g}_{26}\right]+S_2, \nonumber\\
\widetilde{\overline{f}}_{3}&=&\overline{f}_{3}+\left[\widehat{g}_0+\widehat{g}_2-\widehat{g}_7+\widehat{g}_8-\widehat{g}_9
-4\widehat{g}_{11}+2\widehat{g}_{18}-2\widehat{g}_{19}+4\widehat{g}_{24}+4\widehat{g}_{26}\right]\nonumber\\
&&+S_3, \nonumber\\
\widetilde{\overline{f}}_{4}&=&\overline{f}_{4}+\left[\widehat{g}_0-\widehat{g}_2-\widehat{g}_7+\widehat{g}_8-\widehat{g}_9
+4\widehat{g}_{11}+2\widehat{g}_{18}-2\widehat{g}_{19}-4\widehat{g}_{24}+4\widehat{g}_{26}\right]\nonumber\\
&&+S_4, \nonumber\\
\widetilde{\overline{f}}_{5}&=&\overline{f}_{5}+\left[\widehat{g}_0+\widehat{g}_3-2\widehat{g}_8-\widehat{g}_9-4\widehat{g}_{12}
+2\widehat{g}_{18}+2\widehat{g}_{19}+4\widehat{g}_{25}+4\widehat{g}_{26}\right]\nonumber\\
&&+S_5, \nonumber\\
\widetilde{\overline{f}}_{6}&=&\overline{f}_{6}+\left[\widehat{g}_0-\widehat{g}_3-2\widehat{g}_8-\widehat{g}_9+4\widehat{g}_{12}
+2\widehat{g}_{18}+2\widehat{g}_{19}-4\widehat{g}_{25}+4\widehat{g}_{26}\right]\nonumber\\
&&+S_6, \nonumber\\
\widetilde{\overline{f}}_{7}&=&\overline{f}_{7}+\left[\widehat{g}_0+\widehat{g}_1+\widehat{g}_2+\widehat{g}_4+2\widehat{g}_8
-\widehat{g}_{10}-\widehat{g}_{11}+\widehat{g}_{13}+\widehat{g}_{14}-\widehat{g}_{17}+\widehat{g}_{18}+\right.\nonumber\\
&&\left.\widehat{g}_{19}-2\widehat{g}_{22}-2\widehat{g}_{23}-2\widehat{g}_{24}-2\widehat{g}_{26}\right]+S_7,\nonumber\\
\widetilde{\overline{f}}_{8}&=&\overline{f}_{8}+\left[\widehat{g}_0-\widehat{g}_1+\widehat{g}_2-\widehat{g}_4+2\widehat{g}_8
+\widehat{g}_{10}-\widehat{g}_{11}-\widehat{g}_{13}+\widehat{g}_{14}-\widehat{g}_{17}+\widehat{g}_{18}+\right.\nonumber\\
&&\left.\widehat{g}_{19}+2\widehat{g}_{22}+2\widehat{g}_{23}-2\widehat{g}_{24}-2\widehat{g}_{26}\right]+S_8,\nonumber\\
\widetilde{\overline{f}}_{9}&=&\overline{f}_{9}+\left[\widehat{g}_0+\widehat{g}_1-\widehat{g}_2-\widehat{g}_4+2\widehat{g}_8
-\widehat{g}_{10}+\widehat{g}_{11}+\widehat{g}_{13}-\widehat{g}_{14}-\widehat{g}_{17}+\widehat{g}_{18}+\right.\nonumber\\
&&\left.\widehat{g}_{19}+2\widehat{g}_{22}-2\widehat{g}_{23}+2\widehat{g}_{24}-2\widehat{g}_{26}\right]+S_9,\nonumber\\
\widetilde{\overline{f}}_{10}&=&\overline{f}_{10}+\left[\widehat{g}_0-\widehat{g}_1-\widehat{g}_2+\widehat{g}_4+2\widehat{g}_8
+\widehat{g}_{10}+\widehat{g}_{11}-\widehat{g}_{13}-\widehat{g}_{14}-\widehat{g}_{17}+\widehat{g}_{18}+\right.\nonumber\\
&&\left.\widehat{g}_{19}-2\widehat{g}_{22}+2\widehat{g}_{23}+2\widehat{g}_{24}-2\widehat{g}_{26}\right]+S_{10},\nonumber\\
\widetilde{\overline{f}}_{11}&=&\overline{f}_{11}+\left[\widehat{g}_0+\widehat{g}_1+\widehat{g}_3+\widehat{g}_5+\widehat{g}_7
-\widehat{g}_8-\widehat{g}_{10}-\widehat{g}_{12}-\widehat{g}_{13}+\widehat{g}_{15}-\widehat{g}_{17}+\right.\nonumber\\
&&\left.\widehat{g}_{18}-\widehat{g}_{19}-2\widehat{g}_{21}-2\widehat{g}_{23}-2\widehat{g}_{25}-2\widehat{g}_{26}\right]+S_{11},\nonumber\\
\widetilde{\overline{f}}_{12}&=&\overline{f}_{12}+\left[\widehat{g}_0-\widehat{g}_1+\widehat{g}_3-\widehat{g}_5+\widehat{g}_7
-\widehat{g}_8+\widehat{g}_{10}-\widehat{g}_{12}+\widehat{g}_{13}+\widehat{g}_{15}-\widehat{g}_{17}+\right.\nonumber\\
&&\left.\widehat{g}_{18}-\widehat{g}_{19}+2\widehat{g}_{21}+2\widehat{g}_{23}-2\widehat{g}_{25}-2\widehat{g}_{26}\right]+S_{12},\nonumber\\
\widetilde{\overline{f}}_{13}&=&\overline{f}_{13}+\left[\widehat{g}_0+\widehat{g}_1-\widehat{g}_3-\widehat{g}_5+\widehat{g}_7
-\widehat{g}_8-\widehat{g}_{10}+\widehat{g}_{12}-\widehat{g}_{13}-\widehat{g}_{15}-\widehat{g}_{17}+\right.\nonumber\\
&&\left.\widehat{g}_{18}-\widehat{g}_{19}+2\widehat{g}_{21}-2\widehat{g}_{23}+2\widehat{g}_{25}-2\widehat{g}_{26}\right]+S_{13},\nonumber\\
\widetilde{\overline{f}}_{14}&=&\overline{f}_{14}+\left[\widehat{g}_0-\widehat{g}_1-\widehat{g}_3+\widehat{g}_5+\widehat{g}_7
-\widehat{g}_8+\widehat{g}_{10}+\widehat{g}_{12}+\widehat{g}_{13}-\widehat{g}_{15}-\widehat{g}_{17}+\right.\nonumber\\
&&\left.\widehat{g}_{18}-\widehat{g}_{19}-2\widehat{g}_{21}+2\widehat{g}_{23}+2\widehat{g}_{25}-2\widehat{g}_{26}\right]+S_{14},\nonumber\\
\widetilde{\overline{f}}_{15}&=&\overline{f}_{15}+\left[\widehat{g}_0+\widehat{g}_2+\widehat{g}_3+\widehat{g}_6-\widehat{g}_7
-\widehat{g}_8-\widehat{g}_{11}-\widehat{g}_{12}-\widehat{g}_{14}-\widehat{g}_{15}-\widehat{g}_{17}-\right.\nonumber\\
&&\left.2\widehat{g}_{18}-2\widehat{g}_{20}-2\widehat{g}_{24}-2\widehat{g}_{25}-2\widehat{g}_{26}\right]+S_{15},\nonumber\\
\widetilde{\overline{f}}_{16}&=&\overline{f}_{16}+\left[\widehat{g}_0-\widehat{g}_2+\widehat{g}_3-\widehat{g}_6-\widehat{g}_7
-\widehat{g}_8+\widehat{g}_{11}-\widehat{g}_{12}+\widehat{g}_{14}-\widehat{g}_{15}-\widehat{g}_{17}-\right.\nonumber\\
&&\left.2\widehat{g}_{18}+2\widehat{g}_{20}+2\widehat{g}_{24}-2\widehat{g}_{25}-2\widehat{g}_{26}\right]+S_{16},\nonumber\\
\widetilde{\overline{f}}_{17}&=&\overline{f}_{17}+\left[\widehat{g}_0+\widehat{g}_2-\widehat{g}_3-\widehat{g}_6-\widehat{g}_7
-\widehat{g}_8-\widehat{g}_{11}+\widehat{g}_{12}-\widehat{g}_{14}+\widehat{g}_{15}-\widehat{g}_{17}-\right.\nonumber\\
&&\left.2\widehat{g}_{18}+2\widehat{g}_{20}-2\widehat{g}_{24}+2\widehat{g}_{25}-2\widehat{g}_{26}\right]+S_{17},\nonumber\\
\widetilde{\overline{f}}_{18}&=&\overline{f}_{18}+\left[\widehat{g}_0-\widehat{g}_2-\widehat{g}_3+\widehat{g}_6-\widehat{g}_7
-\widehat{g}_8+\widehat{g}_{11}+\widehat{g}_{12}+\widehat{g}_{14}+\widehat{g}_{15}-\widehat{g}_{17}-\right.\nonumber\\
&&\left.2\widehat{g}_{18}-2\widehat{g}_{20}+2\widehat{g}_{24}+2\widehat{g}_{25}-2\widehat{g}_{26}\right]+S_{18},\nonumber\\
\widetilde{\overline{f}}_{19}&=&\overline{f}_{19}+\left[\widehat{g}_0+\widehat{g}_1+\widehat{g}_2+\widehat{g}_3+\widehat{g}_4
+\widehat{g}_5+\widehat{g}_6+\widehat{g}_9+2\widehat{g}_{10}+2\widehat{g}_{11}+2\widehat{g}_{12}+\right.\nonumber\\
&&\left.\widehat{g}_{16}+\widehat{g}_{17}+\widehat{g}_{20}+\widehat{g}_{21}+\widehat{g}_{22}+\widehat{g}_{23}+\widehat{g}_{24}
+\widehat{g}_{25}+\widehat{g}_{26}\right]+S_{19},\nonumber\\
\widetilde{\overline{f}}_{20}&=&\overline{f}_{20}+\left[\widehat{g}_0-\widehat{g}_1+\widehat{g}_2+\widehat{g}_3-\widehat{g}_4
-\widehat{g}_5+\widehat{g}_6+\widehat{g}_9-2\widehat{g}_{10}+2\widehat{g}_{11}+2\widehat{g}_{12}-\right.\nonumber\\
&&\left.\widehat{g}_{16}+\widehat{g}_{17}+\widehat{g}_{20}-\widehat{g}_{21}-\widehat{g}_{22}-\widehat{g}_{23}+\widehat{g}_{24}
+\widehat{g}_{25}+\widehat{g}_{26}\right]+S_{20},\nonumber\\
\widetilde{\overline{f}}_{21}&=&\overline{f}_{21}+\left[\widehat{g}_0+\widehat{g}_1-\widehat{g}_2+\widehat{g}_3-\widehat{g}_4
+\widehat{g}_5-\widehat{g}_6+\widehat{g}_9+2\widehat{g}_{10}-2\widehat{g}_{11}+2\widehat{g}_{12}-\right.\nonumber\\
&&\left.\widehat{g}_{16}+\widehat{g}_{17}-\widehat{g}_{20}+\widehat{g}_{21}-\widehat{g}_{22}+\widehat{g}_{23}-\widehat{g}_{24}
+\widehat{g}_{25}+\widehat{g}_{26}\right]+S_{21},\nonumber\\
\widetilde{\overline{f}}_{22}&=&\overline{f}_{22}+\left[\widehat{g}_0-\widehat{g}_1-\widehat{g}_2+\widehat{g}_3+\widehat{g}_4
-\widehat{g}_5-\widehat{g}_6+\widehat{g}_9-2\widehat{g}_{10}-2\widehat{g}_{11}+2\widehat{g}_{12}+\right.\nonumber\\
&&\left.\widehat{g}_{16}+\widehat{g}_{17}-\widehat{g}_{20}-\widehat{g}_{21}+\widehat{g}_{22}-\widehat{g}_{23}-\widehat{g}_{24}
+\widehat{g}_{25}+\widehat{g}_{26}\right]+S_{22},\nonumber\\
\widetilde{\overline{f}}_{23}&=&\overline{f}_{23}+\left[\widehat{g}_0+\widehat{g}_1+\widehat{g}_2-\widehat{g}_3+\widehat{g}_4
-\widehat{g}_5-\widehat{g}_6+\widehat{g}_9+2\widehat{g}_{10}+2\widehat{g}_{11}-2\widehat{g}_{12}-\right.\nonumber\\
&&\left.\widehat{g}_{16}+\widehat{g}_{17}-\widehat{g}_{20}-\widehat{g}_{21}+\widehat{g}_{22}+\widehat{g}_{23}+\widehat{g}_{24}
-\widehat{g}_{25}+\widehat{g}_{26}\right]+S_{23},\nonumber\\
\widetilde{\overline{f}}_{24}&=&\overline{f}_{24}+\left[\widehat{g}_0-\widehat{g}_1+\widehat{g}_2-\widehat{g}_3-\widehat{g}_4
+\widehat{g}_5-\widehat{g}_6+\widehat{g}_9-2\widehat{g}_{10}+2\widehat{g}_{11}-2\widehat{g}_{12}+\right.\nonumber\\
&&\left.\widehat{g}_{16}+\widehat{g}_{17}-\widehat{g}_{20}+\widehat{g}_{21}-\widehat{g}_{22}-\widehat{g}_{23}+\widehat{g}_{24}
-\widehat{g}_{25}+\widehat{g}_{26}\right]+S_{24},\nonumber\\
\widetilde{\overline{f}}_{25}&=&\overline{f}_{25}+\left[\widehat{g}_0+\widehat{g}_1-\widehat{g}_2-\widehat{g}_3-\widehat{g}_4
-\widehat{g}_5+\widehat{g}_6+\widehat{g}_9+2\widehat{g}_{10}-2\widehat{g}_{11}-2\widehat{g}_{12}+\right.\nonumber\\
&&\left.\widehat{g}_{16}+\widehat{g}_{17}+\widehat{g}_{20}-\widehat{g}_{21}-\widehat{g}_{22}+\widehat{g}_{23}-\widehat{g}_{24}
-\widehat{g}_{25}+\widehat{g}_{26}\right]+S_{25},\nonumber\\
\widetilde{\overline{f}}_{26}&=&\overline{f}_{26}+\left[\widehat{g}_0-\widehat{g}_1-\widehat{g}_2-\widehat{g}_3+\widehat{g}_4
+\widehat{g}_5+\widehat{g}_6+\widehat{g}_9-2\widehat{g}_{10}-2\widehat{g}_{11}-2\widehat{g}_{12}-\right.\nonumber\\
&&\left.\widehat{g}_{16}+\widehat{g}_{17}+\widehat{g}_{20}+\widehat{g}_{21}+\widehat{g}_{22}-\widehat{g}_{23}-\widehat{g}_{24}
-\widehat{g}_{25}+\widehat{g}_{26}\right]+S_{26}.
\end{eqnarray}
The above post-collision state allows completion of the streaming step via Eq.~(\ref{eq:cascadedstreaming1}), following which
frame-independent fields of 3D fluid motion can be obtained from Eqs.~(\ref{eq:densitycalculation})
and (\ref{eq:velocitycalculation}). In the implementation, various optimization strategies such as those discussed in~\cite{dhumieres02} should be fully exploited to minimize the floating point operation count.

Following the general outline of the above derivation, the central moment LBM was also formulated for the three-dimensional, fifteen velocity (D3Q15) lattice, which has a much reduced computational complexity when compared with the D3Q27 lattice. The results are summarized in Appendix~\ref{app:d3q15formulation}.

\section{\label{section_numerical_tests}Numerical Tests}
\begin{figure}[h]
\includegraphics[width =140mm]{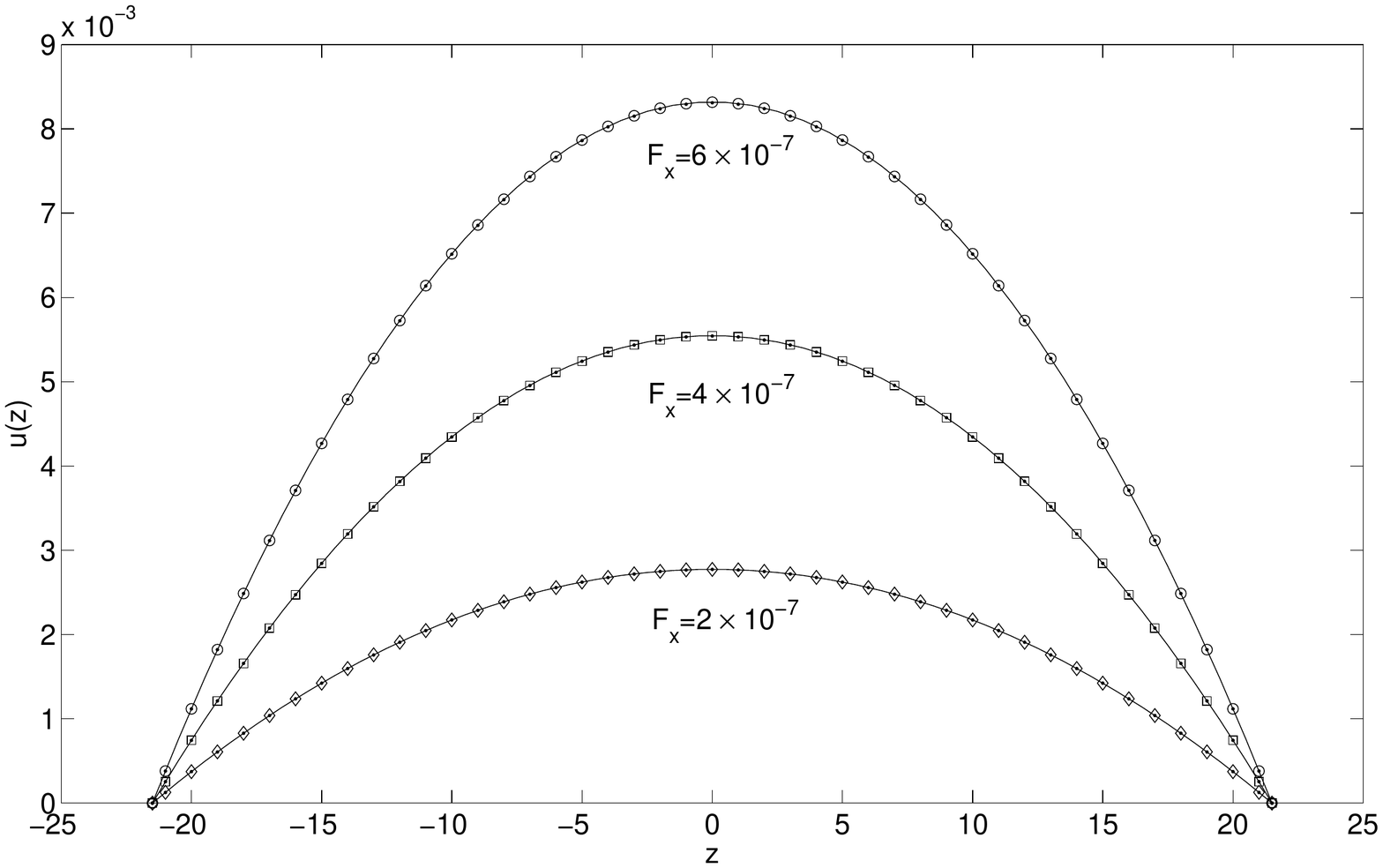}% Here is how to import EPS art
\caption{\label{fig:compare_lattices_velocityprofiles_diffFx} Flow between parallel plates with a constant body force: Comparison of velocity profiles computed by D3Q15 (open symbols) and D3Q27 (filled circles) formulations of the central moment LBM with forcing term with analytical solution (lines) for different values of the body force $F_x$.}
\end{figure}

\begin{figure}[h]
\includegraphics[width = 140mm]{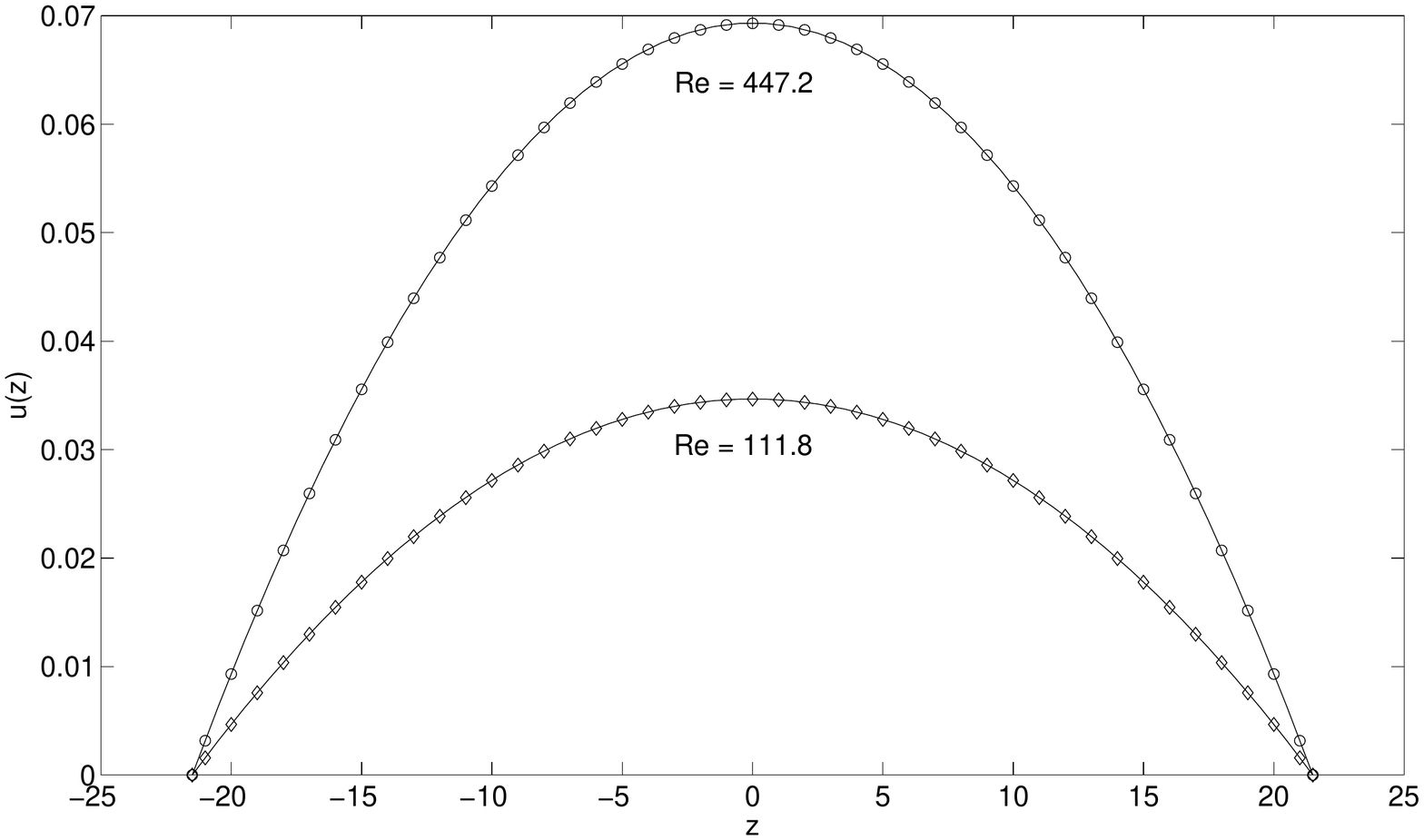}% Here is how to import EPS art
\caption{\label{fig:channel_velocity_profiles_differentRe} Flow between parallel plates with a constant body force: Comparison of velocity profiles computed by D3Q15 formulation of the central moment LBM with forcing term (open symbols) with analytical solution (lines) for different values of Reynolds number $Re$.}
\end{figure}

Both the central moment formulations including forcing terms derived earlier, i.e. for the D3Q15 and D3Q27 lattices, were implemented and assessed. Let us now discuss the validation studies carried out for these computational approaches for a set of canonical problems for which analytical solutions are available. First, we consider a fully developed flow between parallel plates driven by a constant body force. The grid resolution was chosen to be $3\times 3 \times 45$ with relaxation parameter $\omega^{\nu}=1.818$ for the second-order moments ($\omega^{\nu}=\omega_{j}$ where $j=4,5,6,7,8$) that controls the kinematic viscosity $\nu$ ($=0.0167$ here). Rest of the relaxation parameters were set to be unity for this case as well as for all the simple canonical problems considered in our present numerical accuracy study. It may be noted that other values could be used for kinetic modes involving more complex situations (e.g. turbulent flows) and could also be optimized to improve numerical stability. For these parameters, three different values of the body force were considered, i.e. $F_x=2\times 10^{-7}, 4\times 10^{-7}$ and $6\times 10^{-7}$ corresponding to Reynolds numbers (based on the centerline velocity and half-width between plates) $3.6, 7.2$ and $10.7$, respectively. Half-way bounce back boundary condition was implemented to impose the no-slip condition at both the walls. Figure~\ref{fig:compare_lattices_velocityprofiles_diffFx} shows a comparison between the computed results obtained using the central moment LBM implemented for D3Q15 and D3Q27 lattices with the analytical solution ($u(z)=u_0(1-(z/L)^2)$, where $L$ is the half-width and $u_0=F_xL^2/(2\nu)$).
Excellent agreement is seen for both formulations with the benchmark analytical solution. Since the results with D3Q15 and D3Q27 lattices are essentially identical with the former involving considerably lower operation count, henceforth we discuss the numerical performance only with the D3Q15 lattice. It may be noted that the advantage of the central moment formulation for this lattice, over the SRT approach lies in its enhanced numerical stability by independently and carefully adjusting the relaxation parameters for the kinetic modes. This and other assets such as better representation of kinetic layers are similar to the standard (raw moment) MRT approach. Comparison of such different collision models are subjects for future investigations. The central moment LBM using the D3Q15 lattice was further assessed for the channel flow problem at higher Reynolds numbers. By considering the same resolution as before and setting the body force as $F_x=1\times 10^{-6}$, two different Reynolds numbers of $111.8$ and $447.2$ were considered by using $\omega^{\nu}=1.923$ and $1.961$, respectively. Comparisons of computed and analytical solutions were made in Fig.~\ref{fig:channel_velocity_profiles_differentRe}, which again show good agreement.

\begin{figure}[h]
\includegraphics[width = 140mm]{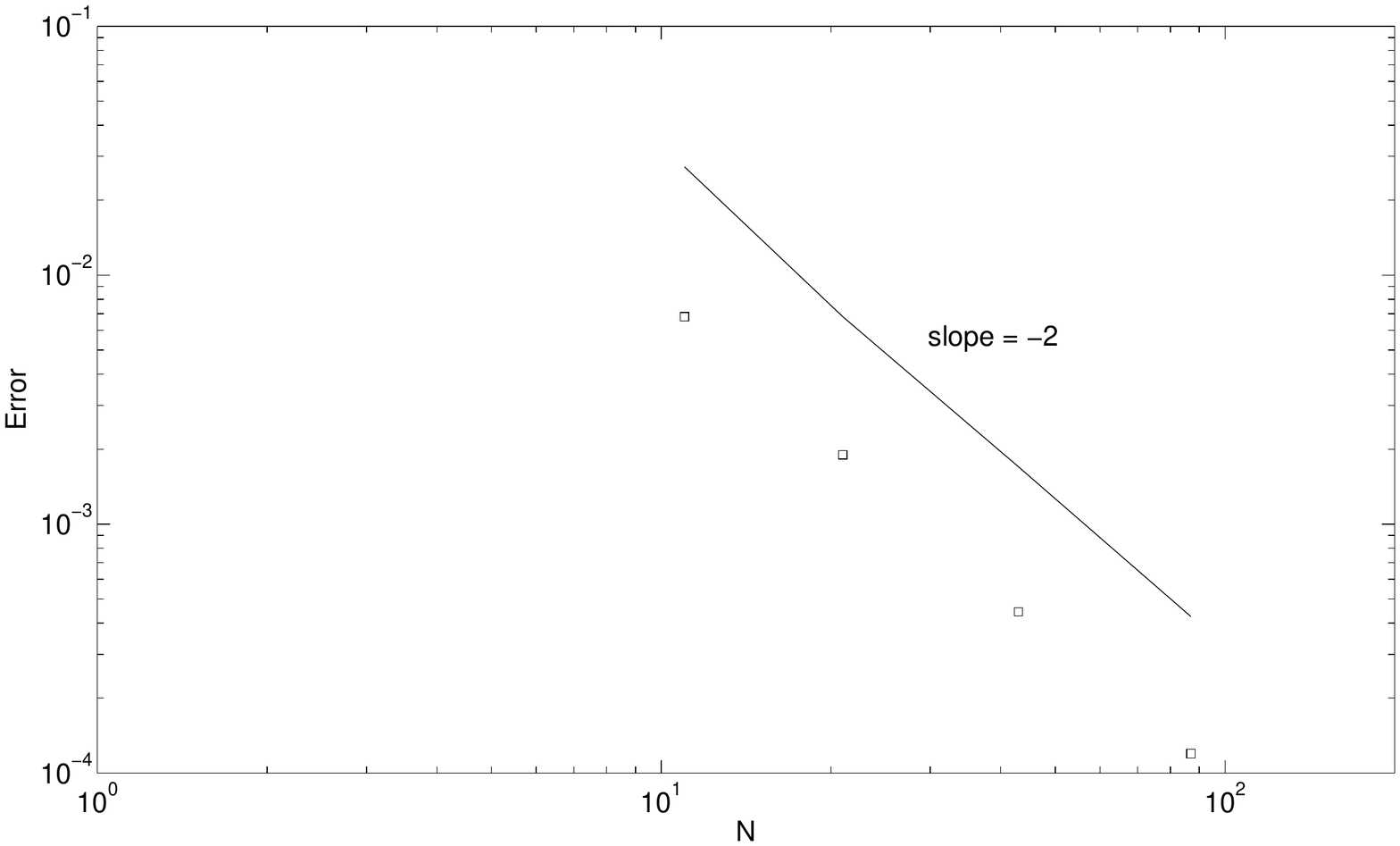}% Here is how to import EPS art
\caption{\label{fig:diffusive_scaling_error} Grid convergence study of the D3Q15 formulation of the central moment LBM with forcing term for channel flow under diffusive scaling. Symbols represent relative (root-mean-square) error between the computed and analytical solution. Best fit slope of computed results is $-1.96$.}
\end{figure}

In order to quantify the error between the computed and analytical solutions and its variation at different resolutions, i.e. to establish the grid convergence of the 3D central moment LBM, the following test was carried out. We again considered channel flow with the computational domain discretized using $3\times 3 \times N$ nodes, where $N$ is the number of nodes in the wall normal direction which was varied. The parameters were chosen so as to satisfy diffusive scaling: the fluid velocity (or the Mach number) was made to scale with the resolution, i.e. $u_0 \sim \Delta x \sim 1/N$. This ensures that the errors associated with compressibility effects also simultaneously reduce with increase in resolution. Thus, with a fixed viscosity $\nu$ to maintain constant Reynolds number ($Re=u_0L/\nu$) for different resolutions, using $u_0=F_xL^2/(2\nu)$ the body force scales as $F_x \sim 1/L^3 \sim 1/N^3$. Setting $\omega^{\nu}=1.818$ and considering $F_x=6.958\times 10^{-6}$ for the coarsest resolution ($N=13$) so that $Re=20.8$, the relative errors in velocity field at different resolutions were computed using $||\delta u||_2=\sum_i||(u_{c,i}-u_{a,i})||_2 / \sum_i ||u_{a,i}||_2$, where $u_{c,i}$ and $u_{a,i}$ are computed and analytical solutions, respectively, $||\cdot ||_2$ is the standard second-norm and the subscript $i$ represents the discrete location of the nodes. Figure~\ref{fig:diffusive_scaling_error} shows a log-log plot of the relative error as a function of the number of grid nodes.
It is evident that quadratic grid convergence is maintained by the 3D cascaded LBM.

\begin{figure}[h!]
\centering
\subfigure[]{\label{fig:Womersley_velocityprofile33}
\includegraphics[width = 140mm]{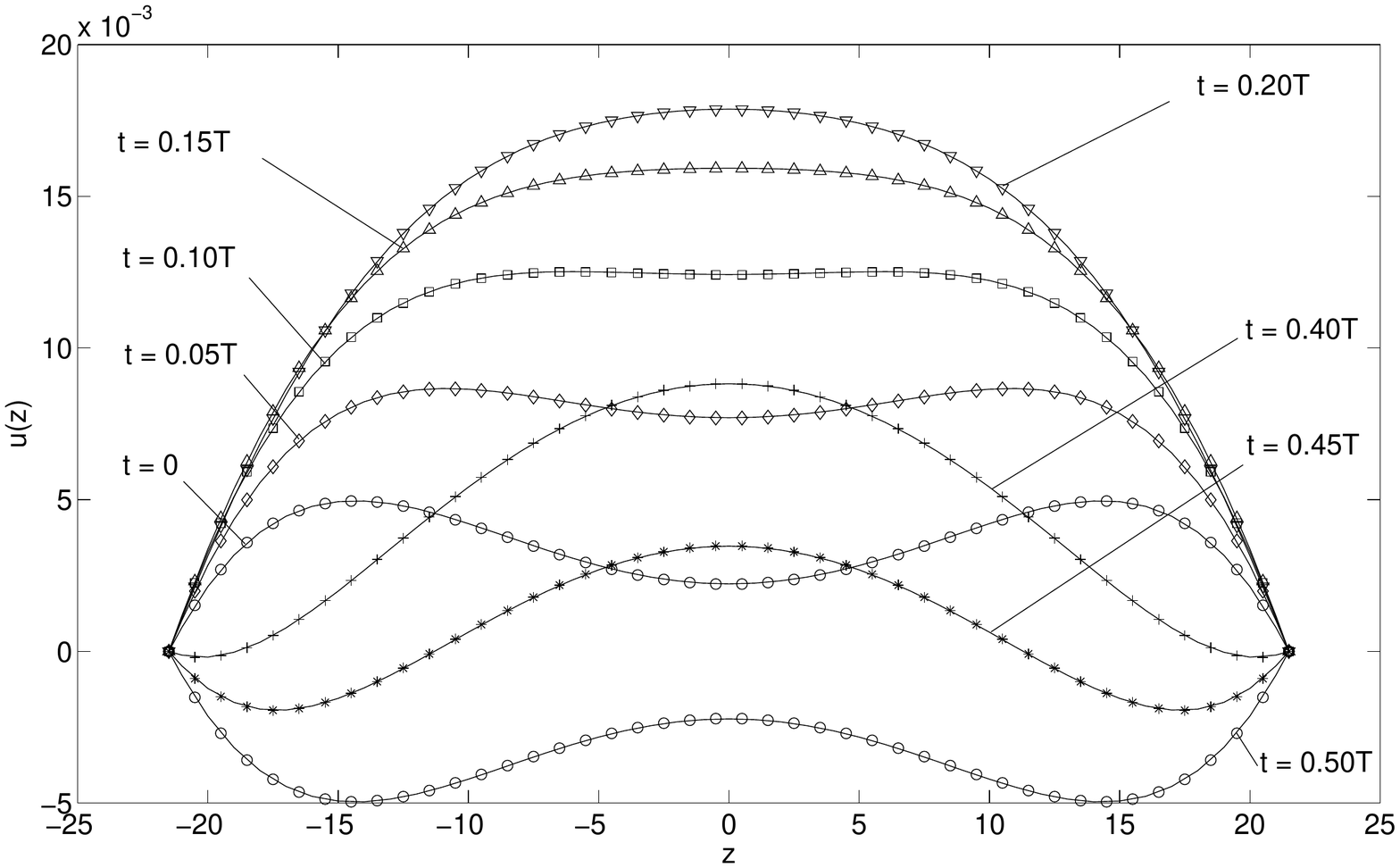}}\\
\subfigure[]{\label{fig:Womersley_velocityprofile66}
\includegraphics[width = 140mm]{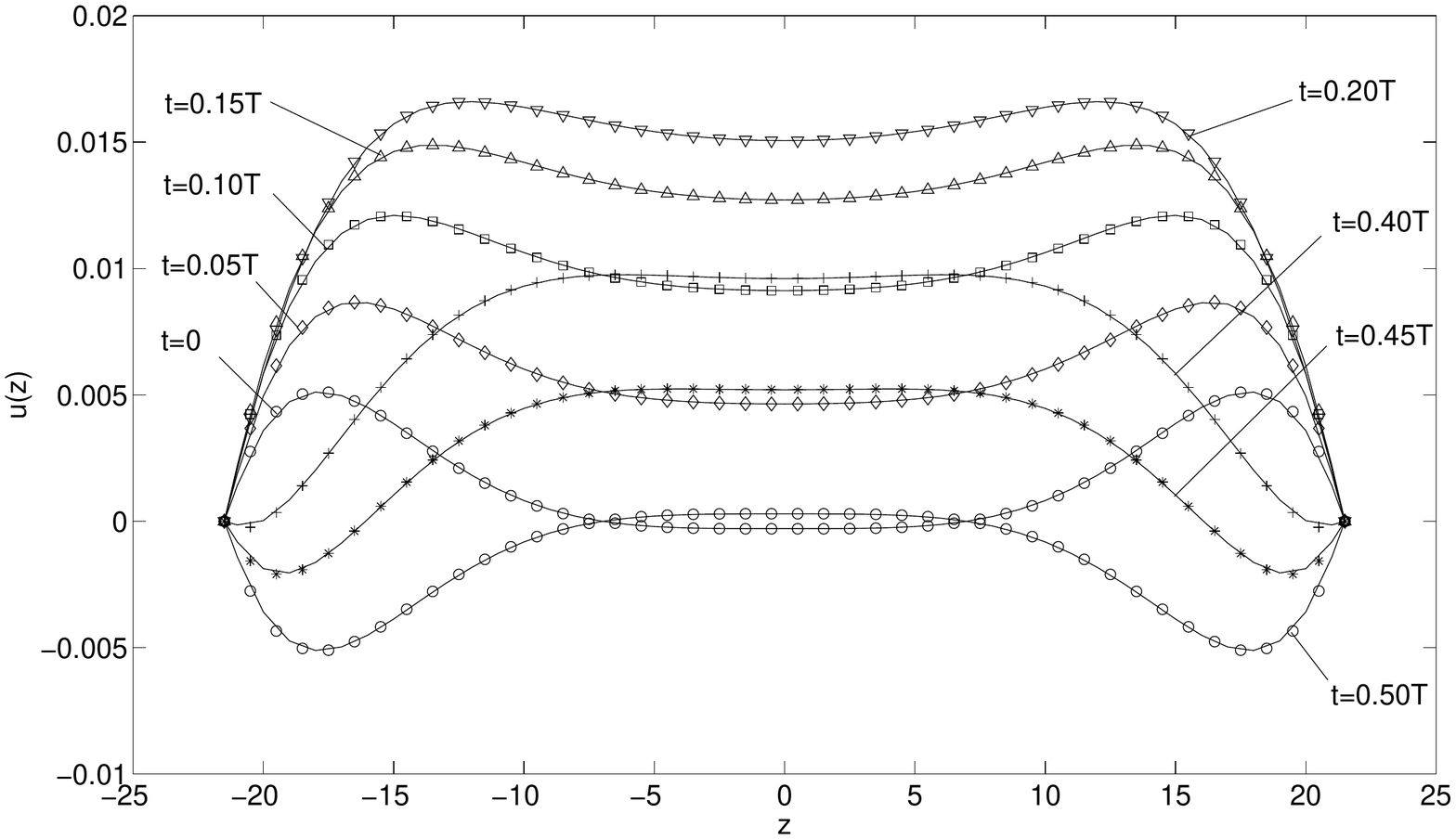}}
\caption{Flow between parallel plates with a temporally varying body force:
Comparison of velocity profiles computed by the D3Q15 formulation of the central moment LBM with forcing term (open symbols) with analytical solution (lines) at different instants within a time period $T$. (a) $Wo=3.3$ and (b) $Wo=6.6$, where $Wo$ is the Womersley number.}
\end{figure}

We will now consider a different canonical problem, where the imposed body force is time dependent and thus represents a more stringent test of the central moment formulation derived in this work. In particular, flow between parallel plates driven by a force which varies sinusoidally in time was computed using this approach. If $\Omega=2\pi/T$ is the angular frequency, where $T$ is the time period of the application of the body force, it may be represented as $F_x=F_mcos(\Omega t)$, where $F_m$ is its maximum amplitude. This problem, generally termed as Womersley flow, is characterized by the dimensionless parameter $\mathrm{Wo}=\sqrt{\frac{\Omega}{\nu}}L$, also called as the Womersley number representing the relative effect of the unsteady response of the fluid flow to the imposed unsteady body force. It has the following analytical solution for the velocity profile
$u_x(z)=\mathcal{R}\left[\frac{iF_m}{\Omega}\left\{1-\frac{cos\left(\beta\frac{z}{L}\right)}{cos(\beta)}\right\}e^{i\Omega t}\right]$, where $\beta=\sqrt{-i\mathrm{Wo}^2}$. Considering $3\times 3 \times 45$ nodes and setting the maximum force amplitude $F_m=1\times 10^{-5}$ with a time period of $T=10,000$, two different values of the relaxation parameter $\omega^{\nu}$, i.e. $1.724$ and $1.923$, were used, which correspond to $Wo$ of $3.3$ and $6.6$, respectively. Figures~\ref{fig:Womersley_velocityprofile33} and~\ref{fig:Womersley_velocityprofile66} show comparisons of the computed velocity profiles with the above analytical solution for different instants within the first half of the time period $T$ at these two Womersley numbers.
It is clear that the central moment LBM reproduces the sharp variations in the velocity profiles at different instants as prescribed by the analytical solution, with very good agreement found between them. Furthermore, the variations in both the amplitude as well as the lag of the response of the fluid flow as seen by its velocity profiles at different Womersley numbers are well reproduced by the computational approach presented in this work.

\begin{figure}[h!]
\centering
\subfigure[]{\label{fig:duct_surf_velocityprofile_computed}
\includegraphics[width = 140mm]{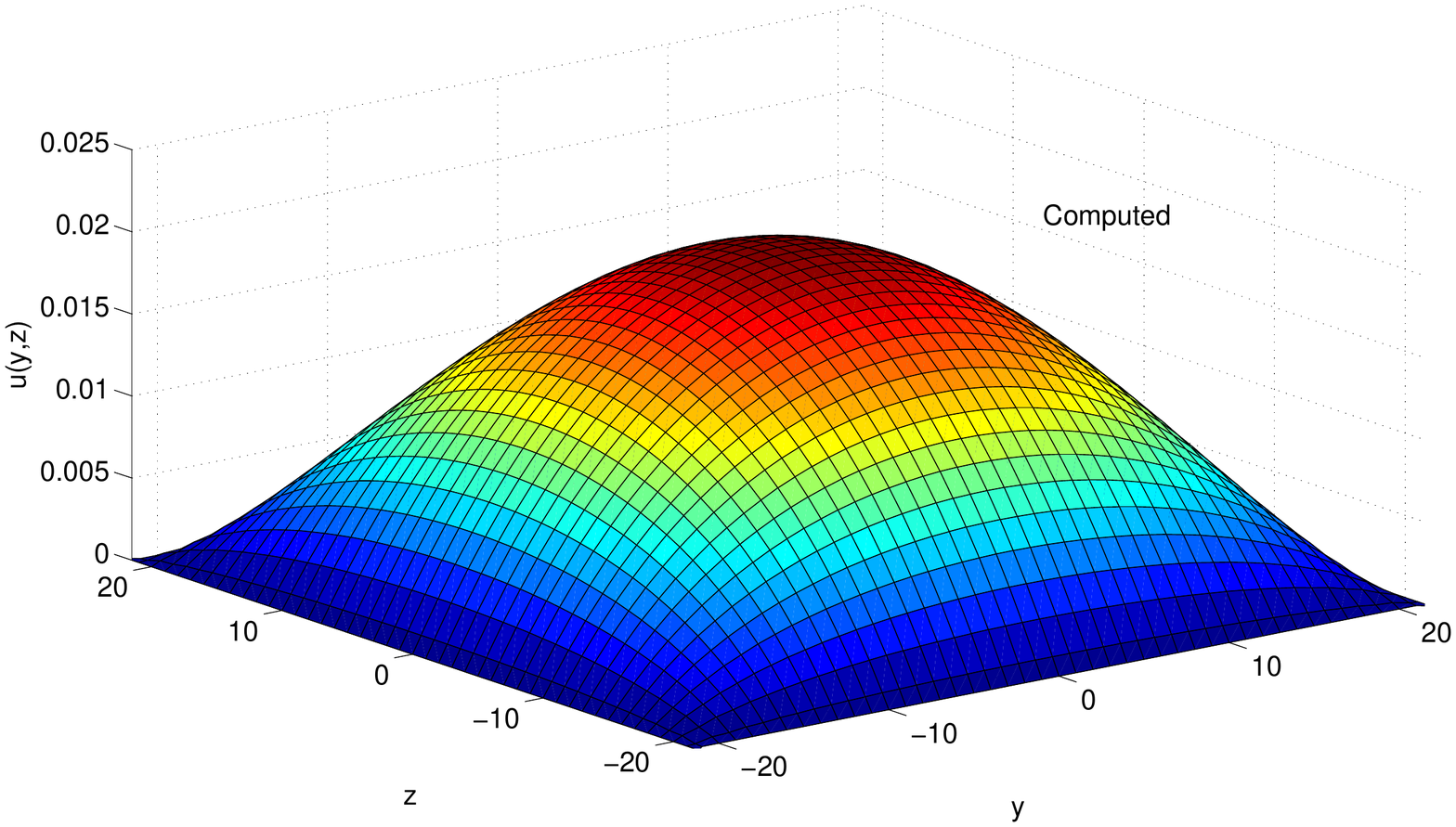}}\\
\subfigure[]{\label{fig:duct_surf_velocityprofile_analytical}
\includegraphics[width = 140mm]{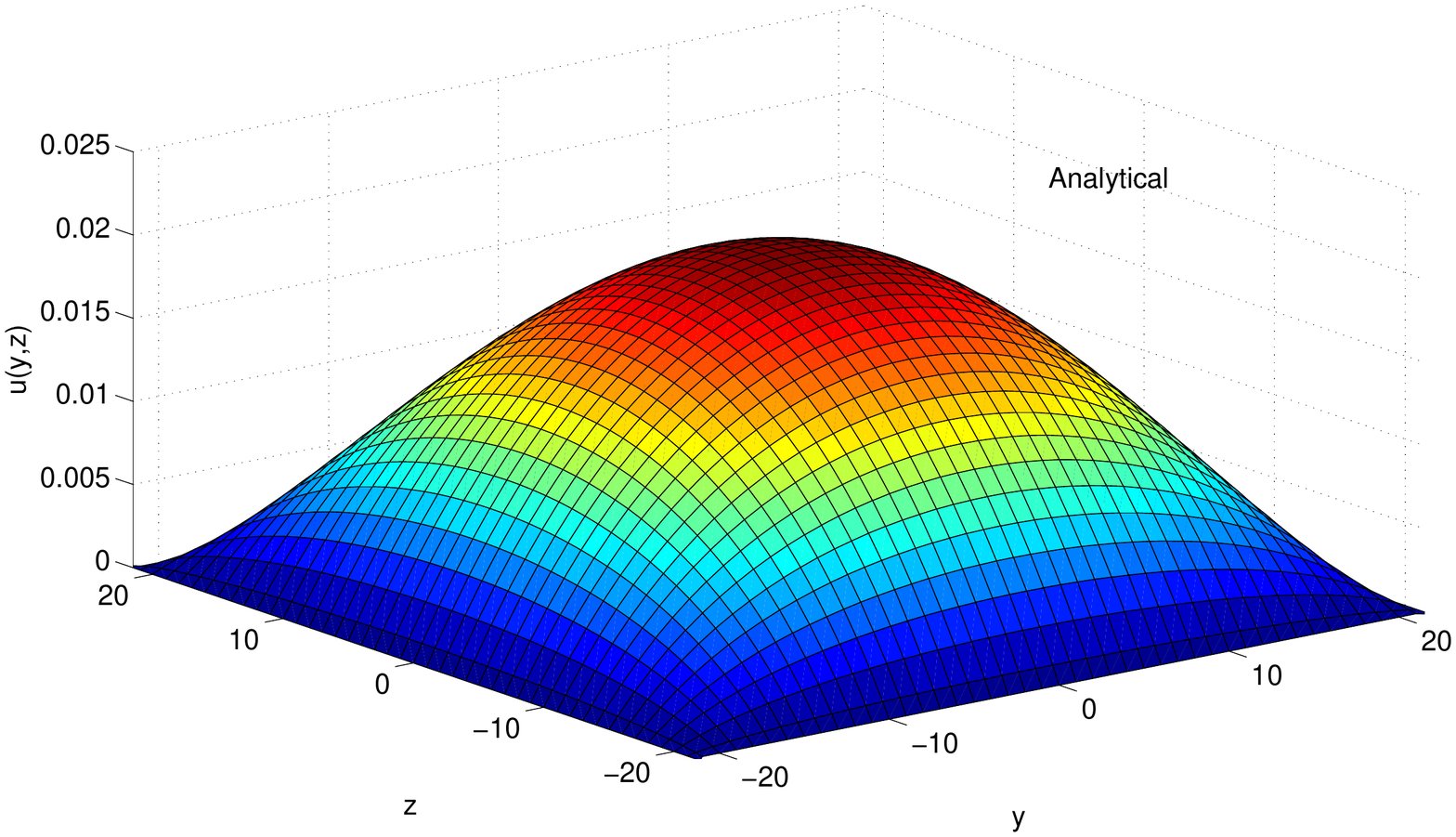}}
\caption{Flow through a square duct with side length $2a$ subjected to a constant body force: Comparison of surface contours of the velocity field for Reynolds number $Re=65.7$ (a) computed by the D3Q15 formulation of the central moment LBM with forcing term with (b) analytical solution (see Eq.~(\ref{eq:ductFourier})).}
\end{figure}

\begin{figure}[h!]
\includegraphics[width = 140mm]{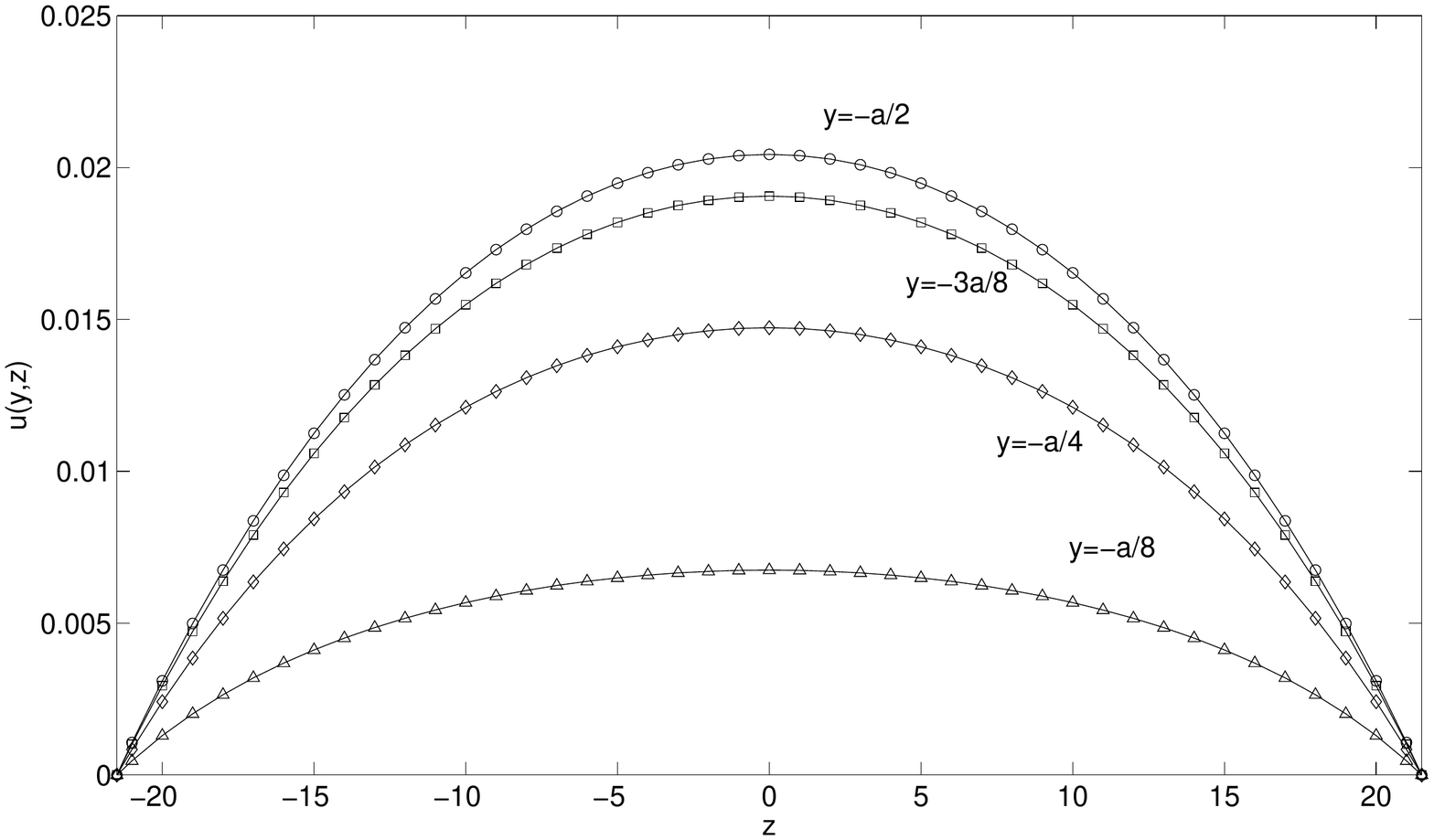}% Here is how to import EPS art
\caption{\label{fig:duct_velocityprofile_lines_comparison} Flow through a square duct with side length $2a$ subjected to a constant body force: Comparison of velocity profiles computed by the D3Q15 formulation of the central moment LBM with forcing term (symbols) with analytical solution (lines) (see Eq.~(\ref{eq:ductFourier})) at different locations in the duct cross-section for Reynolds number $Re=65.7$.}
\end{figure}

It may be noted that in all the problems considered above, the velocity field  has variation along only one coordinate direction normal to the direction of the driving body force. Thus, as a final example, we consider fully developed flow through a square duct in which the flow field has variations in both the coordinate directions normal to the direction of application of the driving force. It has the following analytical solution for the velocity field given in terms of an infinite orthogonal (Fourier) series~\cite{white05}
\begin{equation}
u(y,z)=\frac{16a^2F_x}{\rho \nu \pi^3}\sum_{n=1}^{\infty}(-1)^{(n-1)}\left[1-\frac{\cosh\left(\frac{(2n-1)\pi z}{2a}\right)}{\cosh\left(\frac{(2n-1)\pi}{2}\right)}\right]\frac{\cos\left(\frac{(2n-1)\pi y}{2a}\right)}{(2n-1)^3},\label{eq:ductFourier}
\end{equation}
where $-a \leq y \leq a$ and $-a \leq z \leq a$. Here, $a$ is the duct half-width. We considered the square duct to be resolved by using $3\times 45 \times 45$ nodes. A constant body force of magnitude $F_x=1\times 10^{-6}$ was applied by considering the relaxation parameter $\omega^{\nu}$ equal to $1.923$ such that the Reynolds number (based on maximum or centerline velocity and duct half-width) is equal to $65.7$. As before, the no-slip condition at the walls was imposed using the half-way bounce back approach. Figures~\ref{fig:duct_surf_velocityprofile_computed} and \ref{fig:duct_surf_velocityprofile_analytical} show a comparison between the surface contours of the computed and analytical solution of the velocity field for the above condition.
It is seen that the 3D central moment LBM with forcing term is able to reproduce the distribution of the velocity field over the cross-section of the square duct. In order to more clearly make a quantitative comparison, Fig.~\ref{fig:duct_velocityprofile_lines_comparison} shows plots of the computed velocity profiles at different locations in the cross-section of the duct and their comparison with the corresponding analytical solution (see Eq.~(\ref{eq:ductFourier}))
Evidently, the results computed using the central moment LBM are found to be in excellent agreement with the benchmark solution.

\section{\label{sec:summary}Summary and Conclusions}
A derivation of the 3D central moment lattice Boltzmann method (LBM) in the presence of forcing terms is presented. Suitable orthogonal moment basis for the D3Q27 and D3Q15 lattices are chosen for the specification of the local attractors and
source terms in terms of central moments. In particular, recently proposed factorized form of local attractors for higher moments
and de-aliased source terms that influence only conserved moments, which are obtained from modifications of the properties of the Maxwellian are considered in the construction of the approach. A Galilean invariance matching principle is invoked that exactly preserves the continuous central moments of the attractor and the source terms at the discrete level for all orders supported by the particle velocity set. Based on these, expressions for the temporally semi-implicit and second-order accurate sources are derived through an exact inversion due to the orthogonal properties of the moment basis. The central moment LBM, whose elements are equivalently expressed in terms of raw moments using the binominal theorem, represents frame independent fluid motion in the presence of general external or self-consistent internal forces. A set of numerical tests was carried out for problems involving channel flow driven by constant and temporally varying (periodic) body forces, and flow through a square duct to assess the accuracy of the central moment LBM with forcing term derived in this paper. It is demonstrated that the method maintains second-order grid convergence under diffusive scaling. Comparisons of the computed results are found to be in excellent agreement with analytical solutions for all the benchmark problems considered.

%\newpage
\clearpage

\appendix
\section{\label{app:d3q27matrix}Appendix: Orthogonal Matrix of the Moment Basis $\mathcal{K}$ for the D3Q27 Lattice}

A main element of the central moment method is the moment basis. The components of the orthogonal matrix of the the moment basis
derived in Sec.~\ref{sec:discreteparticlevelocity} (see Eq.~(\ref{eq:collisionmatrix1})) can be written as

$\mathcal{K}=$

\scriptsize
$
\left(
\begin{array}{rrrrrrrrrrrrrrrrrrrrrrrrrrr}
1	&0	&0	&0	&0	&0	&0	&0	&0	&-2	&0	&0	&0	&0	&0	&0	&0	&4	&0	&0	&0	&0	&0	&0	&0	&0	&-8\\
1	&1	&0	&0	&0	&0	&0	&1	&1	&-1	&-4	&0	&0	&0	&0	&0	&0	&0	&-4	&0	&0	&0	&0	&4	&0	&0	&4\\
1	&-1	&0	&0	&0	&0	&0	&1	&1	&-1	&4	&0	&0	&0	&0	&0	&0	&0	&-4	&0	&0	&0	&0	&-4	&0	&0	&4\\
1	&0	&1	&0	&0	&0	&0	&-1	&1	&-1	&0	&-4	&0	&0	&0	&0	&0	&0	&2	&-2	&0	&0	&0	&0	&4	&0	&4\\
1	&0	&-1	&0	&0	&0	&0	&-1	&1	&-1	&0	&4	&0	&0	&0	&0	&0	&0	&2	&-2	&0	&0	&0	&0	&-4	&0	&4\\
1	&0	&0	&1	&0	&0	&0	&0	&-2	&-1	&0	&0	&-4	&0	&0	&0	&0	&0	&2	&2	&0	&0	&0	&0	&0	&4	&4\\
1	&0	&0	&-1	&0	&0	&0	&0	&-2	&-1	&0	&0	&4	&0	&0	&0	&0	&0	&2	&2	&0	&0	&0	&0	&0	&-4	&4\\
1	&1	&1	&0	&1	&0	&0	&0	&2	&0	&-1	&-1	&0	&1	&1	&0	&0	&-1	&1	&1	&0	&0	&-2	&-2	&-2	&0	&-2\\
1	&-1	&1	&0	&-1	&0	&0	&0	&2	&0	&1	&-1	&0	&-1	&1	&0	&0	&-1	&1	&1	&0	&0	&2	&2	&-2	&0	&-2\\
1	&1	&-1	&0	&-1	&0	&0	&0	&2	&0	&-1	&1	&0	&1	&-1	&0	&0	&-1	&1	&1	&0	&0	&2	&-2	&2	&0	&-2\\
1	&-1	&-1	&0	&1	&0	&0	&0	&2	&0	&1	&1	&0	&-1	&-1	&0	&0	&-1	&1	&1	&0	&0	&-2	&2	&2	&0	&-2\\
1	&1	&0	&1	&0	&1	&0	&1	&-1	&0	&-1	&0	&-1	&-1	&0	&1	&0	&-1	&1	&-1	&0	&-2	&0	&-2	&0	&-2	&-2\\
1	&-1	&0	&1	&0	&-1	&0	&1	&-1	&0	&1	&0	&-1	&1	&0	&1	&0	&-1	&1	&-1	&0	&2	&0	&2	&0	&-2	&-2\\
1	&1	&0	&-1	&0	&-1	&0	&1	&-1	&0	&-1	&0	&1	&-1	&0	&-1	&0	&-1	&1	&-1	&0	&2	&0	&-2	&0	&2	&-2\\
1	&-1	&0	&-1	&0	&1	&0	&1	&-1	&0	&1	&0	&1	&1	&0	&-1	&0	&-1	&1	&-1	&0	&-2	&0	&2	&0	&2	&-2\\
1	&0	&1	&1	&0	&0	&1	&-1	&-1	&0	&0	&-1	&-1	&0	&-1	&-1	&0	&-1	&-2	&0	&-2	&0	&0	&0	&-2	&-2	&-2\\
1	&0	&-1	&1	&0	&0	&-1	&-1	&-1	&0	&0	&1	&-1	&0	&1	&-1	&0	&-1	&-2	&0	&2	&0	&0	&0	&2	&-2	&-2\\
1	&0	&1	&-1	&0	&0	&-1	&-1	&-1	&0	&0	&-1	&1	&0	&-1	&1	&0	&-1	&-2	&0	&2	&0	&0	&0	&-2	&2	&-2\\
1	&0	&-1	&-1	&0	&0	&1	&-1	&-1	&0	&0	&1	&1	&0	&1	&1	&0	&-1	&-2	&0	&-2	&0	&0	&0	&2	&2	&-2\\
1	&1	&1	&1	&1	&1	&1	&0	&0	&1	&2	&2	&2	&0	&0	&0	&1	&1	&0	&0	&1	&1	&1	&1	&1	&1	&1\\
1	&-1	&1	&1	&-1	&-1	&1	&0	&0	&1	&-2	&2	&2	&0	&0	&0	&-1	&1	&0	&0	&1	&-1	&-1	&-1	&1	&1	&1\\
1	&1	&-1	&1	&-1	&1	&-1	&0	&0	&1	&2	&-2	&2	&0	&0	&0	&-1	&1	&0	&0	&-1	&1	&-1	&1	&-1	&1	&1\\
1	&-1	&-1	&1	&1	&-1	&-1	&0	&0	&1	&-2	&-2	&2	&0	&0	&0	&1	&1	&0	&0	&-1	&-1	&1	&-1	&-1	&1	&1\\
1	&1	&1	&-1	&1	&-1	&-1	&0	&0	&1	&2	&2	&-2	&0	&0	&0	&-1	&1	&0	&0	&-1	&-1	&1	&1	&1	&-1	&1\\
1	&-1	&1	&-1	&-1	&1	&-1	&0	&0	&1	&-2	&2	&-2	&0	&0	&0	&1	&1	&0	&0	&-1	&1	&-1	&-1	&1	&-1	&1\\
1	&1	&-1	&-1	&-1	&-1	&1	&0	&0	&1	&2	&-2	&-2	&0	&0	&0	&1	&1	&0	&0	&1	&-1	&-1	&1	&-1	&-1	&1\\
1	&-1	&-1	&-1	&1	&1	&1	&0	&0	&1	&-2	&-2	&-2	&0	&0	&0	&-1	&1	&0	&0	&1	&1	&1	&-1	&-1	&-1	&1\\
\end{array}
\right)
$
\normalsize

\section{\label{app:nonconservedrawmoments}Appendix: Non-conserved Transformed Raw Moments for the D3Q27 Lattice}
The non-conserved transformed raw moments of various orders are given in terms of the subsets of the particle velocity directions
as
\begin{eqnarray}
\widehat{\overline{\kappa}}_{xy}^{'}&=&\braket{\overline{f}_{\alpha}|e_{\alpha x}e_{\alpha y}}=\sum_{\alpha =
0}^{26}\overline{f}_{\alpha} e_{\alpha x} e_{\alpha y}=\left(\sum_{\alpha}^{A_4}-\sum_{\alpha}^{B_4}\right)\otimes
\overline{f}_{\alpha},\nonumber\\
\widehat{\overline{\kappa}}_{xz}^{'}&=&\braket{\overline{f}_{\alpha}|e_{\alpha x}e_{\alpha z}}=\sum_{\alpha =
0}^{26}\overline{f}_{\alpha} e_{\alpha x} e_{\alpha z}=\left(\sum_{\alpha}^{A_5}-\sum_{\alpha}^{B_5}\right)\otimes
\overline{f}_{\alpha},\nonumber\\
\widehat{\overline{\kappa}}_{yz}^{'}&=&\braket{\overline{f}_{\alpha}|e_{\alpha y}e_{\alpha z}}=\sum_{\alpha =
0}^{26}\overline{f}_{\alpha} e_{\alpha y} e_{\alpha z}=\left(\sum_{\alpha}^{A_6}-\sum_{\alpha}^{B_6}\right)\otimes
\overline{f}_{\alpha},\nonumber\\
\widehat{\overline{\kappa}}_{xx}^{'}&=&\braket{\overline{f}_{\alpha}|e_{\alpha x}^2}=\sum_{\alpha = 0}^{26}\overline{f}_{\alpha} e_{\alpha x}^2=\left(\sum_{\alpha}^{A_7}\right)\otimes \overline{f}_{\alpha},\nonumber\\
\widehat{\overline{\kappa}}_{yy}^{'}&=&\braket{\overline{f}_{\alpha}|e_{\alpha y}^2}=\sum_{\alpha = 0}^{26}\overline{f}_{\alpha}
e_{\alpha y}^2=\left(\sum_{\alpha}^{A_8}\right)\otimes \overline{f}_{\alpha},\nonumber\\
\widehat{\overline{\kappa}}_{zz}^{'}&=&\braket{\overline{f}_{\alpha}|e_{\alpha z}^2}=\sum_{\alpha = 0}^{26}\overline{f}_{\alpha}
e_{\alpha z}^2=\left(\sum_{\alpha}^{A_9}\right)\otimes \overline{f}_{\alpha},\nonumber\\
\widehat{\overline{\kappa}}_{xyy}^{'}&=&\braket{\overline{f}_{\alpha}|e_{\alpha x}e_{\alpha y}^2}=\sum_{\alpha =
0}^{26}\overline{f}_{\alpha} e_{\alpha x} e_{\alpha y}^2=\left(\sum_{\alpha}^{A_{10}}-\sum_{\alpha}^{B_{10}}\right)\otimes     \overline{f}_{\alpha},\nonumber\\
\widehat{\overline{\kappa}}_{xzz}^{'}&=&\braket{\overline{f}_{\alpha}|e_{\alpha x}e_{\alpha z}^2}=\sum_{\alpha =
0}^{26}\overline{f}_{\alpha} e_{\alpha x} e_{\alpha z}^2=\left(\sum_{\alpha}^{A_{11}}-\sum_{\alpha}^{B_{11}}\right)\otimes
\overline{f}_{\alpha},\nonumber\\
\widehat{\overline{\kappa}}_{xxy}^{'}&=&\braket{\overline{f}_{\alpha}|e_{\alpha x}^2e_{\alpha y}}=\sum_{\alpha =
0}^{26}\overline{f}_{\alpha} e_{\alpha x}^2 e_{\alpha y}=\left(\sum_{\alpha}^{A_{12}}-\sum_{\alpha}^{B_{12}}\right)\otimes
\overline{f}_{\alpha},\nonumber\\
\widehat{\overline{\kappa}}_{yzz}^{'}&=&\braket{\overline{f}_{\alpha}|e_{\alpha y}e_{\alpha z}^2}=\sum_{\alpha =
0}^{26}\overline{f}_{\alpha} e_{\alpha y} e_{\alpha z}^2=\left(\sum_{\alpha}^{A_{13}}-\sum_{\alpha}^{B_{13}}\right)\otimes
\overline{f}_{\alpha},\nonumber\\
\widehat{\overline{\kappa}}_{xxz}^{'}&=&\braket{\overline{f}_{\alpha}|e_{\alpha x}^2 e_{\alpha z}}=\sum_{\alpha =
0}^{26}\overline{f}_{\alpha} e_{\alpha x}^2 e_{\alpha z}=\left(\sum_{\alpha}^{A_{14}}-\sum_{\alpha}^{B_{14}}\right)\otimes
\overline{f}_{\alpha},\nonumber\\
\widehat{\overline{\kappa}}_{yyz}^{'}&=&\braket{\overline{f}_{\alpha}|e_{\alpha y}^2 e_{\alpha z}}=\sum_{\alpha =
0}^{26}\overline{f}_{\alpha} e_{\alpha y}^2 e_{\alpha z}=\left(\sum_{\alpha}^{A_{15}}-\sum_{\alpha}^{B_{15}}\right)\otimes
\overline{f}_{\alpha},\nonumber\\
\widehat{\overline{\kappa}}_{xyz}^{'}&=&\braket{\overline{f}_{\alpha}|e_{\alpha x}e_{\alpha y}e_{\alpha z}}=\sum_{\alpha =
0}^{26}\overline{f}_{\alpha} e_{\alpha x} e_{\alpha y} e_{\alpha z}=\left(\sum_{\alpha}^{A_{16}}-\sum_{\alpha}^{B_{16}}\right)\otimes
\overline{f}_{\alpha},\nonumber\\
\widehat{\overline{\kappa}}_{xxyy}^{'}&=&\braket{\overline{f}_{\alpha}|e_{\alpha x}^2 e_{\alpha y}^2}=\sum_{\alpha = 0}^{26}\overline{f}_{\alpha}
e_{\alpha x}^2 e_{\alpha y}^2=\left(\sum_{\alpha}^{A_{17}}\right)\otimes \overline{f}_{\alpha},\nonumber\\
\widehat{\overline{\kappa}}_{xxzz}^{'}&=&\braket{\overline{f}_{\alpha}|e_{\alpha x}^2 e_{\alpha z}^2}=\sum_{\alpha = 0}^{26}\overline{f}_{\alpha}
e_{\alpha x}^2 e_{\alpha z}^2=\left(\sum_{\alpha}^{A_{18}}\right)\otimes \overline{f}_{\alpha},\nonumber\\
\widehat{\overline{\kappa}}_{yyzz}^{'}&=&\braket{\overline{f}_{\alpha}|e_{\alpha y}^2 e_{\alpha z}^2}=\sum_{\alpha = 0}^{26}\overline{f}_{\alpha}
e_{\alpha y}^2 e_{\alpha z}^2=\left(\sum_{\alpha}^{A_{19}}\right)\otimes \overline{f}_{\alpha},\nonumber\\
\widehat{\overline{\kappa}}_{xxyz}^{'}&=&\braket{\overline{f}_{\alpha}|e_{\alpha x}^2 e_{\alpha y}e_{\alpha z}}=\sum_{\alpha = 0}^{26}\overline{f}_{\alpha} e_{\alpha x}^2 e_{\alpha y} e_{\alpha z}=\left(\sum_{\alpha}^{A_{20}}-\sum_{\alpha}^{B_{20}}\right)\otimes \overline{f}_{\alpha},\nonumber\\
\widehat{\overline{\kappa}}_{xyyz}^{'}&=&\braket{\overline{f}_{\alpha}|e_{\alpha x} e_{\alpha y}^2 e_{\alpha z}}=\sum_{\alpha =
0}^{26}\overline{f}_{\alpha} e_{\alpha x} e_{\alpha y}^2 e_{\alpha z}=\left(\sum_{\alpha}^{A_{21}}-\sum_{\alpha}^{B_{21}}\right)\otimes
\overline{f}_{\alpha},\nonumber\\
\widehat{\overline{\kappa}}_{xyzz}^{'}&=&\braket{\overline{f}_{\alpha}|e_{\alpha x} e_{\alpha y} e_{\alpha z}^2}=\sum_{\alpha =
0}^{26}\overline{f}_{\alpha} e_{\alpha x} e_{\alpha y} e_{\alpha z}^2=\left(\sum_{\alpha}^{A_{22}}-\sum_{\alpha}^{B_{22}}\right)\otimes
\overline{f}_{\alpha},\nonumber\\
\widehat{\overline{\kappa}}_{xyyzz}^{'}&=&\braket{\overline{f}_{\alpha}|e_{\alpha x} e_{\alpha y}^2 e_{\alpha z}^2}=\sum_{\alpha =
0}^{26}\overline{f}_{\alpha} e_{\alpha x} e_{\alpha y}^2 e_{\alpha z}^2=\left(\sum_{\alpha}^{A_{23}}-\sum_{\alpha}^{B_{23}}\right)\otimes
\overline{f}_{\alpha},\nonumber\\
\widehat{\overline{\kappa}}_{xxyzz}^{'}&=&\braket{\overline{f}_{\alpha}|e_{\alpha x}^2 e_{\alpha y} e_{\alpha z}^2}=\sum_{\alpha =
0}^{26}\overline{f}_{\alpha} e_{\alpha x}^2 e_{\alpha y} e_{\alpha z}^2=\left(\sum_{\alpha}^{A_{24}}-\sum_{\alpha}^{B_{24}}\right)\otimes
\overline{f}_{\alpha},\nonumber\\
\widehat{\overline{\kappa}}_{xxyyz}^{'}&=&\braket{\overline{f}_{\alpha}|e_{\alpha x}^2 e_{\alpha y}^2 e_{\alpha z}}=\sum_{\alpha =
0}^{26}\overline{f}_{\alpha} e_{\alpha x}^2 e_{\alpha y}^2 e_{\alpha z}=\left(\sum_{\alpha}^{A_{25}}-\sum_{\alpha}^{B_{25}}\right)\otimes
\overline{f}_{\alpha},\nonumber\\
\widehat{\overline{\kappa}}_{xxyyzz}^{'}&=&\braket{\overline{f}_{\alpha}|e_{\alpha x}^2 e_{\alpha y}^2 e_{\alpha z}^2}=\sum_{\alpha =
0}^{26}\overline{f}_{\alpha} e_{\alpha x}^2 e_{\alpha y}^2 e_{\alpha z}^2=\left(\sum_{\alpha}^{A_{26}}-\sum_{\alpha}^{B_{26}}\right)\otimes
\overline{f}_{\alpha},\label{eq:rawmomenttransformednonconserved}
\end{eqnarray}
where
\begin{eqnarray*}
A_4&=&\left\{7,10,19,22,23,26\right\}, B_4=\left\{8,9,20,21,24,25\right\},\\ A_5&=&\left\{11,14,19,21,24,26\right\}, B_5=\left\{12,13,20,22,23,25\right\},\\
A_6&=&\left\{15,18,19,20,25,26\right\}, B_6=\left\{16,17,21,22,23,24\right\},\\
A_7&=&\left\{1,2,7,8,9,10,11,12,13,14,19,20,21,22,23,24,25,26\right\},\\
A_8&=&\left\{3,4,7,8,9,10,15,16,17,18,19,20,21,22,23,24,25,26\right\},\\
A_9&=&\left\{5,6,11,12,13,14,15,16,17,18,19,20,21,22,23,24,25,26\right\},\\
A_{10}&=&\left\{7,9,19,21,23,25\right\}, B_{10}=\left\{8,10,20,22,24,26\right\},\\
A_{11}&=&\left\{11,13,19,21,23,25\right\}, B_{11}=\left\{12,14,20,22,24,26\right\},\\
A_{12}&=&\left\{7,8,19,20,23,24\right\}, B_{12}=\left\{9,10,21,22,25,26\right\},\\ A_{13}&=&\left\{15,17,19,20,23,24\right\}, B_{13}=\left\{16,18,21,22,25,26\right\},\\
A_{14}&=&\left\{11,12,19,20,21,22\right\}, B_{14}=\left\{13,14,23,24,25,26\right\},\\
A_{15}&=&\left\{15,16,19,20,21,22\right\}, B_{15}=\left\{17,18,23,24,25,26\right\},\\
A_{16}&=&\left\{19,22,24,25\right\}, B_{16}=\left\{20,21,23,26\right\},\\
A_{17}&=&\left\{7,8,9,10,19,20,21,22,23,24,25,26\right\},\\
A_{18}&=&\left\{11,12,13,14,19,20,21,22,23,24,25,26\right\},\\ A_{19}&=&\left\{15,16,17,18,19,20,21,22,23,24,25,26\right\},\\
A_{20}&=&\left\{19,20,25,26\right\}, B_{20}=\left\{21,22,23,24\right\},\\
A_{21}&=&\left\{19,21,24,26\right\}, B_{21}=\left\{20,22,23,25\right\},\\
A_{22}&=&\left\{19,22,23,26\right\}, B_{22}=\left\{20,21,24,25\right\},\\
A_{23}&=&\left\{19,21,23,25\right\}, B_{23}=\left\{20,22,24,26\right\},\\
A_{24}&=&\left\{19,20,23,24\right\}, B_{24}=\left\{21,22,25,26\right\},\\
A_{25}&=&\left\{19,20,21,22\right\}, B_{25}=\left\{23,24,25,26\right\},\\
A_{26}&=&\left\{19,20,21,22,23,24,25,26\right\}.
\end{eqnarray*}

\section{\label{app:rawsourcemoments}Appendix: Raw Source Moments for the D3Q27 Lattice}
The raw source moments of various orders are given in terms of the Cartesian components of the force field as
\begin{eqnarray}
&&\widehat{\sigma}_{0}^{'}=\braket{S_{\alpha}|\rho}=0,\nonumber\\
&&\widehat{\sigma}_{x}^{'}=\braket{S_{\alpha}|e_{\alpha x}}=F_x,\nonumber\\
&&\widehat{\sigma}_{y}^{'}=\braket{S_{\alpha}|e_{\alpha y}}=F_y,\nonumber\\
&&\widehat{\sigma}_{z}^{'}=\braket{S_{\alpha}|e_{\alpha z}}=F_z,\nonumber\\
&&\widehat{\sigma}_{xx}^{'}=\braket{S_{\alpha}|e_{\alpha x}^2}=2F_xu_x,\nonumber\\
&&\widehat{\sigma}_{yy}^{'}=\braket{S_{\alpha}|e_{\alpha y}^2}=2F_yu_y,\nonumber\\
&&\widehat{\sigma}_{zz}^{'}=\braket{S_{\alpha}|e_{\alpha z}^2}=2F_zu_z,\nonumber\\
&&\widehat{\sigma}_{xy}^{'}=\braket{S_{\alpha}|e_{\alpha x}e_{\alpha y}}=F_xu_y+F_yu_x,\nonumber\\
&&\widehat{\sigma}_{xz}^{'}=\braket{S_{\alpha}|e_{\alpha x}e_{\alpha z}}=F_xu_z+F_zu_x,\nonumber\\
&&\widehat{\sigma}_{yz}^{'}=\braket{S_{\alpha}|e_{\alpha y}e_{\alpha z}}=F_yu_z+F_zu_y,\nonumber\\
&&\widehat{\sigma}_{xyy}^{'}=\braket{S_{\alpha}|e_{\alpha x}e_{\alpha y}^2}=F_xu_y^2+2F_yu_yu_x,\nonumber\\
&&\widehat{\sigma}_{xzz}^{'}=\braket{S_{\alpha}|e_{\alpha x}e_{\alpha z}^2}=F_xu_z^2+2F_zu_zu_x,\nonumber\\
&&\widehat{\sigma}_{xxy}^{'}=\braket{S_{\alpha}|e_{\alpha x}^2e_{\alpha y}}=F_yu_x^2+2F_xu_xu_y,\nonumber\\
&&\widehat{\sigma}_{yzz}^{'}=\braket{S_{\alpha}|e_{\alpha y}e_{\alpha z}^2}=F_yu_z^2+2F_zu_zu_y,\nonumber\\
&&\widehat{\sigma}_{xxz}^{'}=\braket{S_{\alpha}|e_{\alpha x}^2e_{\alpha y}}=F_zu_x^2+2F_xu_xu_z,\nonumber\\
&&\widehat{\sigma}_{yyz}^{'}=\braket{S_{\alpha}|e_{\alpha y}^2e_{\alpha z}}=F_zu_y^2+2F_yu_yu_z,\nonumber\\
&&\widehat{\sigma}_{xyz}^{'}=\braket{S_{\alpha}|e_{\alpha x}e_{\alpha y}e_{\alpha z}}=F_xu_yu_z+F_yu_xu_z+F_zu_xu_y,\nonumber\\
&&\widehat{\sigma}_{xxyy}^{'}=\braket{S_{\alpha}|e_{\alpha x}^2e_{\alpha y}^2}=2F_xu_xu_y^2+2F_yu_yu_x^2,\nonumber\\
&&\widehat{\sigma}_{xxzz}^{'}=\braket{S_{\alpha}|e_{\alpha x}^2e_{\alpha z}^2}=2F_xu_xu_z^2+2F_zu_zu_x^2,\nonumber\\
&&\widehat{\sigma}_{yyzz}^{'}=\braket{S_{\alpha}|e_{\alpha y}^2e_{\alpha z}^2}=2F_yu_yu_z^2+2F_zu_zu_y^2,\nonumber\\
&&\widehat{\sigma}_{xxyz}^{'}=\braket{S_{\alpha}|e_{\alpha x}^2e_{\alpha
y}e_{\alpha z}}=u_x^2(F_yu_z+F_zu_y)+2F_xu_xu_yu_z,\nonumber\\
&&\widehat{\sigma}_{xyyz}^{'}=\braket{S_{\alpha}|e_{\alpha x}e_{\alpha
y}^2 e_{\alpha z}}=u_y^2(F_xu_z+F_zu_x)+2F_yu_yu_xu_z,\nonumber\\
&&\widehat{\sigma}_{xyzz}^{'}=\braket{S_{\alpha}|e_{\alpha x}e_{\alpha
y}e_{\alpha z}^2}=u_z^2(F_xu_y+F_yu_x)+2F_zu_zu_xu_y,\nonumber\\
&&\widehat{\sigma}_{xyyzz}^{'}=\braket{S_{\alpha}|e_{\alpha x}e_{\alpha y}^2 e_{\alpha z}^2}=F_xu_y^2u_z^2+2u_xu_yu_z(F_yu_z+F_zu_y),\nonumber\\
&&\widehat{\sigma}_{xxyzz}^{'}=\braket{S_{\alpha}|e_{\alpha x}^2e_{\alpha y} e_{\alpha
z}^2}=F_yu_x^2u_z^2+2u_xu_yu_z(F_xu_z+F_zu_x),\nonumber\\
&&\widehat{\sigma}_{xxyyz}^{'}=\braket{S_{\alpha}|e_{\alpha x}^2e_{\alpha y}^2 e_{\alpha
z}}=F_zu_x^2u_y^2+2u_xu_yu_z(F_xu_y+F_yu_x),\nonumber\\
&&\widehat{\sigma}_{xxyyzz}^{'}=\braket{S_{\alpha}|e_{\alpha x}^2e_{\alpha y}^2 e_{\alpha
z}^2}=2u_xu_yu_z(F_xu_yu_z+F_yu_zu_x+F_zu_xu_y).\label{eq:rawmomentsourceterm}
\end{eqnarray}

\section{\label{app:rawsourcemomentsorthogonalrojections}Appendix: Projections of the Raw Source Moments to the Orthogonal Basis Vectors for the D3Q27 Lattice}
The orthogonal projections of the raw source moments can be written as
\begin{eqnarray}
\widehat{m}^{s}_{0}=\braket{K_0|S_{\alpha}}&=&0, \nonumber\\
\widehat{m}^{s}_{1}=\braket{K_1|S_{\alpha}}&=&F_x, \nonumber\\
\widehat{m}^{s}_{2}=\braket{K_2|S_{\alpha}}&=&F_y, \nonumber\\
\widehat{m}^{s}_{3}=\braket{K_3|S_{\alpha}}&=&F_z, \nonumber\\
\widehat{m}^{s}_{4}=\braket{K_4|S_{\alpha}}&=&(F_xu_y+F_yu_x), \nonumber\\
\widehat{m}^{s}_{5}=\braket{K_5|S_{\alpha}}&=&(F_xu_z+F_zu_x), \nonumber\\
\widehat{m}^{s}_{6}=\braket{K_6|S_{\alpha}}&=&(F_yu_z+F_zu_y), \nonumber\\
\widehat{m}^{s}_{7}=\braket{K_7|S_{\alpha}}&=&2(F_xu_x-F_yu_y), \nonumber\\
\widehat{m}^{s}_{8}=\braket{K_8|S_{\alpha}}&=&2(F_xu_x+F_yu_y-2F_zu_z), \nonumber\\
\widehat{m}^{s}_{9}=\braket{K_9|S_{\alpha}}&=&2(F_xu_x+F_yu_y+F_zu_z), \nonumber\\
\widehat{m}^{s}_{10}=\braket{K_{10}|S_{\alpha}}&=&3\left[F_x(u_y^2+u_z^2)+2u_x(F_yu_y+F_zu_z)\right]-4F_x, \nonumber\\
\widehat{m}^{s}_{11}=\braket{K_{11}|S_{\alpha}}&=&3\left[F_y(u_x^2+u_z^2)+2u_y(F_xu_x+F_zu_z)\right]-4F_y, \nonumber\\
\widehat{m}^{s}_{12}=\braket{K_{12}|S_{\alpha}}&=&3\left[F_z(u_x^2+u_y^2)+2u_z(F_xu_x+F_yu_y)\right]-4F_z, \nonumber\\
\widehat{m}^{s}_{13}=\braket{K_{13}|S_{\alpha}}&=&\left[F_x(u_y^2-u_z^2)+2u_x(F_yu_y-F_zu_z)\right], \nonumber\\
\widehat{m}^{s}_{14}=\braket{K_{14}|S_{\alpha}}&=&\left[F_y(u_x^2-u_z^2)+2u_y(F_xu_x-F_zu_z)\right], \nonumber\\
\widehat{m}^{s}_{15}=\braket{K_{15}|S_{\alpha}}&=&\left[F_z(u_x^2-u_y^2)+2u_z(F_xu_x-F_yu_y)\right], \nonumber\\
\widehat{m}^{s}_{16}=\braket{K_{16}|S_{\alpha}}&=&F_xu_yu_z+F_yu_xu_z+F_zu_xu_y, \nonumber\\
\widehat{m}^{s}_{17}=\braket{K_{17}|S_{\alpha}}&=&6\left[F_xu_x(u_y^2+u_z^2)+F_yu_y(u_x^2+u_z^2)+F_zu_z(u_x^2+u_y^2)\right]\nonumber\\
                                               &&-8(F_xu_x+F_yu_y+F_zu_z),\nonumber\\
\widehat{m}^{s}_{18}=\braket{K_{18}|S_{\alpha}}&=&6\left[F_xu_x(u_y^2+u_z^2)+F_yu_y(u_x^2-2u_z^2)+F_zu_z(u_x^2-2u_y^2)\right]\nonumber\\
                                               &&-4(2F_xu_x-F_yu_y-F_zu_z),\nonumber\\
\widehat{m}^{s}_{19}=\braket{K_{19}|S_{\alpha}}&=&6\left[F_xu_x(u_y^2-u_z^2)+u_x^2(F_yu_y-F_zu_z)\right]\nonumber\\
                                               &&-4(F_yu_y-F_zu_z),\nonumber\\
\widehat{m}^{s}_{20}=\braket{K_{20}|S_{\alpha}}&=&(3u_x^2-2)\left[F_yu_z+F_zu_y\right]+6F_xu_xu_yu_z, \nonumber\\
\widehat{m}^{s}_{21}=\braket{K_{21}|S_{\alpha}}&=&(3u_y^2-2)\left[F_xu_z+F_zu_x\right]+6F_yu_yu_xu_z, \nonumber\\
\widehat{m}^{s}_{22}=\braket{K_{22}|S_{\alpha}}&=&(3u_z^2-2)\left[F_xu_y+F_yu_x\right]+6F_zu_zu_xu_y, \nonumber\\
\widehat{m}^{s}_{23}=\braket{K_{23}|S_{\alpha}}&=&9F_x\left[u_y^2u_z^2-\frac{2}{3}\left((u_y^2+u_z^2)-\frac{2}{3}\right)\right]\nonumber\\
                                               &&+18u_x\left[F_yu_yu_z^2+F_zu_zu_y^2-\frac{2}{3}(F_yu_y+F_zu_z)\right], \nonumber\\
\widehat{m}^{s}_{24}=\braket{K_{24}|S_{\alpha}}&=&9F_y\left[u_x^2u_z^2-\frac{2}{3}\left((u_x^2+u_z^2)-\frac{2}{3}\right)\right]\nonumber\\
                                               &&+18u_y\left[F_xu_xu_z^2+F_zu_zu_x^2-\frac{2}{3}(F_xu_x+F_zu_z)\right], \nonumber\\
\widehat{m}^{s}_{25}=\braket{K_{25}|S_{\alpha}}&=&9F_z\left[u_x^2u_y^2-\frac{2}{3}\left((u_x^2+u_y^2)-\frac{2}{3}\right)\right]\nonumber\\
                                               &&+18u_z\left[F_xu_xu_y^2+F_yu_yu_x^2-\frac{2}{3}(F_xu_x+F_yu_y)\right], \nonumber\\
\widehat{m}^{s}_{26}=\braket{K_{26}|S_{\alpha}}&=&F_xu_x\left[54u_y^2u_z^2-36(u_y^2+u_z^2)+24\right]\nonumber\\
                                               &&+F_yu_y\left[54u_x^2u_z^2-36(u_x^2+u_z^2)+24\right]\nonumber\\
                                               &&+F_zu_z\left[54u_x^2u_y^2-36(u_x^2+u_y^2)+24\right].\label{eq:sourcetermprojection}
\end{eqnarray}

\section{\label{app:sourcetermsvelocityspace}Appendix: Source Terms in Particle Velocity Space for the D3Q27 Lattice}
The source terms in particle velocity space obtained by solving Eq.~(\ref{eq:sourceformulation1}) are given by
\begin{eqnarray}
S_0&=&\frac{1}{216}\left[8\widehat{m}^{s}_{0}-24\widehat{m}^{s}_{9}+24\widehat{m}^{s}_{17}-8\widehat{m}^{s}_{26}\right], \nonumber\\
S_1&=&\frac{1}{216}\left[8\widehat{m}^{s}_{0}+12\widehat{m}^{s}_{1}+18\widehat{m}^{s}_{7}+6\widehat{m}^{s}_{8}-12\widehat{m}^{s}_{9}-12\widehat{m}^{s}_{10}-12\widehat{m}^{s}_{18}\right.\nonumber\\
    &&\left.+12\widehat{m}^{s}_{23}+4\widehat{m}^{s}_{26}\right],\nonumber\\
S_2&=&\frac{1}{216}\left[8\widehat{m}^{s}_{0}-12\widehat{m}^{s}_{1}+18\widehat{m}^{s}_{7}+6\widehat{m}^{s}_{8}-12\widehat{m}^{s}_{9}+12\widehat{m}^{s}_{10}-12\widehat{m}^{s}_{18}\right.\nonumber\\
    &&\left.-12\widehat{m}^{s}_{23}+4\widehat{m}^{s}_{26}\right],\nonumber\\
S_3&=&\frac{1}{216}\left[8\widehat{m}^{s}_{0}+12\widehat{m}^{s}_{2}-18\widehat{m}^{s}_{7}+6\widehat{m}^{s}_{8}-12\widehat{m}^{s}_{9}-12\widehat{m}^{s}_{11}+6\widehat{m}^{s}_{18}\right.\nonumber\\
    &&\left.-18\widehat{m}^{s}_{19}+12\widehat{m}^{s}_{24}+4\widehat{m}^{s}_{26}\right],\nonumber\\
S_4&=&\frac{1}{216}\left[8\widehat{m}^{s}_{0}-12\widehat{m}^{s}_{2}-18\widehat{m}^{s}_{7}+6\widehat{m}^{s}_{8}-12\widehat{m}^{s}_{9}+12\widehat{m}^{s}_{11}+6\widehat{m}^{s}_{18}\right.\nonumber\\
    &&\left.-18\widehat{m}^{s}_{19}-12\widehat{m}^{s}_{24}+4\widehat{m}^{s}_{26}\right],\nonumber\\
S_5&=&\frac{1}{216}\left[8\widehat{m}^{s}_{0}+12\widehat{m}^{s}_{3}-12\widehat{m}^{s}_{8}-12\widehat{m}^{s}_{9}-12\widehat{m}^{s}_{12}+6\widehat{m}^{s}_{18}+18\widehat{m}^{s}_{19}\right.\nonumber\\
    &&\left.+12\widehat{m}^{s}_{25}+4\widehat{m}^{s}_{26}\right],\nonumber\\                                                                                               S_6&=&\frac{1}{216}\left[8\widehat{m}^{s}_{0}-12\widehat{m}^{s}_{3}-12\widehat{m}^{s}_{8}-12\widehat{m}^{s}_{9}+12\widehat{m}^{s}_{12}+6\widehat{m}^{s}_{18}+18\widehat{m}^{s}_{19}\right.\nonumber\\
    &&\left.-12\widehat{m}^{s}_{25}+4\widehat{m}^{s}_{26}\right],\nonumber\\
S_7&=&\frac{1}{216}\left[8\widehat{m}^{s}_{0}+12\widehat{m}^{s}_{1}+12\widehat{m}^{s}_{2}+18\widehat{m}^{s}_{4}+12\widehat{m}^{s}_{8}-3\widehat{m}^{s}_{10}-3\widehat{m}^{s}_{11}\right.\nonumber\\
    &&\left.+27\widehat{m}^{s}_{13}+27\widehat{m}^{s}_{14}-6\widehat{m}^{s}_{17}+3\widehat{m}^{s}_{18}+9\widehat{m}^{s}_{19}-18\widehat{m}^{s}_{22}-6\widehat{m}^{s}_{23}\right.\nonumber\\
    &&\left.-6\widehat{m}^{s}_{24}-2\widehat{m}^{s}_{26}\right],\nonumber\\
S_8&=&\frac{1}{216}\left[8\widehat{m}^{s}_{0}-12\widehat{m}^{s}_{1}+12\widehat{m}^{s}_{2}-18\widehat{m}^{s}_{4}+12\widehat{m}^{s}_{8}+3\widehat{m}^{s}_{10}-3\widehat{m}^{s}_{11}\right.\nonumber\\
    &&\left.-27\widehat{m}^{s}_{13}+27\widehat{m}^{s}_{14}-6\widehat{m}^{s}_{17}+3\widehat{m}^{s}_{18}+9\widehat{m}^{s}_{19}+18\widehat{m}^{s}_{22}+6\widehat{m}^{s}_{23}\right.\nonumber\\
    &&\left.-6\widehat{m}^{s}_{24}-2\widehat{m}^{s}_{26}\right],\nonumber\\
S_9&=&\frac{1}{216}\left[8\widehat{m}^{s}_{0}+12\widehat{m}^{s}_{1}-12\widehat{m}^{s}_{2}-18\widehat{m}^{s}_{4}+12\widehat{m}^{s}_{8}-3\widehat{m}^{s}_{10}+3\widehat{m}^{s}_{11}\right.\nonumber\\
    &&\left.+27\widehat{m}^{s}_{13}-27\widehat{m}^{s}_{14}-6\widehat{m}^{s}_{17}+3\widehat{m}^{s}_{18}+9\widehat{m}^{s}_{19}+18\widehat{m}^{s}_{22}-6\widehat{m}^{s}_{23}\right.\nonumber\\
    &&\left.+6\widehat{m}^{s}_{24}-2\widehat{m}^{s}_{26}\right],\nonumber\\
S_{10}&=&\frac{1}{216}\left[8\widehat{m}^{s}_{0}-12\widehat{m}^{s}_{1}-12\widehat{m}^{s}_{2}+18\widehat{m}^{s}_{4}+12\widehat{m}^{s}_{8}+3\widehat{m}^{s}_{10}+3\widehat{m}^{s}_{11}\right.\nonumber\\
    &&\left.-27\widehat{m}^{s}_{13}-27\widehat{m}^{s}_{14}-6\widehat{m}^{s}_{17}+3\widehat{m}^{s}_{18}+9\widehat{m}^{s}_{19}-18\widehat{m}^{s}_{22}+6\widehat{m}^{s}_{23}\right.\nonumber\\
    &&\left.+6\widehat{m}^{s}_{24}-2\widehat{m}^{s}_{26}\right],\nonumber\\
S_{11}&=&\frac{1}{216}\left[8\widehat{m}^{s}_{0}+12\widehat{m}^{s}_{1}+12\widehat{m}^{s}_{3}+18\widehat{m}^{s}_{5}+18\widehat{m}^{s}_{7}-6\widehat{m}^{s}_{8}-3\widehat{m}^{s}_{10}\right.\nonumber\\
    &&\left.-3\widehat{m}^{s}_{12}-27\widehat{m}^{s}_{13}+27\widehat{m}^{s}_{15}-6\widehat{m}^{s}_{17}+3\widehat{m}^{s}_{18}-9\widehat{m}^{s}_{19}-18\widehat{m}^{s}_{21}\right.\nonumber\\
    &&\left.-6\widehat{m}^{s}_{23}-6\widehat{m}^{s}_{25}-2\widehat{m}^{s}_{26}\right],\nonumber\\                                                                                                                                    S_{12}&=&\frac{1}{216}\left[8\widehat{m}^{s}_{0}-12\widehat{m}^{s}_{1}+12\widehat{m}^{s}_{3}-18\widehat{m}^{s}_{5}+18\widehat{m}^{s}_{7}-6\widehat{m}^{s}_{8}+3\widehat{m}^{s}_{10}\right.\nonumber\\
    &&\left.-3\widehat{m}^{s}_{12}+27\widehat{m}^{s}_{13}+27\widehat{m}^{s}_{15}-6\widehat{m}^{s}_{17}+3\widehat{m}^{s}_{18}-9\widehat{m}^{s}_{19}+18\widehat{m}^{s}_{21}\right.\nonumber\\
    &&\left.+6\widehat{m}^{s}_{23}-6\widehat{m}^{s}_{25}-2\widehat{m}^{s}_{26}\right],\nonumber\\
S_{13}&=&\frac{1}{216}\left[8\widehat{m}^{s}_{0}+12\widehat{m}^{s}_{1}-12\widehat{m}^{s}_{3}-18\widehat{m}^{s}_{5}+18\widehat{m}^{s}_{7}-6\widehat{m}^{s}_{8}-3\widehat{m}^{s}_{10}\right.\nonumber\\
    &&\left.+3\widehat{m}^{s}_{12}-27\widehat{m}^{s}_{13}-27\widehat{m}^{s}_{15}-6\widehat{m}^{s}_{17}+3\widehat{m}^{s}_{18}-9\widehat{m}^{s}_{19}+18\widehat{m}^{s}_{21}\right.\nonumber\\
    &&\left.-6\widehat{m}^{s}_{23}+6\widehat{m}^{s}_{25}-2\widehat{m}^{s}_{26}\right],\nonumber\\
S_{14}&=&\frac{1}{216}\left[8\widehat{m}^{s}_{0}-12\widehat{m}^{s}_{1}-12\widehat{m}^{s}_{3}+18\widehat{m}^{s}_{5}+18\widehat{m}^{s}_{7}-6\widehat{m}^{s}_{8}+3\widehat{m}^{s}_{10}\right.\nonumber\\
    &&\left.+3\widehat{m}^{s}_{12}+27\widehat{m}^{s}_{13}-27\widehat{m}^{s}_{15}-6\widehat{m}^{s}_{17}+3\widehat{m}^{s}_{18}-9\widehat{m}^{s}_{19}-18\widehat{m}^{s}_{21}\right.\nonumber\\
    &&\left.+6\widehat{m}^{s}_{23}+6\widehat{m}^{s}_{25}-2\widehat{m}^{s}_{26}\right],\nonumber\\
S_{15}&=&\frac{1}{216}\left[8\widehat{m}^{s}_{0}+12\widehat{m}^{s}_{2}+12\widehat{m}^{s}_{3}+18\widehat{m}^{s}_{6}-18\widehat{m}^{s}_{7}-6\widehat{m}^{s}_{8}-3\widehat{m}^{s}_{11}\right.\nonumber\\
    &&\left.-3\widehat{m}^{s}_{12}-27\widehat{m}^{s}_{14}-27\widehat{m}^{s}_{15}-6\widehat{m}^{s}_{17}-6\widehat{m}^{s}_{18}-18\widehat{m}^{s}_{20}-6\widehat{m}^{s}_{24}\right.\nonumber\\                            &&\left.-6\widehat{m}^{s}_{25}-2\widehat{m}^{s}_{26}\right],\nonumber\\
S_{16}&=&\frac{1}{216}\left[8\widehat{m}^{s}_{0}-12\widehat{m}^{s}_{2}+12\widehat{m}^{s}_{3}-18\widehat{m}^{s}_{6}-18\widehat{m}^{s}_{7}-6\widehat{m}^{s}_{8}+3\widehat{m}^{s}_{11}\right.\nonumber\\
    &&\left.-3\widehat{m}^{s}_{12}+27\widehat{m}^{s}_{14}-27\widehat{m}^{s}_{15}-6\widehat{m}^{s}_{17}-6\widehat{m}^{s}_{18}+18\widehat{m}^{s}_{20}+6\widehat{m}^{s}_{24}\right.\nonumber\\                            &&\left.-6\widehat{m}^{s}_{25}-2\widehat{m}^{s}_{26}\right],\nonumber\\
S_{17}&=&\frac{1}{216}\left[8\widehat{m}^{s}_{0}+12\widehat{m}^{s}_{2}-12\widehat{m}^{s}_{3}-18\widehat{m}^{s}_{6}-18\widehat{m}^{s}_{7}-6\widehat{m}^{s}_{8}-3\widehat{m}^{s}_{11}\right.\nonumber\\
    &&\left.+3\widehat{m}^{s}_{12}-27\widehat{m}^{s}_{14}+27\widehat{m}^{s}_{15}-6\widehat{m}^{s}_{17}-6\widehat{m}^{s}_{18}+18\widehat{m}^{s}_{20}-6\widehat{m}^{s}_{24}\right.\nonumber\\                            &&\left.+6\widehat{m}^{s}_{25}-2\widehat{m}^{s}_{26}\right],\nonumber\\
S_{18}&=&\frac{1}{216}\left[8\widehat{m}^{s}_{0}-12\widehat{m}^{s}_{2}-12\widehat{m}^{s}_{3}+18\widehat{m}^{s}_{6}-18\widehat{m}^{s}_{7}-6\widehat{m}^{s}_{8}+3\widehat{m}^{s}_{11}\right.\nonumber\\
    &&\left.+3\widehat{m}^{s}_{12}+27\widehat{m}^{s}_{14}+27\widehat{m}^{s}_{15}-6\widehat{m}^{s}_{17}-6\widehat{m}^{s}_{18}-18\widehat{m}^{s}_{20}+6\widehat{m}^{s}_{24}\right.\nonumber\\                            &&\left.+6\widehat{m}^{s}_{25}-2\widehat{m}^{s}_{26}\right],\nonumber\\
S_{19}&=&\frac{1}{216}\left[8\widehat{m}^{s}_{0}+12(\widehat{m}^{s}_{1}+\widehat{m}^{s}_{2}+\widehat{m}^{s}_{3})+18(\widehat{m}^{s}_{4}+\widehat{m}^{s}_{5}+\widehat{m}^{s}_{6})\right.\nonumber\\
    &&\left.\left.+12\widehat{m}^{s}_{9}+6(\widehat{m}^{s}_{10}+\widehat{m}^{s}_{11}+\widehat{m}^{s}_{12})+27\widehat{m}^{s}_{16}+6\widehat{m}^{s}_{17}+9(\widehat{m}^{s}_{20}\right.\right.\nonumber\\                    &&\left.+\widehat{m}^{s}_{21}+\widehat{m}^{s}_{22})+3(\widehat{m}^{s}_{23}+\widehat{m}^{s}_{24}+\widehat{m}^{s}_{25})+\widehat{m}^{s}_{26}\right],\nonumber\\
S_{20}&=&\frac{1}{216}\left[8\widehat{m}^{s}_{0}+12(-\widehat{m}^{s}_{1}+\widehat{m}^{s}_{2}+\widehat{m}^{s}_{3})+18(-\widehat{m}^{s}_{4}-\widehat{m}^{s}_{5}+\widehat{m}^{s}_{6})\right.\nonumber\\
    &&\left.\left.+12\widehat{m}^{s}_{9}+6(-\widehat{m}^{s}_{10}+\widehat{m}^{s}_{11}+\widehat{m}^{s}_{12})-27\widehat{m}^{s}_{16}+6\widehat{m}^{s}_{17}+9(\widehat{m}^{s}_{20}\right.\right.\nonumber\\                    &&\left.-\widehat{m}^{s}_{21}-\widehat{m}^{s}_{22})+3(-\widehat{m}^{s}_{23}+\widehat{m}^{s}_{24}+\widehat{m}^{s}_{25})+\widehat{m}^{s}_{26}\right],\nonumber\\
S_{21}&=&\frac{1}{216}\left[8\widehat{m}^{s}_{0}+12(\widehat{m}^{s}_{1}-\widehat{m}^{s}_{2}+\widehat{m}^{s}_{3})+18(-\widehat{m}^{s}_{4}+\widehat{m}^{s}_{5}-\widehat{m}^{s}_{6})\right.\nonumber\\
    &&\left.\left.+12\widehat{m}^{s}_{9}+6(\widehat{m}^{s}_{10}-\widehat{m}^{s}_{11}+\widehat{m}^{s}_{12})-27\widehat{m}^{s}_{16}+6\widehat{m}^{s}_{17}+9(-\widehat{m}^{s}_{20}\right.\right.\nonumber\\                    &&\left.+\widehat{m}^{s}_{21}-\widehat{m}^{s}_{22})+3(\widehat{m}^{s}_{23}-\widehat{m}^{s}_{24}+\widehat{m}^{s}_{25})+\widehat{m}^{s}_{26}\right],\nonumber\\
S_{22}&=&\frac{1}{216}\left[8\widehat{m}^{s}_{0}+12(-\widehat{m}^{s}_{1}-\widehat{m}^{s}_{2}+\widehat{m}^{s}_{3})+18(\widehat{m}^{s}_{4}-\widehat{m}^{s}_{5}-\widehat{m}^{s}_{6})\right.\nonumber\\
    &&\left.\left.+12\widehat{m}^{s}_{9}+6(-\widehat{m}^{s}_{10}-\widehat{m}^{s}_{11}+\widehat{m}^{s}_{12})+27\widehat{m}^{s}_{16}+6\widehat{m}^{s}_{17}+9(-\widehat{m}^{s}_{20}\right.\right.\nonumber\\                    &&\left.-\widehat{m}^{s}_{21}+\widehat{m}^{s}_{22})+3(-\widehat{m}^{s}_{23}-\widehat{m}^{s}_{24}+\widehat{m}^{s}_{25})+\widehat{m}^{s}_{26}\right],\nonumber\\
S_{23}&=&\frac{1}{216}\left[8\widehat{m}^{s}_{0}+12(\widehat{m}^{s}_{1}+\widehat{m}^{s}_{2}-\widehat{m}^{s}_{3})+18(\widehat{m}^{s}_{4}-\widehat{m}^{s}_{5}-\widehat{m}^{s}_{6})\right.\nonumber\\
    &&\left.\left.+12\widehat{m}^{s}_{9}+6(\widehat{m}^{s}_{10}+\widehat{m}^{s}_{11}-\widehat{m}^{s}_{12})-27\widehat{m}^{s}_{16}+6\widehat{m}^{s}_{17}+9(-\widehat{m}^{s}_{20}\right.\right.\nonumber\\                    &&\left.-\widehat{m}^{s}_{21}+\widehat{m}^{s}_{22})+3(\widehat{m}^{s}_{23}+\widehat{m}^{s}_{24}-\widehat{m}^{s}_{25})+\widehat{m}^{s}_{26}\right],\nonumber\\
S_{24}&=&\frac{1}{216}\left[8\widehat{m}^{s}_{0}+12(-\widehat{m}^{s}_{1}+\widehat{m}^{s}_{2}-\widehat{m}^{s}_{3})+18(-\widehat{m}^{s}_{4}+\widehat{m}^{s}_{5}-\widehat{m}^{s}_{6})\right.\nonumber\\
    &&\left.\left.+12\widehat{m}^{s}_{9}+6(-\widehat{m}^{s}_{10}+\widehat{m}^{s}_{11}-\widehat{m}^{s}_{12})+27\widehat{m}^{s}_{16}+6\widehat{m}^{s}_{17}+9(-\widehat{m}^{s}_{20}\right.\right.\nonumber\\                    &&\left.+\widehat{m}^{s}_{21}-\widehat{m}^{s}_{22})+3(-\widehat{m}^{s}_{23}+\widehat{m}^{s}_{24}-\widehat{m}^{s}_{25})+\widehat{m}^{s}_{26}\right],\nonumber\\
S_{25}&=&\frac{1}{216}\left[8\widehat{m}^{s}_{0}+12(\widehat{m}^{s}_{1}-\widehat{m}^{s}_{2}-\widehat{m}^{s}_{3})+18(-\widehat{m}^{s}_{4}-\widehat{m}^{s}_{5}+\widehat{m}^{s}_{6})\right.\nonumber\\
    &&\left.\left.+12\widehat{m}^{s}_{9}+6(\widehat{m}^{s}_{10}-\widehat{m}^{s}_{11}-\widehat{m}^{s}_{12})+27\widehat{m}^{s}_{16}+6\widehat{m}^{s}_{17}+9(\widehat{m}^{s}_{20}\right.\right.\nonumber\\                    &&\left.-\widehat{m}^{s}_{21}-\widehat{m}^{s}_{22})+3(\widehat{m}^{s}_{23}-\widehat{m}^{s}_{24}-\widehat{m}^{s}_{25})+\widehat{m}^{s}_{26}\right],\nonumber\\
S_{26}&=&\frac{1}{216}\left[8\widehat{m}^{s}_{0}+12(-\widehat{m}^{s}_{1}-\widehat{m}^{s}_{2}-\widehat{m}^{s}_{3})+18(\widehat{m}^{s}_{4}+\widehat{m}^{s}_{5}+\widehat{m}^{s}_{6})\right.\nonumber\\
    &&\left.\left.+12\widehat{m}^{s}_{9}+6(-\widehat{m}^{s}_{10}-\widehat{m}^{s}_{11}-\widehat{m}^{s}_{12})-27\widehat{m}^{s}_{16}+6\widehat{m}^{s}_{17}+9(\widehat{m}^{s}_{20}\right.\right.\nonumber\\                    &&\left.+\widehat{m}^{s}_{21}+\widehat{m}^{s}_{22})+3(-\widehat{m}^{s}_{23}-\widehat{m}^{s}_{24}-\widehat{m}^{s}_{25})+\widehat{m}^{s}_{26}\right].\label{eq:vsourceterms}
\end{eqnarray}

\section{\label{app:momentscollisionkernel}Appendix: Moments of the Collision Kernel for the D3Q27 Lattice}
The moments of the collision kernel follow from the orthogonal property of the moment basis matrix $\mathcal{K}$, which are given by
\begin{eqnarray}
\sum_{\alpha}(\mathcal{K}\cdot \mathbf{\widehat{g}})_{\alpha}=
\sum_{\beta} \braket{K_{\beta}|\rho}\widehat{g}_{\beta}&=&0,\nonumber\\
\sum_{\alpha}(\mathcal{K}\cdot \mathbf{\widehat{g}})_{\alpha}e_{\alpha x}=
\sum_{\beta} \braket{K_{\beta}|e_{\alpha x}}\widehat{g}_{\beta}&=&0,\nonumber\\
\sum_{\alpha}(\mathcal{K}\cdot \mathbf{\widehat{g}})_{\alpha}e_{\alpha y}=
\sum_{\beta} \braket{K_{\beta}|e_{\alpha y}}\widehat{g}_{\beta}&=&0,\nonumber\\
\sum_{\alpha}(\mathcal{K}\cdot \mathbf{\widehat{g}})_{\alpha}e_{\alpha z}=
\sum_{\beta} \braket{K_{\beta}|e_{\alpha z}}\widehat{g}_{\beta}&=&0,\nonumber\\
\sum_{\alpha}(\mathcal{K}\cdot \mathbf{\widehat{g}})_{\alpha}e_{\alpha x}e_{\alpha y}=
\sum_{\beta} \braket{K_{\beta}|e_{\alpha x}e_{\alpha y}}\widehat{g}_{\beta}&=&12\widehat{g}_{4},\nonumber\\
\sum_{\alpha}(\mathcal{K}\cdot \mathbf{\widehat{g}})_{\alpha}e_{\alpha x}e_{\alpha z}=
\sum_{\beta} \braket{K_{\beta}|e_{\alpha x}e_{\alpha z}}\widehat{g}_{\beta}&=&12\widehat{g}_{5},\nonumber\\
\sum_{\alpha}(\mathcal{K}\cdot \mathbf{\widehat{g}})_{\alpha}e_{\alpha y}e_{\alpha z}=
\sum_{\beta} \braket{K_{\beta}|e_{\alpha y}e_{\alpha z}}\widehat{g}_{\beta}&=&12\widehat{g}_{6},\nonumber\\
\sum_{\alpha}(\mathcal{K}\cdot \mathbf{\widehat{g}})_{\alpha}e_{\alpha x}^2=\sum_{\beta} \braket{K_{\beta}|e_{\alpha x}^2}\widehat{g}_{\beta}&=&6\widehat{g}_{7}+6\widehat{g}_{8}+6\widehat{g}_{9},\nonumber\\
\sum_{\alpha}(\mathcal{K}\cdot \mathbf{\widehat{g}})_{\alpha}e_{\alpha y}^2=\sum_{\beta} \braket{K_{\beta}|e_{\alpha y}^2}\widehat{g}_{\beta}&=&-6\widehat{g}_{7}+6\widehat{g}_{8}+6\widehat{g}_{9},\nonumber\\
\sum_{\alpha}(\mathcal{K}\cdot \mathbf{\widehat{g}})_{\alpha}e_{\alpha z}^2=\sum_{\beta} \braket{K_{\beta}|e_{\alpha z}^2}\widehat{g}_{\beta}&=&-12\widehat{g}_{8}+6\widehat{g}_{9},\nonumber\\
\sum_{\alpha}(\mathcal{K}\cdot \mathbf{\widehat{g}})_{\alpha}e_{\alpha x}e_{\alpha y}^2=\sum_{\beta} \braket{K_{\beta}|e_{\alpha x}e_{\alpha y}^2}\widehat{g}_{\beta}&=&12\widehat{g}_{10}+4\widehat{g}_{13},\nonumber\\
\sum_{\alpha}(\mathcal{K}\cdot \mathbf{\widehat{g}})_{\alpha}e_{\alpha x}e_{\alpha z}^2=\sum_{\beta} \braket{K_{\beta}|e_{\alpha x}e_{\alpha z}^2}\widehat{g}_{\beta}&=&12\widehat{g}_{10}-4\widehat{g}_{13},\nonumber\\
\sum_{\alpha}(\mathcal{K}\cdot \mathbf{\widehat{g}})_{\alpha}e_{\alpha x}^2e_{\alpha y}=\sum_{\beta} \braket{K_{\beta}|e_{\alpha x}^2e_{\alpha y}}\widehat{g}_{\beta}&=&12\widehat{g}_{11}+4\widehat{g}_{14},\nonumber\\
\sum_{\alpha}(\mathcal{K}\cdot \mathbf{\widehat{g}})_{\alpha}e_{\alpha y}e_{\alpha z}^2=\sum_{\beta} \braket{K_{\beta}|e_{\alpha y}e_{\alpha z}^2}\widehat{g}_{\beta}&=&12\widehat{g}_{11}-4\widehat{g}_{14},\nonumber\\
\sum_{\alpha}(\mathcal{K}\cdot \mathbf{\widehat{g}})_{\alpha}e_{\alpha x}^2e_{\alpha z}=\sum_{\beta} \braket{K_{\beta}|e_{\alpha x}^2e_{\alpha z}}\widehat{g}_{\beta}&=&12\widehat{g}_{12}+4\widehat{g}_{15},\nonumber\\
\sum_{\alpha}(\mathcal{K}\cdot \mathbf{\widehat{g}})_{\alpha}e_{\alpha y}^2e_{\alpha z}=\sum_{\beta} \braket{K_{\beta}|e_{\alpha y}^2e_{\alpha z}}\widehat{g}_{\beta}&=&12\widehat{g}_{12}-4\widehat{g}_{15},\nonumber\\
\sum_{\alpha}(\mathcal{K}\cdot \mathbf{\widehat{g}})_{\alpha}e_{\alpha x}e_{\alpha y}e_{\alpha z}=\sum_{\beta} \braket{K_{\beta}|e_{\alpha x}e_{\alpha y}e_{\alpha z}}\widehat{g}_{\beta}&=&8\widehat{g}_{16},\nonumber\\
\sum_{\alpha}(\mathcal{K}\cdot \mathbf{\widehat{g}})_{\alpha}e_{\alpha x}^2e_{\alpha y}^2=\sum_{\beta} \braket{K_{\beta}|e_{\alpha x}^2e_{\alpha y}^2}\widehat{g}_{\beta}&=&8\widehat{g}_{8}+8\widehat{g}_{9}+4\widehat{g}_{17}\nonumber\\
&&+4\widehat{g}_{18}+4\widehat{g}_{19},\nonumber\\
\sum_{\alpha}(\mathcal{K}\cdot \mathbf{\widehat{g}})_{\alpha}e_{\alpha x}^2e_{\alpha z}^2=\sum_{\beta} \braket{K_{\beta}|e_{\alpha x}^2e_{\alpha z}^2}\widehat{g}_{\beta}&=&4\widehat{g}_{7}-4\widehat{g}_{8}+8\widehat{g}_{9}\nonumber\\
&&+4\widehat{g}_{17}+4\widehat{g}_{18}-4\widehat{g}_{19},\nonumber\\
\sum_{\alpha}(\mathcal{K}\cdot \mathbf{\widehat{g}})_{\alpha}e_{\alpha y}^2e_{\alpha z}^2=\sum_{\beta} \braket{K_{\beta}|e_{\alpha y}^2e_{\alpha z}^2}\widehat{g}_{\beta}&=&-4\widehat{g}_{7}-4\widehat{g}_{8}+8\widehat{g}_{9}\nonumber\\
&&+4\widehat{g}_{17}-8\widehat{g}_{18},\nonumber\\
\sum_{\alpha}(\mathcal{K}\cdot \mathbf{\widehat{g}})_{\alpha}e_{\alpha x}^2e_{\alpha y}e_{\alpha z}=\sum_{\beta} \braket{K_{\beta}|e_{\alpha x}^2e_{\alpha y}e_{\alpha z}}\widehat{g}_{\beta}&=&8\widehat{g}_{6}+8\widehat{g}_{20},\nonumber\\
\sum_{\alpha}(\mathcal{K}\cdot \mathbf{\widehat{g}})_{\alpha}e_{\alpha x}e_{\alpha y}^2e_{\alpha z}=\sum_{\beta} \braket{K_{\beta}|e_{\alpha x}e_{\alpha y}^2e_{\alpha z}}\widehat{g}_{\beta}&=&8\widehat{g}_{5}+8\widehat{g}_{21},\nonumber\\
\sum_{\alpha}(\mathcal{K}\cdot \mathbf{\widehat{g}})_{\alpha}e_{\alpha x}e_{\alpha y}e_{\alpha z}^2=\sum_{\beta} \braket{K_{\beta}|e_{\alpha x}e_{\alpha y}e_{\alpha z}^2}\widehat{g}_{\beta}&=&8\widehat{g}_{4}+8\widehat{g}_{22},\nonumber\\
\sum_{\alpha}(\mathcal{K}\cdot \mathbf{\widehat{g}})_{\alpha}e_{\alpha x}e_{\alpha y}^2e_{\alpha z}^2=\sum_{\beta} \braket{K_{\beta}|e_{\alpha x}e_{\alpha y}^2e_{\alpha z}^2}\widehat{g}_{\beta}&=&16\widehat{g}_{10}+8\widehat{g}_{23},\nonumber\\
\sum_{\alpha}(\mathcal{K}\cdot \mathbf{\widehat{g}})_{\alpha}e_{\alpha x}^2e_{\alpha y}e_{\alpha z}^2=\sum_{\beta} \braket{K_{\beta}|e_{\alpha x}^2e_{\alpha y}e_{\alpha z}^2}\widehat{g}_{\beta}&=&16\widehat{g}_{11}+8\widehat{g}_{24},\nonumber\\
\sum_{\alpha}(\mathcal{K}\cdot \mathbf{\widehat{g}})_{\alpha}e_{\alpha x}^2e_{\alpha y}^2e_{\alpha z}=\sum_{\beta} \braket{K_{\beta}|e_{\alpha x}^2e_{\alpha y}^2e_{\alpha z}}\widehat{g}_{\beta}&=&16\widehat{g}_{12}+8\widehat{g}_{25},\nonumber\\
\sum_{\alpha}(\mathcal{K}\cdot \mathbf{\widehat{g}})_{\alpha}e_{\alpha x}^2e_{\alpha y}^2e_{\alpha z}^2=\sum_{\beta} \braket{K_{\beta}|e_{\alpha x}^2e_{\alpha y}^2e_{\alpha z}^2}\widehat{g}_{\beta}&=&8\widehat{g}_{9}+8\widehat{g}_{17}+8\widehat{g}_{26}.\label{eq:collisionkernelmoment}
\end{eqnarray}

\section{\label{app:d3q15formulation}Appendix: Formulation of the Central Moment LBM for the Three-dimensional, Fifteen Velocity (D3Q15) Lattice}
\subsection{Moment Basis}
\begin{figure}[h]
\begin{center}
\includegraphics[width = 140mm]{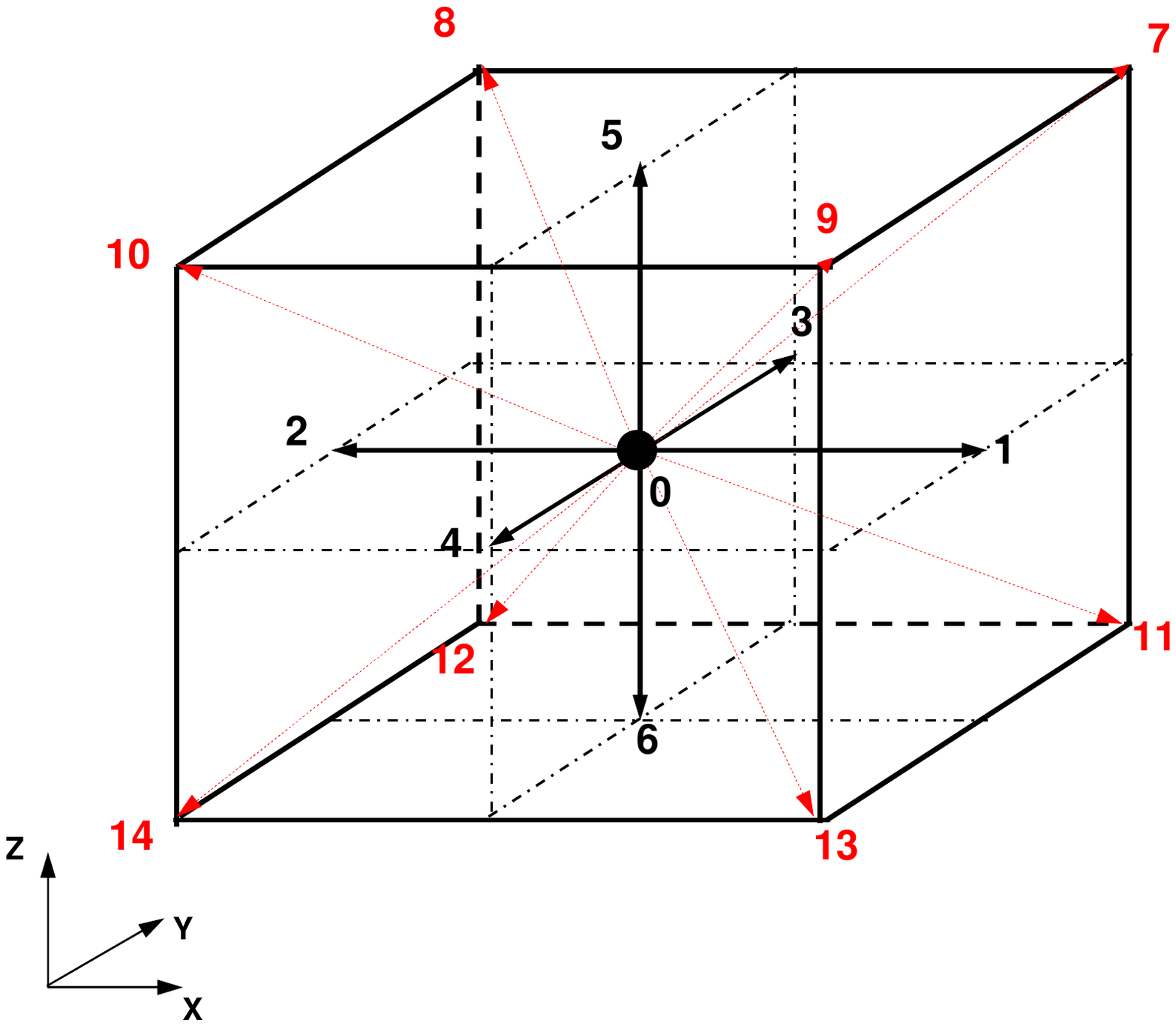}% Here is how to import EPS art
\caption{\label{fig:d3q15} Three-dimensional, fifteen particle velocity (D3Q27) lattice.}
\end{center}
\end{figure}

The particle velocity for the D3Q15 lattice $\overrightarrow{e}_{\alpha}$ (see Fig.~\ref{fig:d3q15}) is given by
\begin{equation}
\overrightarrow{e_{\alpha}} = \left\{\begin{array}{ll}
   {(0,0,0),}&{\alpha=0}\\
   {(\pm 1,0,0), (0,\pm 1,0), (0,0,\pm 1),}&{\alpha=1,\cdots,6}\\
   {(\pm 1,\pm 1,\pm 1),}&{\alpha=7,\cdots,14}
\end{array} \right.
\label{eq:velocityd3q15}
\end{equation}
%%%%% FIGURE %%%%%
%%%%% FIGURE %%%%%
The components of the nominal moment basis chosen are
\begin{eqnarray}
\ket{T_{0}}&=&\ket{\rho}\equiv \ket{|\overrightarrow{e}_{\alpha}|^{0}},\nonumber\\
\ket{T_{1}}&=&\ket{e_{\alpha x}},\nonumber\\
\ket{T_{2}}&=&\ket{e_{\alpha y}},\nonumber\\
\ket{T_{3}}&=&\ket{e_{\alpha z}},\nonumber\\
\ket{T_{4}}&=&\ket{e_{\alpha x}e_{\alpha y}},\nonumber\\
\ket{T_{5}}&=&\ket{e_{\alpha x}e_{\alpha z}},\nonumber\\
\ket{T_{6}}&=&\ket{e_{\alpha y}e_{\alpha z}},\nonumber\\
\ket{T_{7}}&=&\ket{e_{\alpha x}^2-e_{\alpha y}^2},\nonumber\\
\ket{T_{8}}&=&\ket{e_{\alpha x}^2-e_{\alpha z}^2},\nonumber\\
\ket{T_{9}}&=&\ket{e_{\alpha x}^2+e_{\alpha y}^2+e_{\alpha z}^2},\nonumber\\
\ket{T_{10}}&=&\ket{e_{\alpha x}(e_{\alpha x}^2+e_{\alpha y}^2+e_{\alpha z}^2)},\nonumber\\
\ket{T_{11}}&=&\ket{e_{\alpha y}(e_{\alpha x}^2+e_{\alpha y}^2+e_{\alpha z}^2)},\nonumber\\
\ket{T_{12}}&=&\ket{e_{\alpha z}(e_{\alpha x}^2+e_{\alpha y}^2+e_{\alpha z}^2)},\nonumber\\
\ket{T_{13}}&=&\ket{e_{\alpha x}e_{\alpha y}e_{\alpha z}},\nonumber\\
\ket{T_{14}}&=&\ket{e_{\alpha x}^2e_{\alpha y}^2+e_{\alpha x}^2e_{\alpha z}^2+e_{\alpha y}^2e_{\alpha z}^2}.\nonumber
\end{eqnarray}

Based on the above set, the components of the orthogonal basis vectors are obtained by means of the Gram-Schmidt procedure, which are given by
\begin{eqnarray}
\ket{K_{0}}&=&\ket{\rho}\equiv \ket{|\overrightarrow{e}_{\alpha}|^{0}},\nonumber\\
\ket{K_{1}}&=&\ket{e_{\alpha x}},\nonumber\\
\ket{K_{2}}&=&\ket{e_{\alpha y}},\nonumber\\
\ket{K_{3}}&=&\ket{e_{\alpha z}},\nonumber\\
\ket{K_{4}}&=&\ket{e_{\alpha x}e_{\alpha y}},\nonumber\\
\ket{K_{5}}&=&\ket{e_{\alpha x}e_{\alpha z}},\nonumber\\
\ket{K_{6}}&=&\ket{e_{\alpha y}e_{\alpha z}},\nonumber\\
\ket{K_{7}}&=&\ket{e_{\alpha x}^2-e_{\alpha y}^2},\nonumber\\
\ket{K_{8}}&=&\ket{e_{\alpha x}^2+e_{\alpha y}^2+e_{\alpha z}^2}-3\ket{e_{\alpha z}^2},\nonumber\\
\ket{K_{9}}&=&\ket{e_{\alpha x}^2+e_{\alpha y}^2+e_{\alpha z}^2}-2\ket{\rho},\nonumber\\
\ket{K_{10}}&=&5\ket{e_{\alpha x}(e_{\alpha x}^2+e_{\alpha y}^2+e_{\alpha z}^2)}-13\ket{e_{\alpha x}},\nonumber\\
\ket{K_{11}}&=&5\ket{e_{\alpha y}(e_{\alpha x}^2+e_{\alpha y}^2+e_{\alpha z}^2)}-13\ket{e_{\alpha y}},\nonumber\\
\ket{K_{12}}&=&5\ket{e_{\alpha z}(e_{\alpha x}^2+e_{\alpha y}^2+e_{\alpha z}^2)}-13\ket{e_{\alpha z}},\nonumber\\
\ket{K_{13}}&=&\ket{e_{\alpha x}e_{\alpha y}e_{\alpha z}},\nonumber\\
\ket{K_{14}}&=&30\ket{e_{\alpha x}^2e_{\alpha y}^2+e_{\alpha x}^2e_{\alpha z}^2+e_{\alpha y}^2e_{\alpha z}^2}-40\ket{e_{\alpha x}^2+e_{\alpha y}^2+e_{\alpha z}^2}+32\ket{\rho}.
\end{eqnarray}
They can be written in a corresponding matrix form as
$\mathcal{K}=$

\scriptsize
$
\left(
\begin{array}{rrrrrrrrrrrrrrr}
1	&0	&0	&0	&0	&0	&0	&0	&0	&-2	&0	&0	&0	&0	&32\\
1	&1	&0	&0	&0	&0	&0	&1	&1	&-1	&-8	&0	&0	&0	&-8\\
1	&-1	&0	&0	&0	&0	&0	&1	&1	&-1	&8	&0	&0	&0	&-8\\
1	&0	&1	&0	&0	&0	&0	&-1	&1	&-1	&0	&-8	&0	&0	&-8\\
1	&0	&-1	&0	&0	&0	&0	&-1	&1	&-1	&0	&8	&0	&0	&-8\\
1	&0	&0	&1	&0	&0	&0	&0	&-2	&-1	&0	&0	&-8	&0	&-8\\
1	&0	&0	&-1	&0	&0	&0	&0	&-2	&-1	&0	&0	&8	&0	&-8\\
1	&1	&1	&1	&1	&1	&1	&0	&0	&1	&2	&2	&2	&1	&2\\
1	&-1	&1	&1	&-1	&-1	&1	&0	&0	&1	&-2	&2	&2	&-1	&2\\
1	&1	&-1	&1	&-1	&1	&-1	&0	&0	&1	&2	&-2	&2	&-1	&2\\
1	&-1	&-1	&1	&1	&-1	&-1	&0	&0	&1	&-2	&-2	&2	&1	&2\\
1	&1	&1	&-1	&1	&-1	&-1	&0	&0	&1	&2	&2	&-2	&-1	&2\\
1	&-1	&1	&-1	&-1	&1	&-1	&0	&0	&1	&-2	&2	&-2	&1	&2\\
1	&1	&-1	&-1	&-1	&-1	&1	&0	&0	&1	&2	&-2	&-2	&1	&2\\
1	&-1	&-1	&-1	&1	&1	&1	&0	&0	&1	&-2	&-2	&-2	&-1	&2\\
\end{array}
\right)
$
\normalsize

where
\begin{eqnarray}
\mathcal{K}&=&\left[\ket{K_{0}},\ket{K_{1}},\ket{K_{2}},\ket{K_{3}},\ket{K_{4}},\ket{K_{5}},\ket{K_{6}},\ket{K_{7}},\ket{K_{8}}\right.\nonumber\\
           &&\left.\ket{K_{9}},\ket{K_{10}},\ket{K_{11}},\ket{K_{12}},\ket{K_{13}},\ket{K_{14}}\right].
\end{eqnarray}

\subsection{Various Raw Moments and Source Terms in Velocity Space}
The above orthogonal matrix results in a set of moments of the collision kernel, which are needed in the construction of the collision operator, and are given by
\begin{eqnarray}
\sum_{\alpha}(\mathcal{K}\cdot \mathbf{\widehat{g}})_{\alpha}=
\sum_{\beta} \braket{K_{\beta}|\rho}\widehat{g}_{\beta}&=&0,\nonumber\\
\sum_{\alpha}(\mathcal{K}\cdot \mathbf{\widehat{g}})_{\alpha}e_{\alpha x}=
\sum_{\beta} \braket{K_{\beta}|e_{\alpha x}}\widehat{g}_{\beta}&=&0,\nonumber\\
\sum_{\alpha}(\mathcal{K}\cdot \mathbf{\widehat{g}})_{\alpha}e_{\alpha y}=
\sum_{\beta} \braket{K_{\beta}|e_{\alpha y}}\widehat{g}_{\beta}&=&0,\nonumber\\
\sum_{\alpha}(\mathcal{K}\cdot \mathbf{\widehat{g}})_{\alpha}e_{\alpha z}=
\sum_{\beta} \braket{K_{\beta}|e_{\alpha z}}\widehat{g}_{\beta}&=&0,\nonumber\\
\sum_{\alpha}(\mathcal{K}\cdot \mathbf{\widehat{g}})_{\alpha}e_{\alpha x}e_{\alpha y}=
\sum_{\beta} \braket{K_{\beta}|e_{\alpha x}e_{\alpha y}}\widehat{g}_{\beta}&=&8\widehat{g}_{4},\nonumber\\
\sum_{\alpha}(\mathcal{K}\cdot \mathbf{\widehat{g}})_{\alpha}e_{\alpha x}e_{\alpha z}=
\sum_{\beta} \braket{K_{\beta}|e_{\alpha x}e_{\alpha z}}\widehat{g}_{\beta}&=&8\widehat{g}_{5},\nonumber\\
\sum_{\alpha}(\mathcal{K}\cdot \mathbf{\widehat{g}})_{\alpha}e_{\alpha y}e_{\alpha z}=
\sum_{\beta} \braket{K_{\beta}|e_{\alpha y}e_{\alpha z}}\widehat{g}_{\beta}&=&8\widehat{g}_{6},\nonumber\\
\sum_{\alpha}(\mathcal{K}\cdot \mathbf{\widehat{g}})_{\alpha}e_{\alpha x}^2=\sum_{\beta} \braket{K_{\beta}|e_{\alpha x}^2}\widehat{g}_{\beta}&=&2\widehat{g}_{7}+2\widehat{g}_{8}+6\widehat{g}_{9},\nonumber\\
\sum_{\alpha}(\mathcal{K}\cdot \mathbf{\widehat{g}})_{\alpha}e_{\alpha y}^2=\sum_{\beta} \braket{K_{\beta}|e_{\alpha y}^2}\widehat{g}_{\beta}&=&-2\widehat{g}_{7}+2\widehat{g}_{8}+6\widehat{g}_{9},\nonumber\\
\sum_{\alpha}(\mathcal{K}\cdot \mathbf{\widehat{g}})_{\alpha}e_{\alpha z}^2=\sum_{\beta} \braket{K_{\beta}|e_{\alpha z}^2}\widehat{g}_{\beta}&=&-4\widehat{g}_{8}+6\widehat{g}_{9},\nonumber\\
\sum_{\alpha}(\mathcal{K}\cdot \mathbf{\widehat{g}})_{\alpha}e_{\alpha x}e_{\alpha y}^2=\sum_{\beta} \braket{K_{\beta}|e_{\alpha x}e_{\alpha y}^2}\widehat{g}_{\beta}&=&16\widehat{g}_{10},\nonumber\\
\sum_{\alpha}(\mathcal{K}\cdot \mathbf{\widehat{g}})_{\alpha}e_{\alpha x}e_{\alpha z}^2=\sum_{\beta} \braket{K_{\beta}|e_{\alpha x}e_{\alpha z}^2}\widehat{g}_{\beta}&=&16\widehat{g}_{10},\nonumber\\
\sum_{\alpha}(\mathcal{K}\cdot \mathbf{\widehat{g}})_{\alpha}e_{\alpha x}^2e_{\alpha y}=\sum_{\beta} \braket{K_{\beta}|e_{\alpha x}^2e_{\alpha y}}\widehat{g}_{\beta}&=&16\widehat{g}_{11},\nonumber\\
\sum_{\alpha}(\mathcal{K}\cdot \mathbf{\widehat{g}})_{\alpha}e_{\alpha y}e_{\alpha z}^2=\sum_{\beta} \braket{K_{\beta}|e_{\alpha y}e_{\alpha z}^2}\widehat{g}_{\beta}&=&16\widehat{g}_{11},\nonumber\\
\sum_{\alpha}(\mathcal{K}\cdot \mathbf{\widehat{g}})_{\alpha}e_{\alpha x}^2e_{\alpha z}=\sum_{\beta} \braket{K_{\beta}|e_{\alpha x}^2e_{\alpha z}}\widehat{g}_{\beta}&=&16\widehat{g}_{12},\nonumber\\
\sum_{\alpha}(\mathcal{K}\cdot \mathbf{\widehat{g}})_{\alpha}e_{\alpha y}^2e_{\alpha z}=\sum_{\beta} \braket{K_{\beta}|e_{\alpha y}^2e_{\alpha z}}\widehat{g}_{\beta}&=&16\widehat{g}_{12},\nonumber\\
\sum_{\alpha}(\mathcal{K}\cdot \mathbf{\widehat{g}})_{\alpha}e_{\alpha x}e_{\alpha y}e_{\alpha z}=\sum_{\beta} \braket{K_{\beta}|e_{\alpha x}e_{\alpha y}e_{\alpha z}}\widehat{g}_{\beta}&=&8\widehat{g}_{13},\nonumber\\
\sum_{\alpha}(\mathcal{K}\cdot \mathbf{\widehat{g}})_{\alpha}e_{\alpha x}^2e_{\alpha y}^2=\sum_{\beta} \braket{K_{\beta}|e_{\alpha x}^2e_{\alpha y}^2}\widehat{g}_{\beta}&=&8\widehat{g}_{9}+16\widehat{g}_{14},\nonumber\\
\sum_{\alpha}(\mathcal{K}\cdot \mathbf{\widehat{g}})_{\alpha}e_{\alpha x}^2e_{\alpha z}^2=\sum_{\beta} \braket{K_{\beta}|e_{\alpha x}^2e_{\alpha z}^2}\widehat{g}_{\beta}&=&8\widehat{g}_{9}+16\widehat{g}_{14},\nonumber\\
\sum_{\alpha}(\mathcal{K}\cdot \mathbf{\widehat{g}})_{\alpha}e_{\alpha y}^2e_{\alpha z}^2=\sum_{\beta} \braket{K_{\beta}|e_{\alpha y}^2e_{\alpha z}^2}\widehat{g}_{\beta}&=&8\widehat{g}_{9}+16\widehat{g}_{14}.
\end{eqnarray}
Note that unlike the D3Q27 lattice, additional degeneracies for various third and higher order moment basis vectors exist for the D3Q15 lattice, as it contains a more limited set of independent basis vectors.

It may be noted that the components of the raw moments of the source terms
$\widehat{\sigma}_{x^my^nz^p}^{'}=\braket{S_{\alpha}|e_{\alpha x}^me_{\alpha y}^ne_{\alpha z}^p}$ due to force fields can be obtained in an analogous manner as determined for the D3Q27 lattice (see Appendix~\ref{app:rawsourcemoments}). The projections of the source terms to the orthogonal matrix of the moment basis $\mathcal{K}$, i.e. $\braket{K_{\beta}|S_{\alpha}}$, $\beta=0,1,2,\ldots, 14$ for this lattice yield
\begin{eqnarray}
\widehat{m}^{s}_{0}=\braket{K_0|S_{\alpha}}&=&0, \nonumber\\
\widehat{m}^{s}_{1}=\braket{K_1|S_{\alpha}}&=&F_x, \nonumber\\
\widehat{m}^{s}_{2}=\braket{K_2|S_{\alpha}}&=&F_y, \nonumber\\
\widehat{m}^{s}_{3}=\braket{K_3|S_{\alpha}}&=&F_z, \nonumber\\
\widehat{m}^{s}_{4}=\braket{K_4|S_{\alpha}}&=&(F_xu_y+F_yu_x), \nonumber\\
\widehat{m}^{s}_{5}=\braket{K_5|S_{\alpha}}&=&(F_xu_z+F_zu_x), \nonumber\\
\widehat{m}^{s}_{6}=\braket{K_6|S_{\alpha}}&=&(F_yu_z+F_zu_y), \nonumber\\
\widehat{m}^{s}_{7}=\braket{K_7|S_{\alpha}}&=&2(F_xu_x-F_yu_y), \nonumber\\
\widehat{m}^{s}_{8}=\braket{K_8|S_{\alpha}}&=&2(F_xu_x+F_yu_y-2F_zu_z), \nonumber\\
\widehat{m}^{s}_{9}=\braket{K_9|S_{\alpha}}&=&2(F_xu_x+F_yu_y+F_zu_z), \nonumber\\
\widehat{m}^{s}_{10}=\braket{K_{10}|S_{\alpha}}&=&5\left[F_x(3u_x^2+u_y^2+u_z^2)+2u_x(F_yu_y+F_zu_z)\right]-13F_x, \nonumber\\
\widehat{m}^{s}_{11}=\braket{K_{11}|S_{\alpha}}&=&5\left[F_y(u_x^2+3u_y^2+u_z^2)+2u_y(F_xu_x+F_zu_z)\right]-13F_y, \nonumber\\
\widehat{m}^{s}_{12}=\braket{K_{12}|S_{\alpha}}&=&5\left[F_z(u_x^2+u_y^2+3u_z^2)+2u_z(F_xu_x+F_yu_y)\right]-13F_z, \nonumber\\
\widehat{m}^{s}_{13}=\braket{K_{13}|S_{\alpha}}&=&F_xu_yu_z+F_yu_xu_z+F_zu_xu_y, \nonumber\\
\widehat{m}^{s}_{14}=\braket{K_{14}|S_{\alpha}}&=&20\left[F_xu_x\left(3(u_y^2+u_z^2)-4\right)+F_yu_y\left(3(u_x^2+u_z^2)-4\right)\right.\nonumber\\
                                               &&\left.+F_zu_z\left(3(u_x^2+u_y^2)-4\right)\right].
\end{eqnarray}
Using $\widehat{m}^{s}_{\beta}$, the source terms in velocity space can be obtained by a procedure involving exact inversion that invokes orthogonal properties of the collision matrix (see the discussion following Eq.~(\ref{eq:sourceformulation1}) for the D3Q27 lattice). The results are summarized as follows:
\begin{eqnarray}
S_0&=&\frac{1}{45}\left[3\widehat{m}^{s}_{0}-5\widehat{m}^{s}_{9}+\widehat{m}^{s}_{14}\right], \nonumber\\
S_1&=&\frac{1}{180}\left[12\widehat{m}^{s}_{0}+18\widehat{m}^{s}_{1}+45\widehat{m}^{s}_{7}+15\widehat{m}^{s}_{8}-10\widehat{m}^{s}_{9}-9\widehat{m}^{s}_{10}-\widehat{m}^{s}_{14}\right],\nonumber\\
S_2&=&\frac{1}{180}\left[12\widehat{m}^{s}_{0}-18\widehat{m}^{s}_{1}+45\widehat{m}^{s}_{7}+15\widehat{m}^{s}_{8}-10\widehat{m}^{s}_{9}+9\widehat{m}^{s}_{10}-\widehat{m}^{s}_{14}\right],\nonumber\\
S_3&=&\frac{1}{180}\left[12\widehat{m}^{s}_{0}+18\widehat{m}^{s}_{2}-45\widehat{m}^{s}_{7}+15\widehat{m}^{s}_{8}-10\widehat{m}^{s}_{9}-9\widehat{m}^{s}_{11}-\widehat{m}^{s}_{14}\right],\nonumber\\
S_4&=&\frac{1}{180}\left[12\widehat{m}^{s}_{0}-18\widehat{m}^{s}_{2}-45\widehat{m}^{s}_{7}+15\widehat{m}^{s}_{8}-10\widehat{m}^{s}_{9}+9\widehat{m}^{s}_{11}-\widehat{m}^{s}_{14}\right],\nonumber\\
S_5&=&\frac{1}{180}\left[12\widehat{m}^{s}_{0}+18\widehat{m}^{s}_{3}-30\widehat{m}^{s}_{8}-10\widehat{m}^{s}_{9}-9\widehat{m}^{s}_{12}-\widehat{m}^{s}_{14}\right],\nonumber\\
S_6&=&\frac{1}{180}\left[12\widehat{m}^{s}_{0}-18\widehat{m}^{s}_{3}-30\widehat{m}^{s}_{8}-10\widehat{m}^{s}_{9}+9\widehat{m}^{s}_{12}-\widehat{m}^{s}_{14}\right],\nonumber\\
S_7&=&\frac{1}{720}\left[48\widehat{m}^{s}_{0}+72\widehat{m}^{s}_{1}+72\widehat{m}^{s}_{2}+72\widehat{m}^{s}_{3}+90\widehat{m}^{s}_{4}+90\widehat{m}^{s}_{5}+90\widehat{m}^{s}_{6}+40\widehat{m}^{s}_{9}\right.\nonumber\\
                 &&\left.+9\widehat{m}^{s}_{10}+9\widehat{m}^{s}_{11}+9\widehat{m}^{s}_{12}+90\widehat{m}^{s}_{13}+\widehat{m}^{s}_{14}\right],\nonumber\\
S_8&=&\frac{1}{720}\left[48\widehat{m}^{s}_{0}-72\widehat{m}^{s}_{1}+72\widehat{m}^{s}_{2}+72\widehat{m}^{s}_{3}-90\widehat{m}^{s}_{4}-90\widehat{m}^{s}_{5}+90\widehat{m}^{s}_{6}+40\widehat{m}^{s}_{9}\right.\nonumber\\
                 &&\left.-9\widehat{m}^{s}_{10}+9\widehat{m}^{s}_{11}+9\widehat{m}^{s}_{12}-90\widehat{m}^{s}_{13}+\widehat{m}^{s}_{14}\right],\nonumber\\
S_9&=&\frac{1}{720}\left[48\widehat{m}^{s}_{0}+72\widehat{m}^{s}_{1}-72\widehat{m}^{s}_{2}+72\widehat{m}^{s}_{3}-90\widehat{m}^{s}_{4}+90\widehat{m}^{s}_{5}-90\widehat{m}^{s}_{6}+40\widehat{m}^{s}_{9}\right.\nonumber\\
                 &&\left.+9\widehat{m}^{s}_{10}-9\widehat{m}^{s}_{11}+9\widehat{m}^{s}_{12}-90\widehat{m}^{s}_{13}+\widehat{m}^{s}_{14}\right],\nonumber\\
S_{10}&=&\frac{1}{720}\left[48\widehat{m}^{s}_{0}-72\widehat{m}^{s}_{1}-72\widehat{m}^{s}_{2}+72\widehat{m}^{s}_{3}+90\widehat{m}^{s}_{4}-90\widehat{m}^{s}_{5}-90\widehat{m}^{s}_{6}+40\widehat{m}^{s}_{9}\right.\nonumber\\
                 &&\left.-9\widehat{m}^{s}_{10}-9\widehat{m}^{s}_{11}+9\widehat{m}^{s}_{12}+90\widehat{m}^{s}_{13}+\widehat{m}^{s}_{14}\right],\nonumber\\
S_{11}&=&\frac{1}{720}\left[48\widehat{m}^{s}_{0}+72\widehat{m}^{s}_{1}+72\widehat{m}^{s}_{2}-72\widehat{m}^{s}_{3}+90\widehat{m}^{s}_{4}-90\widehat{m}^{s}_{5}-90\widehat{m}^{s}_{6}+40\widehat{m}^{s}_{9}\right.\nonumber\\
                 &&\left.+9\widehat{m}^{s}_{10}+9\widehat{m}^{s}_{11}-9\widehat{m}^{s}_{12}-90\widehat{m}^{s}_{13}+\widehat{m}^{s}_{14}\right],\nonumber\\
S_{12}&=&\frac{1}{720}\left[48\widehat{m}^{s}_{0}-72\widehat{m}^{s}_{1}+72\widehat{m}^{s}_{2}-72\widehat{m}^{s}_{3}-90\widehat{m}^{s}_{4}+90\widehat{m}^{s}_{5}-90\widehat{m}^{s}_{6}+40\widehat{m}^{s}_{9}\right.\nonumber\\
                 &&\left.-9\widehat{m}^{s}_{10}+9\widehat{m}^{s}_{11}-9\widehat{m}^{s}_{12}+90\widehat{m}^{s}_{13}+\widehat{m}^{s}_{14}\right],\nonumber\\
S_{13}&=&\frac{1}{720}\left[48\widehat{m}^{s}_{0}+72\widehat{m}^{s}_{1}-72\widehat{m}^{s}_{2}-72\widehat{m}^{s}_{3}-90\widehat{m}^{s}_{4}-90\widehat{m}^{s}_{5}+90\widehat{m}^{s}_{6}+40\widehat{m}^{s}_{9}\right.\nonumber\\
                 &&\left.+9\widehat{m}^{s}_{10}-9\widehat{m}^{s}_{11}-9\widehat{m}^{s}_{12}+90\widehat{m}^{s}_{13}+\widehat{m}^{s}_{14}\right],\nonumber\\
S_{14}&=&\frac{1}{720}\left[48\widehat{m}^{s}_{0}-72\widehat{m}^{s}_{1}-72\widehat{m}^{s}_{2}-72\widehat{m}^{s}_{3}+90\widehat{m}^{s}_{4}+90\widehat{m}^{s}_{5}+90\widehat{m}^{s}_{6}+40\widehat{m}^{s}_{9}\right.\nonumber\\
                 &&\left.-9\widehat{m}^{s}_{10}-9\widehat{m}^{s}_{11}-9\widehat{m}^{s}_{12}-90\widehat{m}^{s}_{13}+\widehat{m}^{s}_{14}\right].
\end{eqnarray}

For obtaining explicit expressions for the collision kernel, it is convenient to express the non-conserved transformed raw moments using the operator notation given in Eq.~(\ref{eq:summationoperator}), which are given as subsets of the particle velocity directions for the D3Q15 lattice. It follows that
\begin{eqnarray}
\widehat{\overline{\kappa}}_{xy}^{'}&=&\braket{\overline{f}_{\alpha}|e_{\alpha x}e_{\alpha y}}=\sum_{\alpha =
0}^{14}\overline{f}_{\alpha} e_{\alpha x} e_{\alpha y}=\left(\sum_{\alpha}^{A_4}-\sum_{\alpha}^{B_4}\right)\otimes
\overline{f}_{\alpha},\nonumber\\
\widehat{\overline{\kappa}}_{xz}^{'}&=&\braket{\overline{f}_{\alpha}|e_{\alpha x}e_{\alpha z}}=\sum_{\alpha =
0}^{14}\overline{f}_{\alpha} e_{\alpha x} e_{\alpha z}=\left(\sum_{\alpha}^{A_5}-\sum_{\alpha}^{B_5}\right)\otimes
\overline{f}_{\alpha},\nonumber\\
\widehat{\overline{\kappa}}_{yz}^{'}&=&\braket{\overline{f}_{\alpha}|e_{\alpha y}e_{\alpha z}}=\sum_{\alpha =
0}^{14}\overline{f}_{\alpha} e_{\alpha y} e_{\alpha z}=\left(\sum_{\alpha}^{A_6}-\sum_{\alpha}^{B_6}\right)\otimes
\overline{f}_{\alpha},\nonumber\\
\widehat{\overline{\kappa}}_{xx}^{'}&=&\braket{\overline{f}_{\alpha}|e_{\alpha x}^2}=\sum_{\alpha = 0}^{14}\overline{f}_{\alpha} e_{\alpha x}^2=\left(\sum_{\alpha}^{A_7}\right)\otimes \overline{f}_{\alpha},\nonumber\\
\widehat{\overline{\kappa}}_{yy}^{'}&=&\braket{\overline{f}_{\alpha}|e_{\alpha y}^2}=\sum_{\alpha = 0}^{14}\overline{f}_{\alpha}
e_{\alpha y}^2=\left(\sum_{\alpha}^{A_8}\right)\otimes \overline{f}_{\alpha},\nonumber\\
\widehat{\overline{\kappa}}_{zz}^{'}&=&\braket{\overline{f}_{\alpha}|e_{\alpha z}^2}=\sum_{\alpha = 0}^{14}\overline{f}_{\alpha}
e_{\alpha z}^2=\left(\sum_{\alpha}^{A_9}\right)\otimes \overline{f}_{\alpha},\nonumber\\
\widehat{\overline{\kappa}}_{xyy}^{'}&=&\braket{\overline{f}_{\alpha}|e_{\alpha x}e_{\alpha y}^2}=\sum_{\alpha =
0}^{14}\overline{f}_{\alpha} e_{\alpha x} e_{\alpha y}^2=\left(\sum_{\alpha}^{A_{10}}-\sum_{\alpha}^{B_{10}}\right)\otimes     \overline{f}_{\alpha},\nonumber\\
\widehat{\overline{\kappa}}_{xxy}^{'}&=&\braket{\overline{f}_{\alpha}|e_{\alpha x}^2e_{\alpha y}}=\sum_{\alpha =
0}^{14}\overline{f}_{\alpha} e_{\alpha x}^2 e_{\alpha y}=\left(\sum_{\alpha}^{A_{11}}-\sum_{\alpha}^{B_{11}}\right)\otimes
\overline{f}_{\alpha},\nonumber\\
\widehat{\overline{\kappa}}_{xxz}^{'}&=&\braket{\overline{f}_{\alpha}|e_{\alpha x}^2 e_{\alpha z}}=\sum_{\alpha =
0}^{14}\overline{f}_{\alpha} e_{\alpha x}^2 e_{\alpha z}=\left(\sum_{\alpha}^{A_{12}}-\sum_{\alpha}^{B_{12}}\right)\otimes
\overline{f}_{\alpha},\nonumber\\
\widehat{\overline{\kappa}}_{xyz}^{'}&=&\braket{\overline{f}_{\alpha}|e_{\alpha x}e_{\alpha y}e_{\alpha z}}=\sum_{\alpha =
0}^{14}\overline{f}_{\alpha} e_{\alpha x} e_{\alpha y} e_{\alpha z}=\left(\sum_{\alpha}^{A_{13}}-\sum_{\alpha}^{B_{13}}\right)\otimes
\overline{f}_{\alpha},\nonumber\\
\widehat{\overline{\kappa}}_{xxyy}^{'}&=&\braket{\overline{f}_{\alpha}|e_{\alpha x}^2 e_{\alpha y}^2}=\sum_{\alpha = 0}^{14}\overline{f}_{\alpha}
e_{\alpha x}^2 e_{\alpha y}^2=\left(\sum_{\alpha}^{A_{14}}\right)\otimes \overline{f}_{\alpha},
\end{eqnarray}
where
\begin{eqnarray*}
A_4&=&\left\{7,10,11,14\right\}, B_4=\left\{8,9,12,13\right\},\\
A_5&=&\left\{7,9,12,14\right\}, B_5=\left\{8,10,11,13\right\},\\
A_6&=&\left\{7,8,13,14\right\}, B_6=\left\{9,10,11,12\right\},\\
A_7&=&\left\{1,2,7,8,9,10,11,12,13,14\right\},\\
A_8&=&\left\{3,4,7,8,9,10,11,12,13,14\right\},\\
A_9&=&\left\{5,6,7,8,9,10,11,12,13,14\right\},\\
A_{10}&=&\left\{7,9,11,13\right\}, B_{10}=\left\{8,10,12,14\right\},\\
A_{11}&=&\left\{7,8,11,12\right\}, B_{11}=\left\{9,10,13,14\right\},\\
A_{12}&=&\left\{7,8,9,10\right\}, B_{12}=\left\{11,12,13,14\right\},\\
A_{13}&=&\left\{7,10,12,13\right\}, B_{13}=\left\{8,9,11,14\right\},\\
A_{14}&=&\left\{7,8,9,10,11,12,13,14\right\}.
\end{eqnarray*}

\subsection{Collision Kernel}
Following the same procedure and the notations as used for the D3Q27 lattice and considering factorized attractors for the higher order moments, the cascaded form of the central moment collision operator in the presence of forcing terms can be constructed. The results are summarized as follows (for collisional invariants, $\widehat{g}_0=\widehat{g}_1=\widehat{g}_2=\widehat{g}_3=0$):
\begin{eqnarray}
\widehat{g}_4&=&\frac{\omega_4}{8}\left[-\widehat{\overline{\eta}}_{xy}^{'}+\rho u_xu_y+\frac{1}{2}(\widehat{\sigma}_{x}^{'}u_y+\widehat{\sigma}_{y}^{'}u_x)\right],\\
\widehat{g}_5&=&\frac{\omega_5}{8}\left[-\widehat{\overline{\eta}}_{xz}^{'}+\rho u_xu_z+\frac{1}{2}(\widehat{\sigma}_{x}^{'}u_z+\widehat{\sigma}_{z}^{'}u_x)\right],\\
\widehat{g}_6&=&\frac{\omega_6}{8}\left[-\widehat{\overline{\eta}}_{yz}^{'}+\rho u_yu_z+\frac{1}{2}(\widehat{\sigma}_{y}^{'}u_z+\widehat{\sigma}_{z}^{'}u_y)\right],\\
\widehat{g}_7&=&\frac{\omega_7}{4}\left[-(\widehat{\overline{\eta}}_{xx}^{'}-\widehat{\overline{\eta}}_{yy}^{'})+\rho (u_x^2-u_y^2) +(\widehat{\sigma}_{x}^{'}u_x-\widehat{\sigma}_{y}^{'}u_y)\right],\\
\widehat{g}_8&=&\frac{\omega_8}{12}\left[-(\widehat{\overline{\eta}}_{xx}^{'}+\widehat{\overline{\eta}}_{yy}^{'}-2\widehat{\overline{\eta}}_{zz}^{'})+\rho (u_x^2+u_y^2-2u_z^2)\right.\nonumber\\
                                      &&\left.+(\widehat{\sigma}_{x}^{'}u_x+\widehat{\sigma}_{y}^{'}u_y-2\widehat{\sigma}_{z}^{'}u_z)\right],\\
\widehat{g}_9&=&\frac{\omega_9}{18}\left[-(\widehat{\overline{\eta}}_{xx}^{'}+\widehat{\overline{\eta}}_{yy}^{'}+\widehat{\overline{\eta}}_{zz}^{'})+\rho (u_x^2+u_y^2+u_z^2)\right.\nonumber\\
                                      &&\left.+(\widehat{\sigma}_{x}^{'}u_x+\widehat{\sigma}_{y}^{'}u_y+\widehat{\sigma}_{z}^{'}u_z)+\rho\right],\\
\widehat{g}_{10}&=&\frac{\omega_{10}}{16}\left[-\widehat{\overline{\eta}}_{xyy}^{'}+2u_y\widehat{\overline{\eta}}_{xy}^{'}+u_x\widehat{\overline{\eta}}_{yy}^{'}-2\rho u_xu_y^2-\frac{1}{2}\widehat{\sigma}_{x}^{'}u_y^2-\widehat{\sigma}_{y}^{'}u_yu_x\right]\nonumber\\
                                      &&+u_y\widehat{g}_4+\frac{1}{8}u_x(-\widehat{g}_7+\widehat{g}_8+3\widehat{g}_9),\\
\widehat{g}_{11}&=&\frac{\omega_{11}}{16}\left[-\widehat{\overline{\eta}}_{xxy}^{'}+2u_x\widehat{\overline{\eta}}_{xy}^{'}+u_y\widehat{\overline{\eta}}_{xx}^{'}-2\rho u_x^2u_y-\frac{1}{2}\widehat{\sigma}_{y}^{'}u_x^2-\widehat{\sigma}_{x}^{'}u_xu_y\right]\nonumber\\
                                      &&+u_x\widehat{g}_4+\frac{1}{8}u_y(\widehat{g}_7+\widehat{g}_8+3\widehat{g}_9),\\
\widehat{g}_{12}&=&\frac{\omega_{12}}{16}\left[-\widehat{\overline{\eta}}_{xxz}^{'}+2u_x\widehat{\overline{\eta}}_{xz}^{'}+u_z\widehat{\overline{\eta}}_{xx}^{'}-2\rho u_x^2u_z-\frac{1}{2}\widehat{\sigma}_{z}^{'}u_x^2-\widehat{\sigma}_{x}^{'}u_xu_z\right]\nonumber\\
                                      &&+u_x\widehat{g}_5+\frac{1}{8}u_z(\widehat{g}_7+\widehat{g}_8+3\widehat{g}_9),\\
\widehat{g}_{13}&=&\frac{\omega_{13}}{8}\left[-\widehat{\overline{\eta}}_{xyz}^{'}+u_x\widehat{\overline{\eta}}_{yz}^{'}+u_y\widehat{\overline{\eta}}_{xz}^{'}+u_z\widehat{\overline{\eta}}_{xy}^{'}-2\rho u_xu_yu_z-\frac{1}{2}\left(\widehat{\sigma}_{x}^{'}u_yu_z\right.\right.\nonumber\\
                 &&\left.\left.+\widehat{\sigma}_{y}^{'}u_xu_z+\widehat{\sigma}_{z}^{'}u_xu_y\right)\right]
                    +u_z\widehat{g}_4+u_y\widehat{g}_5+u_x\widehat{g}_6,\\
\widehat{g}_{14}&=&\frac{\omega_{14}}{16}\left[-\widehat{\overline{\eta}}_{xxyy}^{'}+2u_x\widehat{\overline{\eta}}_{xyy}^{'}+2u_y\widehat{\overline{\eta}}_{xxy}^{'}
                    -u_x^2\widehat{\overline{\eta}}_{yy}^{'}-u_y^2\widehat{\overline{\eta}}_{xx}^{'}-4u_xu_y\widehat{\overline{\eta}}_{xy}^{'}\right.\nonumber\\
                    &&\left.+\widetilde{\widehat{\kappa}}_{xx}\widetilde{\widehat{\kappa}}_{yy}+3\rho u_x^2u_y^2+\widehat{\sigma}_{x}^{'}u_xu_y^2+\widehat{\sigma}_{y}^{'}u_yu_x^2\right]-2u_xu_y\widehat{g}_{4}+\frac{1}{8}(u_x^2-u_y^2)\widehat{g}_{7}\nonumber\\
                    &&+\frac{1}{8}(-u_x^2-u_y^2)\widehat{g}_{8}+\left(\frac{3}{8}(-u_x^2-u_y^2)-\frac{1}{2}\right)\widehat{g}_{9}+2u_x\widehat{g}_{10}+2u_y\widehat{g}_{11},
\end{eqnarray}
where $\omega_4,\omega_5,\ldots,\omega_{14}$ are relaxation parameters. Note that similar to the D3Q27 lattice, we have the following relation for the shear viscosity of the fluid $\nu=c_s^2\left(\frac{1}{\omega^\nu}-\frac{1}{2}\right)$, where $\omega^{\nu}=\omega_{j}$ and $j=4,5,6,7,8$. The remaining parameters can be adjusted independently to control numerical stability.

\subsection{Operational Steps}
Finally, by expanding the elements of the matrix multiplication of $\mathcal{K}$ with $\widehat{\mathbf{g}}$ in Eq.~(\ref{eq:cascadecollision1}), the post-collision values of the distribution function augmented by source terms
corresponding to the D3Q15 lattice are
\begin{eqnarray}
\widetilde{\overline{f}}_{0}&=&\overline{f}_{0}+\left[\widehat{g}_0-2\widehat{g}_9+32\widehat{g}_{14}\right]+S_0, \nonumber\\
\widetilde{\overline{f}}_{1}&=&\overline{f}_{1}+\left[\widehat{g}_0+\widehat{g}_1+\widehat{g}_{7}+\widehat{g}_{8}-\widehat{g}_{9}-8\widehat{g}_{10}-8\widehat{g}_{14}\right]+S_1,\nonumber\\
\widetilde{\overline{f}}_{2}&=&\overline{f}_{2}+\left[\widehat{g}_0-\widehat{g}_1+\widehat{g}_{7}+\widehat{g}_{8}-\widehat{g}_{9}+8\widehat{g}_{10}-8\widehat{g}_{14}\right]+S_2,\nonumber\\
\widetilde{\overline{f}}_{3}&=&\overline{f}_{3}+\left[\widehat{g}_0+\widehat{g}_2-\widehat{g}_{7}+\widehat{g}_{8}-\widehat{g}_{9}-8\widehat{g}_{11}-8\widehat{g}_{14}\right]+S_3,\nonumber\\
\widetilde{\overline{f}}_{4}&=&\overline{f}_{4}+\left[\widehat{g}_0-\widehat{g}_2-\widehat{g}_{7}+\widehat{g}_{8}-\widehat{g}_{9}+8\widehat{g}_{11}-8\widehat{g}_{14}\right]+S_4,\nonumber\\
\widetilde{\overline{f}}_{5}&=&\overline{f}_{5}+\left[\widehat{g}_0+\widehat{g}_3-2\widehat{g}_{8}-\widehat{g}_{9}-8\widehat{g}_{12}-8\widehat{g}_{14}\right]+S_5,\nonumber\\
\widetilde{\overline{f}}_{6}&=&\overline{f}_{6}+\left[\widehat{g}_0-\widehat{g}_3-2\widehat{g}_{8}-\widehat{g}_{9}+8\widehat{g}_{12}-8\widehat{g}_{14}\right]+S_6,\nonumber\\
\widetilde{\overline{f}}_{7}&=&\overline{f}_{7}+\left[\widehat{g}_0+\widehat{g}_1+\widehat{g}_2+\widehat{g}_3+\widehat{g}_4+\widehat{g}_5+\widehat{g}_6+\widehat{g}_9+2\widehat{g}_{10}+2\widehat{g}_{11}+2\widehat{g}_{12}\right.\nonumber\\
                                &&\left.+\widehat{g}_{13}+2\widehat{g}_{14}\right]+S_7,\nonumber\\
\widetilde{\overline{f}}_{8}&=&\overline{f}_{8}+\left[\widehat{g}_0-\widehat{g}_1+\widehat{g}_2+\widehat{g}_3-\widehat{g}_4-\widehat{g}_5+\widehat{g}_6+\widehat{g}_9-2\widehat{g}_{10}+2\widehat{g}_{11}+2\widehat{g}_{12}\right.\nonumber\\
                                &&\left.-\widehat{g}_{13}+2\widehat{g}_{14}\right]+S_8,\nonumber\\
\widetilde{\overline{f}}_{9}&=&\overline{f}_{9}+\left[\widehat{g}_0+\widehat{g}_1-\widehat{g}_2+\widehat{g}_3-\widehat{g}_4+\widehat{g}_5-\widehat{g}_6+\widehat{g}_9+2\widehat{g}_{10}-2\widehat{g}_{11}+2\widehat{g}_{12}\right.\nonumber\\
                                &&\left.-\widehat{g}_{13}+2\widehat{g}_{14}\right]+S_9,\nonumber\\
\widetilde{\overline{f}}_{10}&=&\overline{f}_{10}+\left[\widehat{g}_0-\widehat{g}_1-\widehat{g}_2+\widehat{g}_3+\widehat{g}_4-\widehat{g}_5-\widehat{g}_6+\widehat{g}_9-2\widehat{g}_{10}-2\widehat{g}_{11}+2\widehat{g}_{12}\right.\nonumber\\
                                &&\left.+\widehat{g}_{13}+2\widehat{g}_{14}\right]+S_{10},\nonumber\\
\widetilde{\overline{f}}_{11}&=&\overline{f}_{11}+\left[\widehat{g}_0+\widehat{g}_1+\widehat{g}_2-\widehat{g}_3+\widehat{g}_4-\widehat{g}_5-\widehat{g}_6+\widehat{g}_9+2\widehat{g}_{10}+2\widehat{g}_{11}-2\widehat{g}_{12}\right.\nonumber\\
                                &&\left.-\widehat{g}_{13}+2\widehat{g}_{14}\right]+S_{11},\nonumber\\
\widetilde{\overline{f}}_{12}&=&\overline{f}_{12}+\left[\widehat{g}_0-\widehat{g}_1+\widehat{g}_2-\widehat{g}_3-\widehat{g}_4+\widehat{g}_5-\widehat{g}_6+\widehat{g}_9-2\widehat{g}_{10}+2\widehat{g}_{11}-2\widehat{g}_{12}\right.\nonumber\\
                                &&\left.+\widehat{g}_{13}+2\widehat{g}_{14}\right]+S_{12},\nonumber\\
\widetilde{\overline{f}}_{13}&=&\overline{f}_{13}+\left[\widehat{g}_0+\widehat{g}_1-\widehat{g}_2-\widehat{g}_3-\widehat{g}_4-\widehat{g}_5+\widehat{g}_6+\widehat{g}_9+2\widehat{g}_{10}-2\widehat{g}_{11}-2\widehat{g}_{12}\right.\nonumber\\
                                &&\left.+\widehat{g}_{13}+2\widehat{g}_{14}\right]+S_{13},\nonumber\\
\widetilde{\overline{f}}_{14}&=&\overline{f}_{14}+\left[\widehat{g}_0-\widehat{g}_1-\widehat{g}_2-\widehat{g}_3+\widehat{g}_4+\widehat{g}_5+\widehat{g}_6+\widehat{g}_9-2\widehat{g}_{10}-2\widehat{g}_{11}-2\widehat{g}_{12}\right.\nonumber\\
                                &&\left.-\widehat{g}_{13}+2\widehat{g}_{14}\right]+S_{14}.
\end{eqnarray}

\section*{Acknowledgments}
The authors wish to thank the anonymous referees for their helpful comments. This research was in part supported by the National Science Foundation through Teragrid resources provided by abe-queenbee-steele.teragrid sites under grant number TG-CTS 110023.


\begin{thebibliography}{00}
\bibitem{broadwell64}
J.E. Broadwell, {Study of Rarefied Shear Flow by the Discrete Velocity Method,}
{J.\ Fluid Mech.} {\bf 19} (1964) 401.

\bibitem{frisch86}
U. Frisch, B. Hasslacher and Y. Pomeau, {Lattice-Gas Automata for the Navier-Stokes Equation,}
{Phys.\ Rev.\ Lett.} {\bf 56} (1986) 1505.

\bibitem{mcnamara88}
G.R. McNamara and G. Zanetti, {Use of the Boltzmann Equation to Simulate Lattice-Gas Automata,}
{Phys.\ Rev.\ Lett.} {\bf 61} (1988) 2332.

\bibitem{abe97}
T. Abe, {Derivation of the Lattice Boltzmann Method by means of the
discrete ordinate method for the Boltzmann equation,} {J.\ Comp.\
Phys.} {\bf 131} (1997) 241.

\bibitem{he97}
X. He and L.S. Luo, {A priori derivation of the lattice Boltzmann
equation,} {Phys.\ Rev.\ E} {\bf 55} (1997) 115.

\bibitem{shan98}
X. Shan and X. He, {Discretization of the Velocity Space in the Solution of the Boltzmann Equation,} {Phys.\ Rev.\
Lett.} {\bf 80} (1998) 65.

\bibitem{luo00}
L.-S. Luo, {Theory of the lattice Boltzmann method: Lattice Boltzmann
models for nonideal gases,} {Phys.\ Rev.\ E} {\bf 62} (2000) 4982.

\bibitem{he02}
X. He and G. Doolen, {Thermodynamic Foundations of Kinetic Theory
and Lattice Boltzmann Models for Multiphase Flows,} {J.\ Stat.\
Phys.} {\bf 107} (2002) 1572.

\bibitem{asinari08}
P. Asinari and L.-S. Luo, {A Consistent Lattice Boltzmann Equation with Baroclinic Coupling for Mixtures,}
{J.\ Comput.\ Phys.} {\bf 227} (2008) 3878.

\bibitem{shan06}
X. Shan, X.F. Yuan and H. Chen, {Kinetic Theory Representation of Hydrodynamics: A Way Beyond the Navier-Stokes Equation,}
{J.\ Fluid Mech.} {\bf 550} (2006) 413.

%\bibitem{junk05}
%M. Junk, A. Klar and L.S. Luo, {Asymptotic Analysis of the Lattice
%Boltzmann Equation,} {J.\ Comp.\ Phys.} {\bf 210} (2005) 676.

\bibitem{benzi92}
R. Benzi, S. Succi and M. Vergassola, {The Lattice Boltzmann Equation: Theory and Applications,}
{Phys.\ Rep.} {\bf 222} (1992) 145.

\bibitem{chen98}
S. Chen and G. Doolen, {Lattice Boltzmann Method for Fluid Flows,}
{Annu.\ Rev.\ Fluid Mech.} {\bf 8} (1998) 2527.

\bibitem{succi01}
S. Succi, {The Lattice Boltzmann Equation for Fluid Dynamics and
Beyond,} {Clarendon Press, Oxford}  (2001).

\bibitem{yu03}
D. Yu, R. Mei, L.-S. Luo and W. Shyy, {Viscous Flow Computations with the Method of Lattice Boltzmann Equation,}
{Prog.\ Aero.\ Sci.} {\bf 39} (2003) 329.

\bibitem{qian92}
Y. Qian, D. d'Humi\`eres and P. Lallemand, {Lattice BGK Models for Navier-Stokes Equation,} {Europhys.\ Lett.} {\bf 17} (1992) 479.

\bibitem{chen92}
H. Chen, S. Chen and W.H. Matthaeus, {Recover of the Navier-Stokes Eqauations using the Lattice-Gas Boltzmann Method,} {Phys.\ Rev.~A} {\bf 45} (1992) 5339.

%\bibitem{bhatnagar54}
%P. Bhatnagar, E. Gross and M. Krook, {A Model for Collision Processes in Gases. I. Small Amplitude Processes in Charged and Neutral One-Component Systems,} {Phys.\ Rev.} {\bf 94} (1954) 511.

\bibitem{dhumieres92}
D. d`Humi{\`e}res, {Generalized Lattice Boltzmann Equations,} {Progress in Aeronautics and Astronautics (Eds. B.D. Shigal and D.P Weaver)} {\bf 159} (1992) 450.

\bibitem{higuera89a}
F. J. Higuera and J. Jim\'{e}nez, {Boltzmann Approach to Lattice Gas Simulations,} {Europhys.\ Lett.} {\bf 9} (1989) 663.

\bibitem{higuera89b}
F. J. Higuera, S. Succi and R. Benzi, {Lattice gas-dynamics with
enhanced collisions,} {Europhys.\ Lett.} {\bf 9} (1989) 345.

\bibitem{grad49}
H. Grad, {On the Kinetic Theory of Rarefied Gases,} {Comm.\ Pure Appl.\
Math.} {\bf 2} (1949) 331.

\bibitem{lallemand00}
P. Lallemand and Li-Shi Luo, {Theory of the Lattice Boltzmann Method: Dispersion, Dissipation, Isotropy, Galilean Invariance, and Stability,} {Phys.\ Rev.\ E} {\bf 61} (2000) 6546.

\bibitem{dhumieres02}
D. d`Humi{\`e}res, I. Ginzburg, M. Krafczyk, P. Lallemand and L.-S. Luo, {Multiple-Relaxation-Time Lattice Boltzmann Models in Three Dimensions,} {Phil.\ Trans.\ R. Soc.\ Lond.\ A} {\bf 360} (2002) 437.

\bibitem{adhikari08}
R. Adhikari and S. Succi, {Duality in Matrix Lattice Boltzmann Models,} {Phys.\ Rev.\ E} {\bf 78} (2008) 066701.

%\bibitem{premnath06}
%K.N. Premnath and J. Abraham, {Three-dimensional Multi-Relaxation
%Time (MRT) Lattice-Boltzmann Models for Multiphase Flow,} {J.\
%Comp.\ Phys.} {\bf 224} (2007) 539.
%
%\bibitem{premnath05a}
%K.N. Premnath and J. Abraham, {Simulations of Binary Drop Collisions
%with a Multiple-Relaxation-Time Lattice-Boltzmann Model,} {Phys.\
%Fluids} {\bf 17} (2005) 122105.
%
%\bibitem{premnath05b}
%K.N. Premnath, M.E. McCracken and J. Abraham, {A Review of Lattice Boltzmann Methods Relevant to Engine Sprays,} {Trans.\ of the SAE: J.\ Engines} {\bf 114} (2005) 929.
%
%\bibitem{premnath09a}
%K.N. Premnath, M.J. Pattison and S. Banerjee, {Generalized Lattice Boltzmann Equation with Forcing Term for Computation of Wall Bounded Turbulent Flows,} {Phys.\ Rev.\ E} {\bf 79} (2009) 026703.
%
%\bibitem{premnath09b}
%K.N. Premnath, M.J. Pattison and S. Banerjee, {Dynamic Subgrid Scale Modeling of Turbulent Flows using Lattice-Boltzmann Method,} {Physica A} {\bf 388} (2009) 2640.
%
%\bibitem{pattison08}
%M.J. Pattison, K.N. Premnath, N.B. Morley and M. Abdou, {Progress in
%Lattice Boltzmann Methods for Magnetohydrodynamic Flows Relevant to
%Fusion Applications,} {Fusion Engg.\ Des.} {\bf 83} (2008) 557.

%\bibitem{shan07}
%X. Shan and H. Chen, {A General Multiple-Relaxation Boltzmann Collision Model,} {Int.\ J.\ Mod.\ Phys. C} {\bf 18} (2007) 635.

\bibitem{karlin99}
I.V. Karlin, A. Ferrante H.C.\"{O}ttinger, {Perfect Entropy Functions of the Lattice Boltzmann Method,}
{Euro.\ Phys.\ Lett} {\bf 47} (1999) 182.

%\bibitem{boghosian01}
%B.M. Boghosian, J. Yepez, P.V. Coveney and A. Wagner, {Entropic Lattice Boltzmann Methods,}
%{Proc.\ R.\ Soc.\ Lond.\ A} {\bf 457} (2001) 717.

\bibitem{ansumali02}
S. Ansumali and I.V. Karlin, {Single Relaxation Time Model for Entropic Lattice Boltzmann Methods,}
{Phys.\ Rev.\ E} {\bf 65} (2002) 056312.

\bibitem{succi02}
S. Succi, I. Karlin and H. Chen, {Role of the H theorem in Lattice
Boltzmann Hydrodynamic Simulations,} {Rev.\ Mod.\ Phys.} {\bf 74}
(2002) 1203.

\bibitem{boghosian03}
B.M. Boghosian, P.J. Love, P.V. Coveney, I.V. Karlin, S. Succi  and J. Yepez, {Galilean-Invariant Lattice-Boltzmann Models with H Theorem,} {Phys.\ Rev.\ E} {\bf 68}
(2003) 025103(R).

\bibitem{asinari09}
P. Asinari and I.V. Karlin, {Generalized Maxwell State and H Theorem for Computing Fluid Flows using the Lattice Boltzmann Method,}
{Phys.\ Rev.\ E} {\bf 79} (2009) 036703.

\bibitem{asinari10}
P. Asinari and I.V. Karlin, {Quasiequilibrium Lattice Boltzmann Models with Tunable Bulk Viscosity for Enhancing Stability,}
{Phys.\ Rev.\ E} {\bf 81} (2010) 016702.

%\bibitem{chikatamarla09}
%S.S. Chikatamarla and I.V. Karlin, {Lattices for the Lattice Boltzmann Method,}
%{Phys.\ Rev.\ E} {\bf 79} (2009) 046701.
%
%\bibitem{dubois09}
%F. Dubois and P. Lallemand, {Towards Higher Order Lattice Boltzmann Schemes,}
%{J.\ Stat.\ Mech.: Theory and Experiment} {\bf P06006} (2009) 1.

\bibitem{karlin10}
I.V. Karlin and P. Asinari, {Factorization Symmetry in the Lattice Boltzmann Method,}
{Physica A} {\bf 389} (2010) 1530.

%\bibitem{rubinstein08}
%R. Rubinstein and L.-S. Luo, {Theory of the Lattice Boltzmann Equation: Symmetry Properties of Discrete Velocity Sets,}
%{Phys.\ Rev.\ E} {\bf 77} (2008) 036709.
%
%\bibitem{chen08}
%H. Chen and I. Goldhirsch and S.A. Orszag, {Discrete Rotational Symmetry, Moment Isotropy and Higher Order Lattice Boltzmann Models,}
%{J.\ Sci.\ Comput.} {\bf 34} (2008) 87.
%
%\bibitem{banda06}
%M.K. Banda and W.-A. Yong and A. Klar, {A Stability Notion for Lattice Boltzmann Equations,}
%{SIAM\ J.\ Sci.\ Comput.} {\bf 27} (2006) 2098.
%
%\bibitem{junk09a}
%M. Junk and W.-A. Yong, {Weighted $\mathbb{L}^2$-Stability of the Lattice Boltzmann Method,}
%{SIAM\ J.\ Numer.\ Anal.} {\bf 47} (2009) 1651.

%\bibitem{rheinlander09}
%M. Rheinl\"{a}nder, {On the Stability Structure for Lattice Boltzmann Schemes,}
%{Comp.\ Math.\ Appl.} {\bf 59} (2010) 2150.

%\bibitem{yong09}
%W.-A. Yong, {Onsager-like Relation for the Lattice Boltzmann Method,}
%{Comp.\ Math.\ Appl.} {\bf 58} (2009) 862.
%
%\bibitem{junk09b}
%M. Junk and Z. Yang, {Convergence of Lattice Boltzmann Methods for Navier-Stokes Flows in Periodic and Bounded Domains,}
%{Numer.\ Math.} {\bf 112} (2009) 65.

\bibitem{chen08}
H. Chen and X. Shan, {Fundamental Conditions for N-th-order Accurate Lattice Boltzmann Models,}
{Physica D} {\bf 237} (2008) 2003.

\bibitem{chikatamarla09}
S.S. Chikatamarla and I.V. Karlin, {Lattices for the Lattice Boltzmann Method,}
{Phys.\ Rev.\ E} {\bf 79} (2009) 046701.

\bibitem{shan10}
X. Shan, {General Solution of Lattices for Cartesian Lattice Bhatnagar-Gross-Krook Models,}
{Phys.\ Rev.\ E} {\bf 81} (2010) 036702.

%\bibitem{chikatamarla10}
%S.S. Chikatamarla, C.E. Frouzakis, I.V. Karlin, A.G. Tomboulides and K.B. Boulouchos, {Lattice Boltzmann Method for Direct Numerical Simulation of Turbulent Flows,} {J.\ Fluid Mech.} {\bf 656} (2010) 298.

\bibitem{qian98}
Y.-H. Qian and Y. Zhou, {Complete Galilean-Invariant Lattice BGK Models for the Navier-Stokes Equation,} {Euro.\ Phys.\ Lett.} {\bf 42} (1998) 359.

\bibitem{rubinstein08}
R. Rubinstein and L.-S. Luo, {Theory of the Lattice Boltzmann Equation: Symmetry Properties of Discrete Velocity Sets,}
{Phys.\ Rev.\ E} {\bf 77} (2008) 036709.

\bibitem{geier06}
M. Geier, A. Greiner and J.G. Korvink, {Cascaded Digital Lattice Boltzmann Automata for High Reynolds Number Flow,}
{Phys.\ Rev.\ E} {\bf 73} (2006) 066705.

%\bibitem{struchtrup05}
%H. Struchtrup, {Macroscopic Transport Equations for Rarefied Gas Flows: Approximation Methods in Kinetic Theory,} {Springer, New York} (2005).

\bibitem{geier07}
M. Geier, A. Greiner and J.G. Korvink, {Properties of the Cascaded Lattice Boltzmann Automaton,}
{Int.\ J.\ Mod.\ Phys.\ C} {\bf 18} (2007) 455.

%\bibitem{geier08}
%M. Geier, {De-aliasing and Stabilization Formalism of the Cascaded Lattice Boltzmann Automaton for Under-Resolved High Reynolds Number Flow,} {Int.\ J.\ Num.\ Meth.\ Fluids} {\bf 56} (2008) 1249.

\bibitem{asinari08a}
P. Asinari, {Generalized Local Equilibrium in the Cascaded Lattice Boltzmann Method,}
{Phys.\ Rev.\ E} {\bf 78} (2008) 016701.

\bibitem{geier09}
M. Geier, A. Greiner and J.G. Korvink, {A Factorized Central Moment Lattice Boltzmann Method,} {Eur.\ Phys.\ J.\ Special Topics} {\bf 171} (2009) 55.

%\bibitem{asinari08a}
%P. Asinari, {Generalized Local Equilibrium in the Cascaded Lattice Boltzmann Method,}
%{Phys.\ Rev.\ E} {\bf 78} (2008) 016701.

\bibitem{premnath09d}
K.N. Premnath and S. Banerjee, {Incorporating Forcing Terms in Cascaded Lattice Boltzmann Approach by Method of Central Moments,}
{Phys.\ Rev.\ E} {\bf 80} (2009) 036702.

\bibitem{white05}
F.M. White, {Viscous Fluid Flow,} {McGraw Hill, New York}  (2005).

%\bibitem{chapman64}
%S. Chapman and T.G. Cowling, {Mathematical Theory of Nonuniform
%Gases,} {Cambridge University Press, New York} (1964).
%
%\bibitem{chen98a}
%H. Chen, S. Succi and S. Orszag, {Analysis of Subgrid Scale Turbulence using the Boltzmann Bhatnagar-Gross-Krook Kinetic Equation,} {Phys.\ Rev.\ E} {\bf 59} (1998) R2527.
%
%\bibitem{ansumali04}
%S. Ansumali, I. Karlin and S. Succi, {Kinetic Theory of Turbulence Modelling: Smallness Parameter, Scaling and Microscopic Derivation of Smagorinsky Model,} {Physica A} {\bf 338} (2004) 379.
%
%\bibitem{chen04}
%H. Chen, S. Orszag and I. Staroselsky, {Expanded Analogy between Boltzmann Kinetic Theory of Fluids and Turbulence,} {J.\ Fluid Mech.} {\bf 519} (2004) 301.
%
%\bibitem{teixeira98}
%C. Teixeira, {Incorporating Turbulence Models into the Lattice-Boltzmann Method,} {Int.\ J.\ Mod.\ Phys.\ C} {\bf 9} (1998) 1159.
%
%\bibitem{chen03}
%H. Chen, S. Kandasamy, S. Orszag, R. Shock, S. Succi and V. Yakhot, {Extended Boltzmann Kinetic Equation for Turbulent Flows,} {Science} {\bf 301} (2003) 633.
%
%\bibitem{hou96}
%S. Hou, J. Sterling, S. Chen and G.D. Doolen, {A Lattice Boltzmann Subgrid Model for High Reynolds Number Flows,}
%{Fields Inst.\ Comm.} {\bf 6} (1996) 151.
%
%\bibitem{yu06}
%H. Yu, L.-S. and S. Girimaji, {LES of Turbulent Square Jet Flow using an MRT Lattice Boltzmann Model,} {Comput.\ Fluids} {\bf 35} (2006) 957.
%
%\bibitem{dong08}
%Y.H. Dong, P. Sagaut and S. Marie, {Inertial Consistent Subgrid Model for Large-Eddy Simulation Based on the Lattice Boltzmann Method,} {Phys.\ Fluids}  {\bf 20} (2008) 035104.
%
%\bibitem{sagaut10}
%P. Sagaut, {Toward Advanced Subgrid Models for Lattice-Boltzmann-Based Large-Eddy Simulation: Theoretical Formulations,}
%{Comp.\ Math.\ Appl.} {\bf 59} (2010) 2194.
%
%\bibitem{girimaji07}
%S. Girimaji, {Boltzmann Kinetic Equation for Filtered Fluid Turbulence,} {Phys.\ Rev.\ Lett.}  {\bf 99} (2007) 034501.
%
%\bibitem{stolz99}
%S. Stolz and N.A. Adams, {An Approximate Deconvolution Procedure for Large-Eddy Simulation,} {Phys.\ Fluids} {\bf 11} (1999) 1699.
%
%\bibitem{speziale85}
%C.G. Speziale, {Galilean Invariance of Subgrid-Scale Stress Models in the Large-Eddy Simulation of Turbulence,}
%{J.\ Fluid Mech.} {\bf 156} (1985) 55.
%
%\bibitem{germano86}
%M. Germano, {A Proposal for a Redefinition of the Turbulent Stresses in the Filtered Navier-Stokes Equations,}
%{Phys.\ Fluids} {\bf 29} (1986) 2323.
%
%\bibitem{fureby97}
%C. Fureby and G. Tabor, {Mathematical and Physical Constraints on Large-Eddy Simulations,}
%{Theoret.\ Comput.\ Fluid Dynamics} {\bf 9} (1997) 85.
%
%\bibitem{monin07}
%A.S. Monin and A.M. Yaglom, {Statistical Fluid Mechanics, Vol I: Mechanics of Turbulence,} {Dover Publications} (2007).
%
%\bibitem{germano92}
%M. Germano, {Turbulence: The Filtering Approach,} {J.\ Fluid Mech.}  {\bf 238} (1992) 325.
%
%\bibitem{harris99}
%S. Harris, {An Introduction to the Theory of the Boltzmann Equation,} {Dover Publications, New York} (1999).
%
%\bibitem{kogan69}
%M.N. Kogan, {Rarefied Gas Dynamics,} {Plenum Press, New York} (1969).
%
%\bibitem{muller93}
%I. M\"{u}ller and T. Ruggeri, {Extended Thermodynamics,} {Springer Tracts in Natural Philosophy, New York} (1993).
%
%\bibitem{he97a}
%X. He and L.-S. Luo, {Lattice Boltzmann Model for the Incompressible Navier-Stokes Equation,} {J.\ Stat.\ Phys.} {\bf 88} (1997) 927.
%
%\bibitem{maxwell60}
%J.C. Maxwell, {Illustrations of the Dynamical Theory of Gases,}
%{Phil.\ Mag.} {\bf 19} (1860) 19.
%
%\bibitem{reynolds95}
%O. Reynolds, {On the Dynamical Theory of Incompressible Viscous Fluids and the Determination of the Criterion,}
%{Phil.\ Trans.\ Roy.\ Soc.\ London A} {\bf 186} (1895) 123.
%
%\bibitem{prandtl25}
%L. Prandtl, {Uber die Ausgebildete Turbulenz,}
%{ZAMM} {\bf 5} (1925) 136.

%\bibitem{pope00}
%S.B. Pope, {Turbulent Flows,} {Cambridge University Press}  (2000).
%
%\bibitem{sone02}
%Y. Sone, {Kinetic Theory and Fluid Dynamics,} {Birkh\"{a}user, Boston} (2002).

%\bibitem{chikatamarla10}
%S.S. Chikatamarla, C.E. Frouzakis, I.V. Karlin, A.G. Tomboulides and K.B. Boulouchos, {Lattice Boltzmann Method for Direct Numerical Simulation of Turbulent Flows,} {J.\ Fluid Mech.} {\bf 656} (2010) 298.

%\bibitem{qian93}
%Y.-H. Qian and S.A. Orszag, {Lattice BGK Models for the Navier-Stokes Equation: Nonlinear Deviation in Compressible Regimes,} {Euro.\ Phys.\ Lett.} {\bf 21} (1993) 255.

%\bibitem{qian98}
%Y.-H. Qian and Y. Zhou, {Complete Galilean-Invariant Lattice BGK Models for the Navier-Stokes Equation,} {Euro.\ Phys.\ Lett.} {\bf 42} (1998) 359.

\end{thebibliography}
\end{document}